\documentclass[longauth]{aa}  

\usepackage{graphicx}
\usepackage{txfonts}
\usepackage{hyperref}
\hypersetup{
    colorlinks=true,
    linkcolor=blue,
    citecolor=blue,
    filecolor=magenta,      
    urlcolor=cyan,
    }
\usepackage{textgreek}
\usepackage{pdflscape}
\usepackage{booktabs}
\usepackage{amsmath}
\usepackage{caption}
\usepackage{placeins}
\usepackage{capt-of}

\newcommand{\lya}{Ly\textalpha}
\newcommand{\ha}{H\textalpha}
\newcommand{\hb}{H\textbeta}
\newcommand{\hg}{H\textgamma}
\newcommand{\oiii}{[O\,\textsc{iii}]}
\newcommand{\oii}{[O\,\textsc{ii}]}

\newcommand{\xiion}{$\xi_\mathrm{ion}$}
\newcommand{\xiioncorr}{$\xi_\mathrm{ion}^\mathrm{corr}$}
\newcommand{\fesc}{$f_\mathrm{esc}$}
\newcommand{\muv}{$M_\mathrm{UV}$}

\newcommand{\flux}{erg\,s$^{-1}$\,cm$^{-2}$}

\newcommand{\kms}{km\,s$^{-1}$}

\makeatletter
\renewcommand*\aa@pageof{, page \thepage{} of \pageref*{LastPage}}
\makeatother

\begin{document} 

\title{JADES: The production and escape of ionizing photons from faint Lyman-alpha emitters in the epoch of reionization}
\titlerunning{Faint Ly$\alpha$ emitters at $z\gtrsim6$}

\author{
Aayush Saxena\inst{1,2}\thanks{E-mail: aayush.saxena@physics.ox.ac.uk}
\and Andrew J. Bunker\inst{1}
\and Gareth C. Jones\inst{1}
\and Daniel P. Stark\inst{3}
\and Alex J. Cameron\inst{1}
\and Joris Witstok\inst{4,5}
\and Santiago Arribas\inst{6}
\and William M. Baker\inst{4,5}
\and Stefi Baum\inst{7}
\and Rachana Bhatawdekar\inst{8,9}
\and Rebecca Bowler\inst{10}
\and Kristan Boyett\inst{11,12}
\and Stefano Carniani\inst{13}
\and Stephane Charlot\inst{14}
\and Jacopo Chevallard\inst{1}
\and Mirko Curti\inst{15,4,5}
\and Emma Curtis-Lake\inst{16}
\and Daniel J. Eisenstein\inst{17}
\and Ryan Endsley\inst{18}
\and Kevin Hainline\inst{3}
\and Jakob M. Helton\inst{3}
\and Benjamin D. Johnson\inst{17}
\and Nimisha Kumari\inst{19}
\and Tobias J. Looser\inst{4,5}
\and Roberto Maiolino\inst{4,5,2}
\and Marcia Rieke\inst{3}
\and Hans-Walter Rix\inst{20}
\and Brant E. Robertson\inst{21}
\and Lester Sandles\inst{4,5}
\and Charlotte Simmonds\inst{4,5}
\and Renske Smit\inst{22}
\and Sandro Tacchella\inst{4,5}
\and Christina C. Williams\inst{23}
\and Christopher N. A. Willmer\inst{3}
\and Chris Willott\inst{24}
}

\institute{
Department of Physics, University of Oxford, Denys Wilkinson Building, Keble Road, Oxford OX1 3RH, UK
\and Department of Physics and Astronomy, University College London, Gower Street, London WC1E 6BT, UK
\and Steward Observatory, University of Arizona, 933 N. Cherry Ave., Tucson, AZ 85721, USA
\and Kavli Institute for Cosmology, University of Cambridge, Madingley Road, Cambridge CB3 0HA, UK
\and Cavendish Laboratory, University of Cambridge, 19 JJ Thomson Avenue, Cambridge CB3 0HE, UK
\and Centro de Astrobiolog\'{i}a (CAB), CSIC-INTA, Cra. de Ajalvir Km.~4, 28850 Torrej\'{o}n de Ardoz, Madrid, Spain
\and Department of Physics and Astronomy, University of Manitoba, Winnipeg, MB R3T 2N2, Canada
\and European Space Agency (ESA), European Space Astronomy Centre (ESAC), Camino Bajo del Castillo s/n, 28692 Villanueva de la Cañada, Madrid, Spain
\and European Space Agency, ESA/ESTEC, Keplerlaan 1, 2201 AZ Noordwijk, NL
\and Jodrell Bank Centre for Astrophysics, Department of Physics and Astronomy, School of Natural Sciences, The University of Manchester, Manchester, M13 9PL, UK
\and School of Physics, University of Melbourne, Parkville 3010, VIC, Australia
\and ARC Centre of Excellence for All Sky Astrophysics in 3 Dimensions (ASTRO 3D), Australia
\and Scuola Normale Superiore, Piazza dei Cavalieri 7, I-56126 Pisa, Italy
\and Sorbonne Universit\'e, CNRS, UMR 7095, Institut d'Astrophysique de Paris, 98 bis bd Arago, 75014 Paris, France
\and European Southern Observatory, Karl-Schwarzschild-Strasse 2, D-85748 Garching bei Muenchen, Germany
\and Centre for Astrophysics Research, Department of Physics, Astronomy and Mathematics, University of Hertfordshire, Hatfield AL10 9AB, UK
\and centre for Astrophysics $|$ Harvard \& Smithsonian, 60 Garden St., Cambridge MA 02138 USA
\and Department of Astronomy, University of Texas, Austin, TX 78712, USA
\and AURA for European Space Agency, Space Telescope Science Institute, 3700 San Martin Drive. Baltimore, MD, 21210
\and Max-Planck-Institut f\"ur Astronomie, K\"onigstuhl 17, D-69117, Heidelberg, Germany
\and Department of Astronomy and Astrophysics, University of California, Santa Cruz, 1156 High Street, Santa Cruz, CA 95064, USA
\and Astrophysics Research Institute, Liverpool John Moores University, 146 Brownlow Hill, Liverpool L3 5RF, UK
\and NSF's National Optical-Infrared Astronomy Research Laboratory, 950 North Cherry Avenue, Tucson, AZ 85719, USA
\and NRC Herzberg, 5071 West Saanich Rd, Victoria, BC V9E 2E7, Canada
}

\authorrunning{A. Saxena et al.}

\date{}


\abstract{We present the properties of 17 faint Lyman-$\alpha$ emitting galaxies (LAEs) at $z>5.8$ from the \emph{JWST} Advanced Deep Extragalactic Survey (JADES) spectroscopic data in the \emph{Hubble} Ultra Deep Field/GOODS-S. These LAEs span a redshift range $z\approx5.8-8.0$ and a UV magnitude range $M_{\textrm{UV}} \approx -17$ to $-20.6$, with the Ly$\alpha$ equivalent width (EW) in the range $\approx 25-350$\,\AA. The detection of other rest-optical emission lines in the spectra of these LAEs enables the determination of accurate systemic redshifts and Ly$\alpha$ velocity offsets, as well as the physical and chemical composition of their stars and interstellar media. These faint LAEs are consistent with metal-poor systems with high ionization parameters, similar to the general galaxy population at $z>6$. We measured an average ionizing photon production efficiency, log($\xi_{\textrm{ion}}$/erg$^{-1}$\,Hz) $\approx 25.57$ across our LAEs, which does not evolve strongly with redshift. We report an anti-correlation between the Ly$\alpha$ escape fraction and the velocity offset from systemic redshift, consistent with model expectations. We further find that the strength and velocity offset of Ly$\alpha$ are neither correlated with galaxy spectroscopic properties nor with $\xi_{\textrm{ion}}$. We find a decrease in Ly$\alpha$ escape fractions with redshift, indicative of decreasing sizes of ionized bubbles around LAEs at high redshifts. We used a range of galaxy properties to predict Lyman continuum escape fractions for our LAEs, finding that the ionizing photon output into the intergalactic medium from our LAEs remains roughly constant across the observed Ly$\alpha$ EW, showing a mild increase at fainter UV magnitudes and at higher redshifts. We derived correlations between the ionizing photon output from LAEs and their UV magnitudes, Ly$\alpha$ strengths and redshifts, which can be used to constrain the ionizing photon contribution of LAEs at $z>6$ towards cosmic reionization.}


\maketitle

\section{Introduction} 
\label{sec:intro}

Cosmic reionization is a crucial phase transition in the Universe's history, and the understanding of which is an important challenge in observational astronomy \citep[see recent review by][]{rob22a}. The emergence of ionizing UV photons from the first structures to form in the Universe began interacting with the neutral intergalactic medium (IGM), gradually ionizing it to near completion by $z\sim6$ \citep[e.g.][]{fan06}, although certain studies have favoured a later end to reionization \citep[e.g.][]{wei19, kea20, bos22}. To quantify the contribution towards the cosmic reionization budget from ionizing photon sources in the early Universe, a good understanding is needed of the space density of sources, the efficiency of hydrogen ionizing Lyman continuum (LyC; $\lambda_0 < 912$\,\AA) photon production, and crucially, the fraction of LyC photons that manage to escape into the IGM \citep[e.g.][]{day18}.

\emph{JWST} spectroscopy has offered ground-breaking insights into the state of the interstellar medium (ISM), chemical enrichment of the gas and stars as well as ionizing photon production in galaxies at $z>6$, pushing towards fainter UV magnitudes than were previously possible from the ground \citep{arr22, end23b,  tac23, tru23, sun23, cam23, curtis23, cur23, fuj23,  kat23b, rob23, san23}. However, accurately measuring the escape fraction of LyC photons (\fesc) becomes hard already at $z>4$ mainly due to the increasing neutrality of the IGM \citep[e.g.][]{ino14}, which efficiently absorbs LyC photons along the line of sight. A further complication is introduced by the fact that no clear dependence between \fesc(LyC) and galaxy properties has been observationally established \citep[e.g.][]{nai18, fle19, nak20, pah21, sax22a} 

Therefore, in order to constrain the all-important escape fraction of ionizing photons from reionization era galaxies, it is of the utmost importance to find reliable indirect indicators of \fesc(LyC). The presence of young, actively forming stars as well as gas and dust-free environments is thought to enable significant \fesc(LyC) from galaxies \citep[e.g.][]{zac13}, and spectroscopic and/or photometric indicators probing such conditions can be explored as indirect indicators of LyC photon escape \citep[e.g.][]{flu22a, flu22b, top22, mas23}. Important insights can also be gained from high-resolution simulations of reionization era galaxies, where a good handle on the escaping LyC radiation can be correlated with prevalent galaxy conditions \citep[e.g.][]{bar20, maj22} that can then be converted into observables \citep[e.g.][]{cho23} and used to predict \fesc\ from galaxies.

Uniquely, galaxies at $z\gtrsim6$ that show strong \lya\ emission in their spectra, typically with equivalent widths (EW)\,$>20$\,\AA; e.g. \citealt{aji03}), also known as \lya\ emitters (or LAEs) can be excellent probes of studying how reionization unfolds over redshifts. The presence of strong \lya\ emission at $z\gtrsim6$ often traces the existence of large ionized bubbles in an otherwise neutral IGM (\citealt{mir98, fur06, rob16, cas22, tra23, tan23, jun23, sax23a}, c.f. \citealt{bun23}), offering direct observational insights into reionized regions of the early Universe. Further, as intrinsic \lya\ luminosities are expected to increase with star formation rates, the fraction of galaxies that appear to be strong LAEs can be an important diagnostic of the ionizing photon production capabilities of reionization era galaxies \citep[e.g.][]{smi19, gar21, mat22} as well as the evolving state of the IGM neutral fraction \citep[e.g.][]{car12, car14, star16, pen18, hoa19, kus20, ful20, jon23}.

Considerable information about the neutral gas and dust content within a galaxy can be gained by the observed strength and emission line profile of \lya\ \citep[e.g.][]{hay23a}. The separation between the blue and red peaks in the emission as well as the offset from systemic redshift in the absence of a double-peaked profile can be used to infer neutral gas densities \citep{ver15, orl18} and dust \citep{hay13}, although it has been shown that the neutral gas distribution may play the more dominant role in controlling \lya\ escape \citep[e.g.][]{ate08}. At $z>6$, both the number density of LAEs \citep{hai02, mal06} and the shape of the \lya\ line originating from star-forming galaxies residing within ionized bubbles can further be used to estimate the size of those bubbles \citep[e.g.][]{mason20, hay23b, witstok23}.

With a plethora of models available to link the observed \lya\ properties to both galaxy properties and the state of the IGM at $z>6$, it is imperative to expand samples of observed LAEs in the reionization era, pushing to fainter magnitudes. Probing \lya\ emission from UV-faint galaxies has the added advantage of providing much tighter constraints on both bubble sizes as well as the ionized fraction of the IGM \citep[e.g.][]{mas18, bol22}. Importantly, \lya\ emission from fainter galaxies can provide additional sightlines from which the impact of galaxy associations on the production efficiency of ionizing photons \citep[e.g.][]{wit23} and their transmission through the IGM \citep[e.g.][]{tra23} can be studied in detail.

Perhaps most importantly, detailed studies of faint LAEs can inform our understanding of the key drivers of cosmic reionization, particularly testing whether compact star-forming galaxies are indeed contributing the bulk of ionizing photons towards the reionization budget \citep[e.g.][]{rob15}, which are often expected to produce large intrinsic \lya\ luminosities \citep[e.g.][]{mat22}. LAEs that have their \lya\ emission peaking close to systemic redshifts are also expected to have high LyC escape fractions \citep{ver15, dij16, nai22}. With signatures of hard radiation fields \citep{stark15a, mai17, fel20, sax22b, roy23} and elevated ionizing photon production efficiencies \citep[e.g.][]{mat17b, har18, nin23, sim23a, sim24} measured from LAEs across redshifts, \lya\ emitting galaxies in the reionization era are exciting laboratories to both test and constrain reionization models. With access to stellar and ISM properties of LAEs at high redshifts thanks to \emph{JWST}, it is now finally possible to study the potential role of LAEs in driving cosmic reionization.

In an attempt to quantify the production and escape of both \lya\ and LyC photons from LAEs in the reionization era, in this study we dramatically increase the number of faint LAEs detected at $z\gtrsim6$ using exquisitely deep spectra from the \emph{JWST} Advanced Deep Extragalactic Survey (JADES; \citealt{eis23}). The main aim of this work is to explore the physical properties of faint LAEs in the reionization era, while also investigating the physical mechanisms within the galaxy that control the visibility of \lya\ emission. We further assess the impact of an increasingly neutral IGM on the emergent \lya\ emission at the highest redshifts. Finally, using all available spectroscopic and photometric information about our faint LAEs, we estimate their ionizing photon contribution towards the global reionization budget. In companion papers, we also measure the LAE fraction \citep{jon23} as well as the size of ionized bubbles around our LAEs and their clustering \citep{witstok23}.

The layout of this paper is as follows: Section \ref{sec:data} describes the \emph{JWST} data used in this study as well as the measurement of key spectroscopic quantities that are used in this study. Section \ref{sec:properties} presents the chemical enrichment and ionization state inferred from the spectra of our LAEs compared with other reionization era galaxies in the literature. Section \ref{sec:lya_escape} explores the mechanisms within galaxies that control the escape of \lya\ photons along the line of sight. Section \ref{sec:reionization} discusses the implications for the reionization of the Universe from these new LAE observations and presents quantities that would help build realistic reionization models. The main conclusions of this study are presented in Section \ref{sec:conclusions}.

Throughout this paper, we use the \citet{planck} cosmology. Magnitudes are in the AB system \citep{oke83} and all distances used are proper distances, unless otherwise stated.

\section{Data and measurements} 
\label{sec:data}

\subsection{NIRSpec data}
The \emph{JWST} observations used in this study are part of JADES, which is a collaboration between the Near-Infrared Camera (NIRCam; \citealt{rie22}) and Near-Infrared Spectrograph (NIRSpec; \citealt{fer22, jak22}) Instrument Science teams with an aim of using over 750 hours of guaranteed time observations (GTO) to study the evolution of galaxies in the Great Observatories Origins Deep Survey (GOODS)-South and GOODS-North fields \citep{gia04}. We describe the NIRSpec and NIRCam observations and data reduction steps below.

Spectroscopic data presented in this work was obtained using the Micro-Shutter Assembly (MSA; \citealt{fer22}) on the NIRSpec instrument on board \emph{JWST}. Two `Tiers' of JADES data was utilized in this study: the Deep Tier NIRSpec observations are part of the GTO program ID: 1210 (PI: L\"utzgendorf) and in GOODS-S centred near the Hubble Ultra Deep Field (HUDF), obtained between 22 October and 25 October 2022 over 3 visits, and the Medium Tier observations are part of GTO program 1180 (PI: Eisenstein) obtained over a larger area in GOODS-S \citep[see][for an overview of the field layout]{eis23}).

For Deep observations, the PRISM/CLEAR setup, which gives wavelength coverage in the range $0.6-5.3$ $\mu$m with a spectral resolution of $R\sim100$ \citep{bok23}, and G140M/F070LP, G235M/F170LP, G395M/F290LP, and G395H/F290LP filter/grating setups were used, whereas for Medium observations all of the above but the G395H/F290LP filter/grating setup were used. For the Deep Tier, three sub-pointings were planned in the same field (although each sub-pointing had minor pointing differences), with each visit having a total of 33.613\,ks of exposure in PRISM/CLEAR and 8.4\,ks of exposure in each of the gratings. The Medium Tier observations were carried out in parallel to NIRCam observations, and therefore, consisted of several single pointings covering a larger sky area, with 3.8\,ks of exposure time in PRISM/CLEAR and 3.1\,ks of exposure time in the gratings per pointing. We note that as the sources targeted were generally high-priority targets owing to their possible high redshift nature, it was possible for one target to be covered over multiple Medium tier pointings. We refer the readers to \citet{bun23b} and \citet{eis23} for further details about the observational setup, strategy and challenges. 

The targets for spectroscopy were selected from existing deep \textit{HST}-based catalogueues as well as JADES NIRCam catalogueues \citep{rie23}. Candidate high redshift galaxies with photometric redshifts $z>5.7$, identified via the classic photometric \lq drop-out' technique \citep[e.g.][]{ste96}, whereby the Lyman break in the spectrum of a galaxy is captured in adjacent broad-band filters, were assigned higher priorities. Full details of the target selection and priority classes can be found in the accompanying paper by Bunker et al. (2023c).

The data reduction was carried out using pipelines developed by the ESA NIRSpec Science Operations Team (SOT) and the NIRSpec GTO Team (\citealt{fer22}, Carniani et al. in prep). Some of the  main data reduction steps implemented by the pipeline are pixel-level background subtraction, pixel-to-pixel flat-field correction, absolute flux calibration, slit-loss correction, and eventually 2-dimensional (2D) and 1-dimensional (1D) spectra extraction and co-addition. In this version of the reduction, the final 1D spectra are not extracted from the 2D spectra, but result from the weighted averaging of 1D spectra from all integrations \citep[see][]{curtis23}. Due to the compact size of our LAEs, slit-loss corrections were applied by modelling it as point-like source. A nominal 3-pixel extraction aperture was used to produce the co-added 1D spectra. A detailed description of the data reduction and spectral extraction methods is given in \citet{bun23b} (but see also \citealt{curtis23} and \citealt{cam23})).

\subsection{Identification of Lyman-alpha emitters}
\lya\ emission in the spectra of galaxies in the parent sample was identified through a combination of template fitting \citep{jon23} of the R100 spectra as well as visual inspection of both the R100 and R1000 (G140M) spectra of all confirmed high-redshift galaxies in the parent sample. Using both these methods, we identified 9 candidate LAEs in Deep and 7 candidate LAEs in Medium at $z>5.8$. We then measure the \lya\ line flux by fitting a single Gaussian function to the emission in both R100 and R1000 spectra. 
\begin{figure*}
    \centering
    \includegraphics[width=\linewidth]{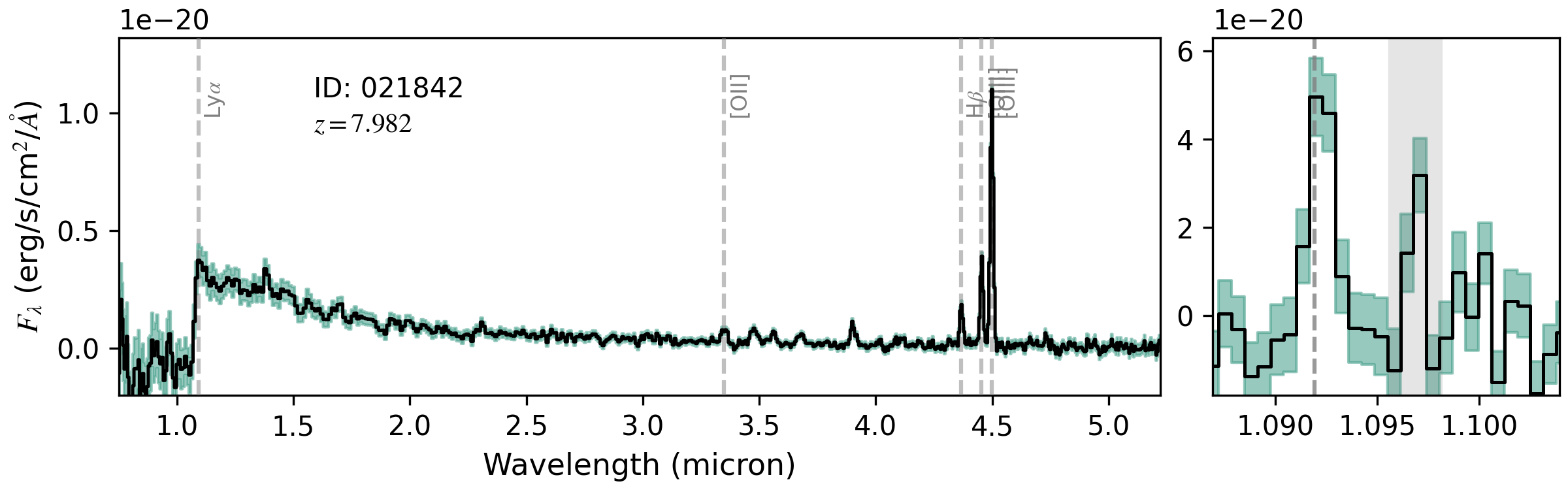}
    \includegraphics[width=\linewidth]{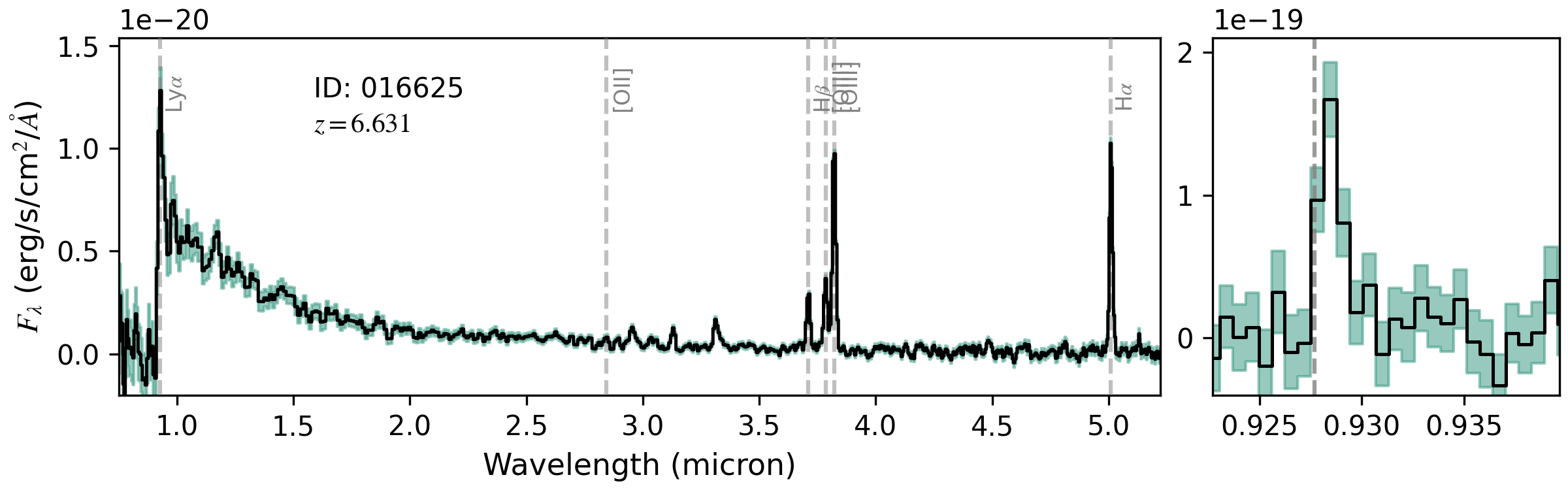}
    \includegraphics[width=\linewidth]{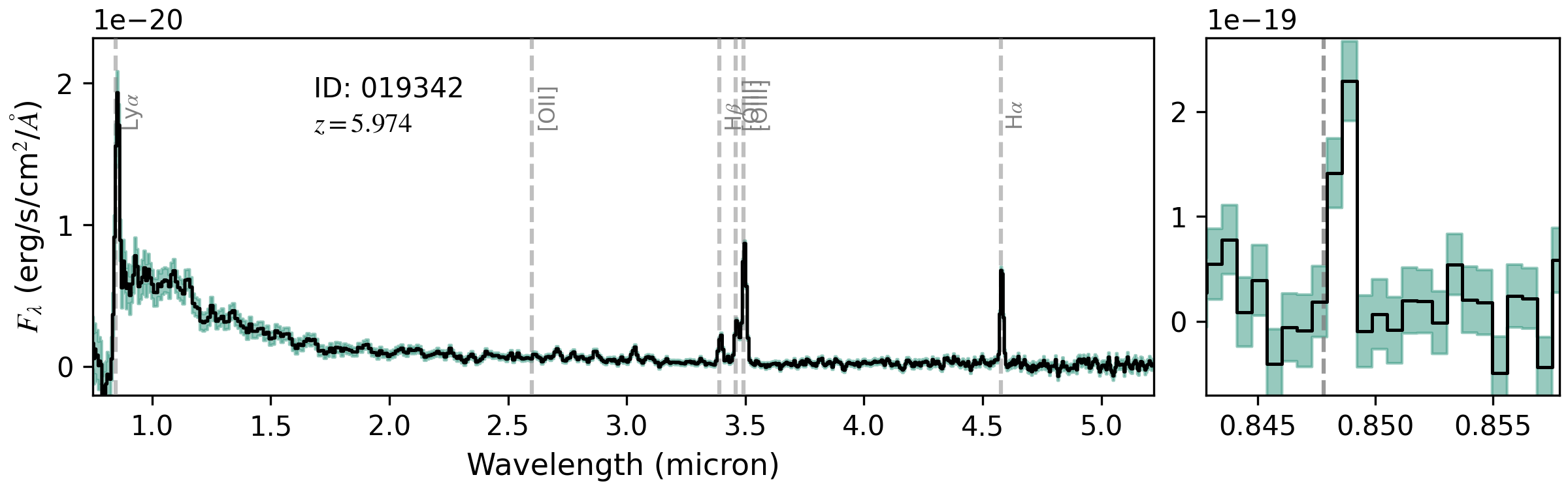}
    \caption{Example spectra showing the PRISM (left) and G140M grating zoom-in on \lya\ emission (right) of faint LAEs selected from the Deep Tier spectra in JADES, with $1\sigma$ noise shown as green shaded region. The \lya\ line sensitivity in the G140M spectra is considerably higher compared to PRISM, where the decreasing spectral resolution at bluer wavelengths results in diminished emission line sensitivity. Therefore, a combination of both PRISM and Grating spectra is key to identifying \lya\ emission.}
    \label{fig:spectra}
\end{figure*}

The \lya\ line in one of the 17 LAEs at $z>5.8$ presented in this work fell in the detector gap in R1000. With the exception of this galaxy, all visually identified LAEs encouragingly showed clear \lya\ emission both in the PRISM and in G140M spectra. In Figure \ref{fig:spectra} we show the full 1D spectrum from PRISM (R100) as well as a zoom-in on the \lya\ emission identified in the G140M grating (R1000) from a selection of LAEs in our sample and the spectra of all LAEs are shown in Appendix \ref{app:a}. In Table \ref{tab:sources} we list the spectroscopic redshifts, the complete JADES Source name/ID and the effective exposure times in both PRISM and Gratings for the LAEs identified in this work.

Overall, we find that the line fluxes we measure from the medium resolution grating are systematically higher than the ones measured from PRISM, as can also be seen in Figure \ref{fig:spectra}, which is not surprising given the degradation in spectral resolution that PRISM spectra suffer from at shorter wavelengths. Therefore, going forward we use \lya\ measurements from the G140M grating, with the exception of one source for which \lya\ was in the detector gap.  

\subsection{Systemic redshifts}
Accurate `systemic' redshifts were measured by identifying strong emission lines in the higher resolution Grating spectra, which generally consisted of \oii, \hb, \oiii\ and \ha. The redshift was derived by fitting single Gaussian functions to the strongest emission lines and using a signal to noise ratio (S/N) weighted combination of the centroids of the fits to obtain the best redshift solution. For the redshift range of our sources, the \hb, \oiii\ and \ha\ lines fell in the G395M grating and the \oii\ line fell in the G235M grating spectra. Vacuum wavelengths for all of these strong rest-frame optical lines were used for redshift determination.  

We found that on average, the difference between the redshifts derived from R100 and R1000 spectra were of the order $\Delta z \sim 0.004$, but the redshifts derived from lines in the medium dispersion gratings were found to be consistent. Therefore, the redshifts that we derive and use further in the study are from the medium dispersion gratings, which also have a much narrower line-spread function (LSF) and are more sensitive to narrow emission lines. The source IDs, JADES source names, redshifts and exposure times are given in Table \ref{tab:sources}. From here on, we use the IDs to refer to the objects presented in this paper. The references for the discovery papers of these targets can be found in \citet{bun23b}.
\begin{table*}
    \centering
    \caption{IDs, redshifts, and exposure times for the \lya\ emitting galaxies identified in this study.}
    \begin{tabular}{l c c c c}
    \toprule
    ID & JADES Source name & $z_{\mathrm{spec}}$ & $T_\mathrm{exp}^\mathrm{PRISM}$ & $T_\mathrm{exp}^\mathrm{Grating}$ \\
     & & & (ks) & (ks) \\
     \midrule
    \emph{Deep Tier} \\
  21842 & JADES-GS$+$53.15682$-$27.76716 & $7.982$ & $100.0$ & $25.0$ \\
  10013682* & JADES-GS$+$53.16746$-$27.77201 & $7.276$ & $66.6$ & $16.7$ \\
  4297 & JADES-GS$+$53.15579$-$27.81520 & $6.712$ & $33.3$ & $8.3$ \\
  16625 & JADES-GS$+$53.16904$-$27.77884 & $6.631$ & $100.0$ & $25.0$ \\
  18846 & JADES-GS$+$53.13492$-$27.77271 & $6.336$ & $100.0$ & $25.0$ \\
  19342 & JADES-GS$+$53.16062$-$27.77161 & $5.974$ & $100.0$ & $25.0$  \\
  9422 & JADES-GS$+$53.12175$-$27.79763 & $5.937$ & $100.0$ & $25.0$  \\
  6002 & JADES-GS$+$53.11041$-$27.80892 & $5.937$ & $100.0$ & $25.0$  \\
  19606 & JADES-GS$+$53.17655$-$27.77111 & $5.889$ & $33.3$ & $8.3$ \\
  10056849 & JADES-GS$+$53.11351$-$27.77284 & $5.814$ & $100.0$ & $8.3$ \\

  \emph{Medium Tier} \\
  12637 & JADES-GS$+$53.13347$-$27.76037 & $7.660$ & $19.0$ & $15.5$ \\
  15362 & JADES-GS$+$53.11634$-$27.76194 & $6.794$ & $15.2$ & $12.4$ \\  
  13607 & JADES-GS$+$53.13743$-$27.76519 & $6.622$ & $7.6$ & $6.2$ \\
  14123 & JADES-GS$+$53.17836$-$27.80098 & $6.327$ & $7.6$ & $6.2$ \\
  58850 & JADES-GS$+$53.09517$-$27.76061 & $6.263$ & $3.8$ & $3.1$ \\
  17138 & JADES-GS$+$53.08604$-$27.74760 & $6.204$ & $3.8$ & $3.1$ \\
  9365 & JADES-GS$+$53.16280$-$27.76084 & $5.917$ & $7.6$ & $6.2$ \\
\bottomrule
 \end{tabular}
    \label{tab:sources}
\end{table*}

\subsection{UV magnitudes and slopes}
\begin{figure*}
    \centering
    \includegraphics[width=0.49\linewidth]{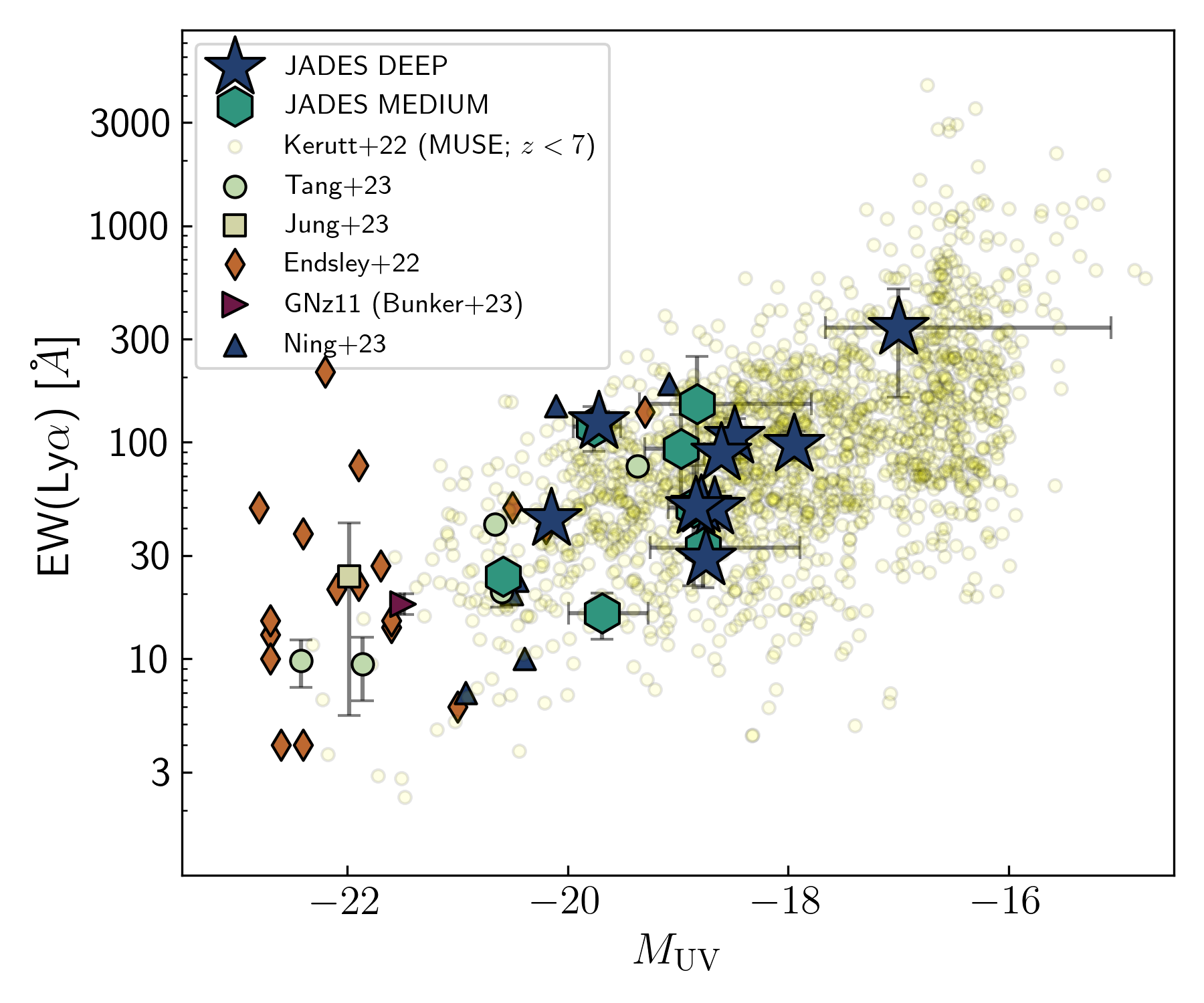}
    \includegraphics[width=0.49\linewidth]{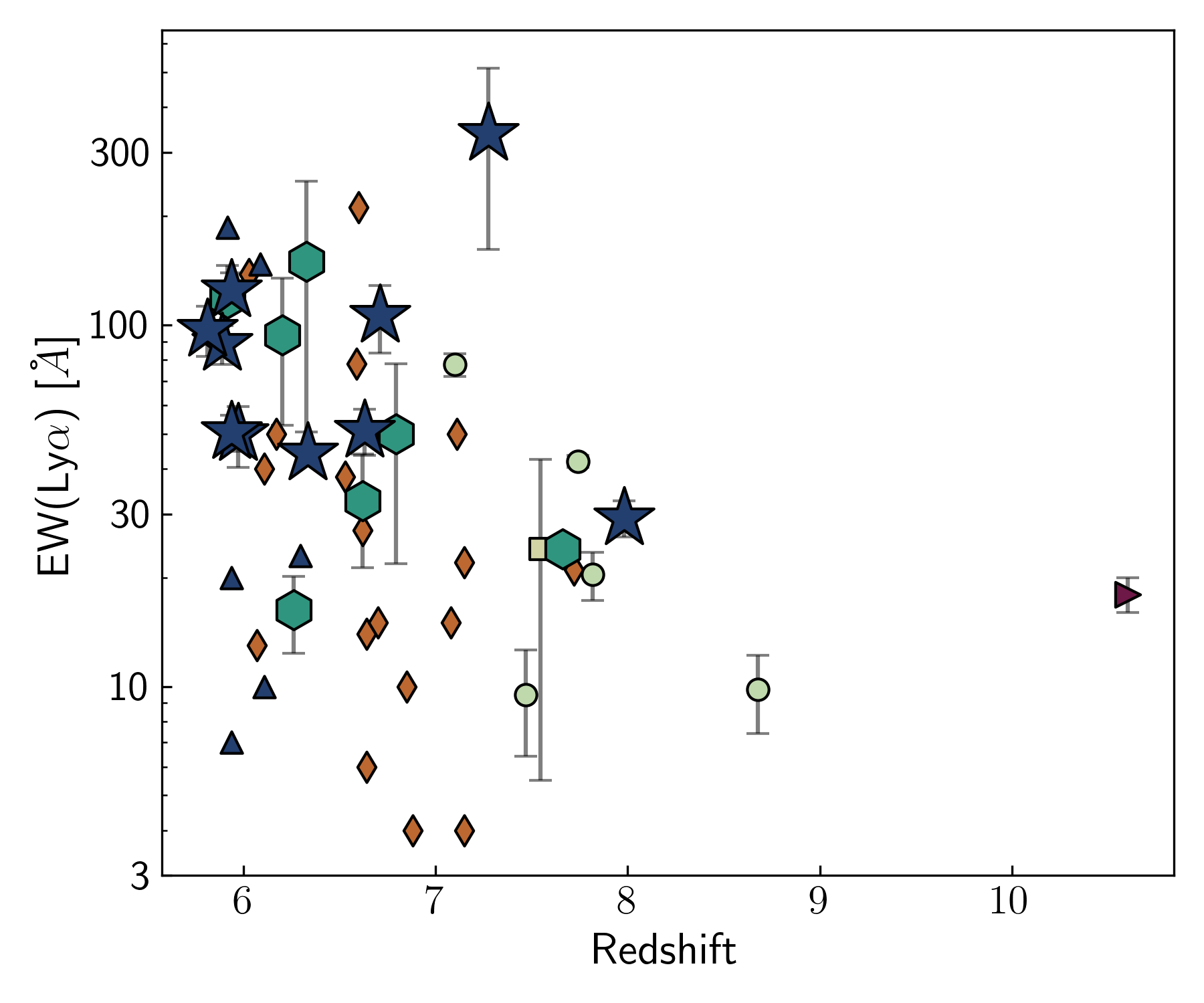}
    \caption{Distribution of EW(\lya) and \muv\ (left) and redshift (right) of galaxies from our Deep and Medium observations. Also shown for comparison are measurements from the literature of LAEs at $z\gtrsim6$ from \citet{tan23}, \citet{jun23}, \citet{nin23} and \citet{end23b}, as well as from GNz11 at $z=10.603$ \citep{bun23} (see Section \ref{sec:literature} for a brief explanation about each of these data sets). In the left panel, we additionally show measurements from LAEs identified from the MUSE WIDE and DEEP surveys \citep{ker22} at $z<5.7$, which serve as a relevant `reference' sample at redshifts where the IGM is not expected to impact the emergent \lya\ emission significantly. We note here that several lensed LAEs also overlap with the UV magnitude distribution of our JADES sample \citep[e.g.][]{hoa19, ful20, bol22}, which we do not show here. Our new LAE sample appears to be homogeneously mixed with the lower redshift LAE sample, indicating that the IGM may not be playing a significant role in affecting the \lya\ emission reported in this paper, further stressing the presence of highly ionized regions surrounding these LAEs \citep[see][]{witstok23}. The LAEs presented in this work are clearly much fainter in the UV than those that have been previously analysed in the literature at $z\gtrsim6$. Our sample also includes the extremely high EW LAE with $M_\mathrm{UV} \sim -17.0$ that was recently reported by \citet{sax23a}.}
    \label{fig:sample}
\end{figure*}

UV magnitudes at rest-frame 1500\,\AA\ ($M_\mathrm{UV}$) were measured directly from the R100 PRISM spectra. To do this, the spectra were shifted from observed to rest-frame using the spectroscopic redshifts and a 50\,\AA-wide boxcar filter centred on 1500\,\AA\ to measure the median flux and error. The measured fluxes and errors were then used to calculate absolute magnitudes and errors. The distribution of the UV magnitudes and \lya\ equivalent widths from our sample of LAEs is shown in Figure \ref{fig:sample}. The UV-faint galaxies in our sample show systematically high EW(\lya), which is likely due to the flux-limited nature of spectroscopic observations, only enabling high EW LAEs to be identified at fainter UV magnitudes. 

To put our sample into perspective, we also show measurements from other LAEs at $z\gtrsim6$ identified using \emph{JWST} \citep{tan23, jun23} or ground-based observations \citep{nin23, end23b} in the Figure. The LAEs presented in this work have on average fainter UV magnitudes compared to the majority of other strong LAEs at $z\gtrsim6$ identified from ground-based telescopes in the literature. We do note, however, ground-based spectroscopic surveys around lensing clusters have been able to identify a handful of faint $z>6$ LAEs \citep[e.g.][]{hoa19, ful20, bol22}, which have comparable UV magnitudes to the galaxies in our sample. However, the intrinsic UV magnitudes of lensed galaxies are prone to uncertainties from the lensing models.

We further compare our faint LAEs to those with similar UV magnitudes identified at $z<5.7$ from the MUSE DEEP and WIDE surveys \citep{ker22}, which are also shown in Figure \ref{fig:sample}. Interestingly, we find that the distribution of \lya\ EW is highly comparable to the MUSE LAEs at a given UV magnitude. Since the MUSE LAEs shown here were selected to lie at redshifts where the IGM is not expected to significantly impact the emergent \lya\ emission along the line of sight, the fact that our sample at $z\gtrsim6$ appears to be homogeneously mixed with the MUSE sample indicates that IGM attenuation is likely not playing a very strong role on the emergent \lya\ line from our galaxies. This reinforces the idea that our LAEs must be surrounded by highly ionized regions/bubbles \citep[e.g.][]{witstok23}, and the shape and strength of the observed \lya\ emission is likely controlled mainly by the ISM/CGM around the galaxies.

UV slopes ($\beta$, where $f_\lambda \propto \lambda^\beta$) are also measured directly from the R100 spectra by fitting a power-law function using chi-squared minimization to the flux density in the wavelength range 1340\,\AA\ to 2600\,\AA, using the \citet{cal94} spectral windows to avoid strong emission and/or absorption features at rest-UV wavelengths.  The redshifts, UV magnitudes at $1500$\,\AA\ and observed UV slopes are given in Table~\ref{tab:uv}.
\begin{table}
    \centering
    \caption{Observed rest-frame UV magnitude ($M_\mathrm{UV}$) and slope ($\beta$) measured for LAEs in this study.}
    \begin{tabular}{l c c c}
    \toprule
    ID & $z_{\mathrm{spec}}$ & $M_{\mathrm{UV}}$ & $\beta$ \\
     \midrule
    \emph{Deep Tier} \\
  21842 & $7.982$ & $-18.75^{-0.05}_{+0.05}$ & $-2.52 \pm 0.03$ \\
  10013682* & $7.276$ & $-17.00^{-0.66}_{+1.93}$ & $-2.17 \pm 0.60$ \\
  4297 & $6.712$ & $-18.49^{-0.06}_{+0.06}$ & $-2.39 \pm 0.09$ \\
  16625 & $6.631$ & $-18.79^{-0.10}_{+0.10}$ & $-2.59 \pm 0.02$ \\
  18846 & $6.336$ & $-20.15^{-0.05}_{+0.05}$ & $-2.43 \pm 0.01$ \\
  19342 & $5.974$ & $-18.67^{-0.03}_{+0.03}$ & $-2.75 \pm 0.04$ \\
  9422 & $5.937$ & $-19.72^{-0.04}_{+0.04}$ & $-2.33 \pm 0.04$ \\
  6002 & $5.937$ & $-18.84^{-0.07}_{+0.07}$ & $-2.59 \pm 0.01$ \\
  19606 & $5.889$ & $-18.61^{-0.17}_{+0.15}$ & $-2.70 \pm 0.06$ \\
  10056849 & $5.814$ & $-17.95^{-0.07}_{+0.06}$ & $-2.49 \pm 0.04$ \\

  \emph{Medium Tier} \\
  12637 & $7.660$ & $-20.59^{-0.07}_{+0.07}$ & $-2.20 \pm 0.02$ \\
  15362 & $6.794$ & $-18.86^{-0.23}_{+0.29}$ & $-2.14 \pm 0.15$ \\  
  13607 & $6.622$ & $-18.77^{-0.88}_{+0.48}$ & $-1.79 \pm 0.29$ \\
  14123 & $6.327$ & $-18.83^{-1.04}_{+0.52}$ & $-2.26 \pm 0.21$ \\
  58850 & $6.263$ & $-19.96^{-0.42}_{+0.30}$ & $-1.93 \pm 0.06$ \\
  17138 & $6.204$ & $-18.97^{-0.47}_{+0.33}$ & $-2.26 \pm 0.54$ \\
  9365 & $5.917$ & $-19.76^{-0.24}_{+0.19}$ & $-2.52 \pm 0.09$ \\
\bottomrule
 \end{tabular}
    \label{tab:uv}
\end{table}

\subsection{Lyman-alpha velocity offsets}
Using accurate systemic redshifts from the medium resolution grating spectra, we then use the peak of the \lya\ line detected in the G140M grating spectra of our LAEs to calculate velocity offsets from the expected \lya\ emission (vacuum wavelength) at systemic redshift. As mentioned earlier, the wavelength calibrations between the different gratings were compared against the lower resolution PRISM spectra were noted to be slightly inconsistent, but the wavelengths across the grating spectra were all consistent with each other \citep[see][]{bun23b}. Therefore, inferring the observed velocity offset of \lya\ from G140M spectra should not be affected by systematic offsets.

The \lya\ velocity offsets were measured as follows: we implemented a Monte Carlo (MC) based approach to determine both the peak of the emission line using 100 trials, where the emission line was fitted by either a singly symmetric or asymmetric Gaussian, depending on whichever function yielded a lower chi-squared statistic and hence, a better fit to the data. The spectral resolution of our $R\sim1000$ spectra is unfortunately not high enough to properly characterize the line shape, therefore we do not enforce a symmetric or asymmetric Gaussian fit, but simply pick the model that returns a better fit. 

For each MC trial, we perturbed the line fluxes by randomly sampling from the error spectrum. The line fitting was performed for each trial, with the median of the Gaussian centre then used to measure the velocity offset, and one sigma dispersion on the line centre distribution used to derive the error on this measurement. Due to the nature of this method, emission lines with high S/N naturally have lower errors on the velocity offsets.

\subsection{Other emission line measurements}
The rest-frame optical emission line fluxes for all LAEs are measured from the higher spectral resolution grating spectra, unless the lines are not clearly detected in the grating. In that case we measure and report the line fluxes from the PRISM spectra. The main emission lines that we measure for our sample of LAEs are \oii\,$\lambda\lambda3726,3729$ (which appear to be blended), \hb, \oiii\,$\lambda\lambda4959,5007$ and \ha.

We once again fit single Gaussian functions to all of these lines, measuring the local continuum from a wavelength region adjacent to the emission line. Using these line fluxes we also calculate line ratios such as \oiii\,$\lambda5007$/\oii\,$\lambda \lambda 3726,3729$ (O32) and (\oii\,$\lambda \lambda 3726,3729$ + \oiii\,$\lambda \lambda 4959,5007$)/\hb\ (R23).

\subsection{Dust measurements from Balmer decrements}
Here we use the Balmer emission line decrements calculated from \ha/\hb\ (or \hb/\hg\ when \ha\ is not within the spectral coverage). We calculate the intrinsic ratio of these lines using \textsc{pyneb} \citep{pyneb}, assuming a temperature of $10^4$\,K and electron density of $100$\,cm$^{-3}$. This gives an intrinsic \ha/\hb\ ratio of $2.863$. We assume the dust attenuation curve for the Small Magellanic Cloud (SMC; \citealt{gor03}), which has been shown to be the most appropriate for high redshift galaxies \citep[e.g.][]{shi20}. 

Dust attenuation, $E(B-V)$ is then calculated by comparing the observed Balmer line ratios with the intrinsic. We note that the \ha\ line is detected for all but one galaxy in our sample, and therefore, we primarily use the observed \ha/\hb\ ratios to calculate dust attenuation across our sample, but for $z>7$ LAEs, where \ha\ moves out of NIRSpec coverage, we use \hb/\hg. 

\subsection{Ionizing photon production efficiency}
\label{sec:xiion}
We use the \ha\ flux (or \hb\ when \ha\ is not within the spectral coverage, using an intrinsic ratio of $2.863$ under the assumptions that were mentioned in the previous section) and the monochromatic luminosity at 1500\,\AA\ to calculate the ionizing photon production efficiency, or \xiion, given by 
\begin{equation}
    \left(\frac{\xi_{\mathrm{ion}}}{\mathrm{erg}^{-1}}\,\mathrm{Hz}\right) = \frac{N(H^0)}{L_\mathrm{1500, int}}
\end{equation}
where $N(H^0)$ is the intrinsic hydrogen ionizing photon production rate in units of s$^{-1}$ and $L_\mathrm{1500, int}$ is the intrinsic (dust-corrected) luminosity density at rest-frame 1500\,\AA\ in units of erg\,s$^{-1}$\,Hz$^{-1}$. 

The \ha\ line (or other Balmer lines) luminosity can be used to calculate the intrinsic ionizing photon production rate. Assuming the same physical conditions as before of $T_e = 10^4$\,K and $n_e = 100$\,cm$^{-3}$,
\begin{equation}
    N(H^0) \times (1-f_\mathrm{esc}) = 7.3 \times 10^{11} L(\mathrm{H}\alpha) 
    \label{eq:xiion}
\end{equation}
where \fesc\ is the escape fraction of LyC photons out of the galaxy \citep[e.g.][]{mas20, sim23a}. To calculate \xiion\ for our LAEs, we assume Case-B recombination, that is \fesc(LyC) $=0$ (see Section \ref{sec:reionization}, however, for a discussion about non-zero \fesc(LyC)).

\subsection{Lyman-alpha escape fractions}
We now use the strength of Balmer emission lines seen in the spectrum together with the inferred dust attenuation to derive a \lya\ escape fraction for all LAEs in our sample. Assuming Case-B recombination, $n_e = 100$\,cm$^{-3}$ and $T_e = 10,000$\,K, the intrinsic \lya/\ha\ ratio is $8.2$ \citep[e.g.][]{ost89}. We then calculate \fesc(\lya) as the ratio of the observed (dust-corrected) \lya\ to Balmer line emission to the intrinsic ratio, which for the \ha\ emission line looks like: \fesc(\lya) = L(\lya)/($8.2\,\times\,$L(\ha)). 

The observed \lya\ properties, which include line fluxes, equivalent widths, velocity offset from systemic redshift and the \lya\ escape fraction (\fesc) are given in Table \ref{tab:lya}.
\begin{table*}
    \centering
    \caption{Observed \lya\ emission properties of LAEs in this study. The flux units are \flux\ (cgs).}
    \begin{tabular}{l c c c c c c c}
    \toprule
    ID & $z_{\mathrm{spec}}$ & $F^{\mathrm{Ly}\alpha}_{\mathrm{R100}}$ & EW$^{\mathrm{Ly}\alpha}_{\mathrm{R100}}$ & $F^{\mathrm{Ly}\alpha}_{\mathrm{R1000}}$ & EW$^{\mathrm{Ly}\alpha}_{\mathrm{R1000}}$ & $\Delta v^{\mathrm{Ly}\alpha}_{\mathrm{sys}}$ & \fesc(\lya) \\
     & & ($\times10^{-18}$\,cgs) & ($\AA$) & ($\times10^{-18}$\,cgs) & ($\AA$) & (km\,s$^{-1}$) &  \\
     \midrule
    \emph{Deep Tier} \\
  21842 & $7.982$ & $0.3 \pm 0.2$ & $14.1 \pm 5.7$ & $0.7 \pm 0.1$ & $29.2 \pm 3.3$ & $166.5 \pm 30.5$ & $0.09 \pm 0.01$ \\
  10013682* & $7.276$ & $1.5 \pm 0.2$ & $258.5 \pm 43.0$ & $2.2 \pm 0.5$ & $337.2 \pm 175.5$ & $178.4 \pm 21.1$ & $0.93 \pm 0.12$ \\
  4297 & $6.712$ & $0.9 \pm 0.3$ & $35.3 \pm 10.5$ & $2.7 \pm 0.3$ & $106.2 \pm 22.5$ & $153.0 \pm 76.1$ & $0.55 \pm 0.04$ \\
  16625 & $6.631$ & $13.7 \pm 2.0$ & $32.0 \pm 4.8$ & $21.8 \pm 4.8$ & $51.0 \pm 7.4$ & $244.2 \pm 25.8$ & $0.14 \pm 0.02$ \\
  18846 & $6.336$ & $4.2 \pm 2.1$ & $24.1 \pm 0.9$ & $7.7 \pm 1.3$ & $44.5 \pm 1.7$ & $139.4 \pm 5.1$ & $0.31 \pm 0.01$ \\
  19342 & $5.974$ & $2.4 \pm 0.2$ & $55.1 \pm 6.3$ & $2.2 \pm 0.5$ & $49.9 \pm 9.6$ & $257.0 \pm 21.2$ & $0.24 \pm 0.04$ \\
  9422 & $5.937$ & $9.3 \pm 2.7$ & $109.2 \pm 14.7$ & $10.6 \pm 0.9$ & $124.4 \pm 17.2$ & $147.6 \pm 15.1$ & $0.26 \pm 0.01$ \\
  6002 & $5.937$ & $2.0 \pm 0.2$ & $35.3 \pm 2.8$ & $2.9 \pm 0.6$ & $50.5 \pm 5.8$ & $181.0 \pm 42.1$ & $0.43 \pm 0.04$ \\
  19606 & $5.889$ & $6.3 \pm 0.4$ & $111.2 \pm 26.3$ & $-$ & $-$ & $-$ & $0.50 \pm 0.03$ \\
  10056849 & $5.814$ & $4.9 \pm 0.3$ & $127.0 \pm 10.5$ & $3.8 \pm 0.5$ & $97.2 \pm 15.2$ & $233.0 \pm 36.3$ & $0.42 \pm 0.06$ \\
  \emph{Medium Tier} \\
  12637 & $7.660$ & $0.6 \pm 0.2$ & $3.5 \pm 0.9$ & $4.2 \pm 0.3$ & $24.0 \pm 1.9$ & $277.2 \pm 34.4$ & $0.13 \pm 0.01$ \\
  15362 & $6.794$ & $-$ & $-$ & $1.5 \pm 0.7$ & $50.0 \pm 28.2$ & $27.0 \pm 124.6$ & $0.20 \pm 0.07$ \\
  13607 & $6.622$ & $0.2 \pm 0.2$ & $2.9 \pm 2.1$ & $2.4 \pm 0.9$ & $29.4 \pm 9.0$ & $116.8 \pm 61.1$ & $0.26 \pm 0.08$ \\
  14123 & $6.327$ & $7.0 \pm 2.0$ & $241.2 \pm 160.8$ & $4.3 \pm 1.2$ & $150.1 \pm 99.6$ & $194.2 \pm 43.4$ & $0.35 \pm 0.07$ \\
  58850 & $6.263$ & $9.1 \pm 0.8$ & $4.8 \pm 2.9$ & $3.1 \pm 0.8$ & $16.3 \pm 3.9$ & $254.6 \pm 43.1$ & $0.06 \pm 0.01$ \\
  17138 & $6.204$ & $3.0 \pm 1.1$ & $65.0 \pm 22.8$ & $4.6 \pm 2.1$ & $93.6 \pm 40.9$ & $0.0 \pm 71.9$ & $0.40 \pm 0.10$ \\
  9365 & $5.917$ & $12.2 \pm 1.4$ & $109.5 \pm 16.6$ & $11.2 \pm 1.9$ & $118.4 \pm 27.6$ & $256.7 \pm 60.9$ & $0.28 \pm 0.04$ \\  
\bottomrule
 \end{tabular}
    \label{tab:lya}
\end{table*}

\subsection{Comparison samples from the literature}
\label{sec:literature}
To put our results into a more global context while also increasing the baseline of several physical parameters that were also measured from our sample of faint LAEs, we describe here a selection of literature samples of LAEs at $z\gtrsim6$ with which we compare our results. Perhaps the most immediate comparison is offered by LAEs identified by \citet{tan23} using \emph{JWST} spectroscopy through the CEERS survey \citep[see also][]{fuj23}. We also include CEERS results from \citet{jun23} in this study. Since CEERS is shallower and wider than JADES, it more efficiently selects the rarer UV-bright galaxies by probing a much larger volume at high redshifts.

We also use the compilation of $z\gtrsim6$ LAEs from \citet{end22c} that have ALMA emission line measurements, enabling robust measurements of the \lya\ velocity offsets. This compilation includes LAEs from the ALMA REBELS survey \citep{bou22b} as well as other LAEs at $z>6$: CLM1 \citep{cub03, wil15}, WMH5 \citep{wil15}, B14-65666 \citep{fur16, has19}, EGS-zs8-1 \citep{oes15, sta17}, COS-z7-1 \citep{pen16, lap17, sta17}, COSMOS24108 \citep{pen16, pen18b}, NTTDF6345 \citep{pen11, pen16}, UDS16291 \citep{pen16, pen18b}, BDF-3299 \citep{van11, mai15, car17}, RXJ2248-ID3 \citep{mai17}, A383-5.2 \citep{stark15a, knu16}, VR7 \citep{mat19, mat20}, CR7 \citep{sob15, mat17} and Himiko \citep{ouc13, carn18}.

We also include \lya\ and \ha\ based measurements from \citet{nin23} as well as \citet{sim23a}, which use narrow/medium band photometry to infer \ha\ strengths in spectroscopically confirmed LAEs at $z\sim6$ identified from MUSE data, enabling the determination of \fesc(\lya) and \xiion.

Finally, we also use the \lya\ emission measurements from GNz-11, spectroscopically confirmed to lie at $z=10.60$ with weak \lya\ emission detected in the medium band NIRSpec gratings \citep{bun23}.

\section{Spectroscopic properties of Lyman-alpha emitters at $z\gtrsim5.8$}
\label{sec:properties}
In this section we explore the general spectroscopic properties of LAEs identified in the JADES Deep and Medium Tier surveys, with the aim of comparing the ionization and chemical enrichment of LAEs with the general galaxy population at $z\gtrsim6$ as well as evaluating the ionizing photon production efficiencies of LAEs across cosmic time.

\subsection{Chemical enrichment and dust}

\begin{figure}
    \centering
    \includegraphics[width=\linewidth]{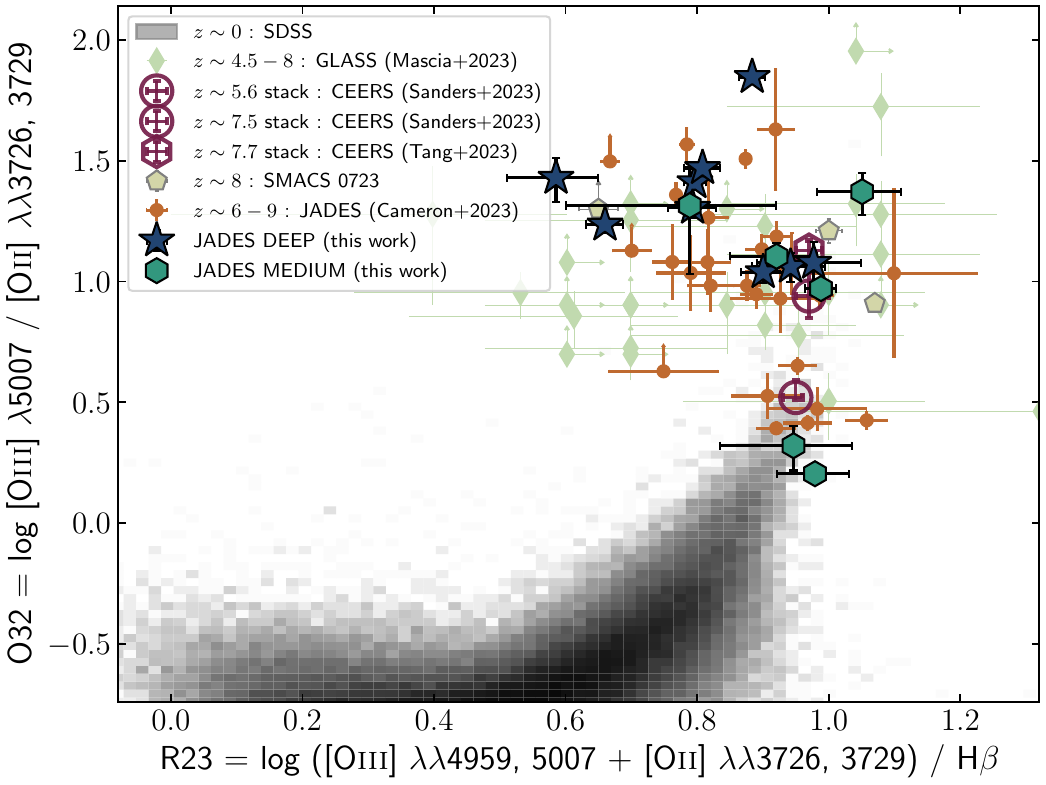}
    \caption{Comparison of R23 and O32 ratios of LAEs with the general galaxy populations inferred from NIRSpec observations at $z\gtrsim6$.
    Orange circles show JADES Deep galaxies from \citep{cam23} which are {\em not} LAEs. Diamonds and pentagons show measurements from individual lensed galaxies at $z\gtrsim5$ \citep{mas23, nak23}. Measurements from stacked galaxies at $z\sim5.6-7.7$ are shown as hollow purple markers \citep{san23, tan23}. Also shown are measurements from SDSS for $z<0.1$ star-forming galaxies \citep{aih11}. Overall, we find that our faint LAEs occupy a similar parameter to that occupied by the general star-forming galaxy population at $z\gtrsim6$. This also highlights that the observed presence of \lya\ emission in reionization era galaxies is driven by external factors, and not necessarily by the ISM/stellar properties of the galaxies. Interestingly, UV-faint LAEs from our sample tend to show higher O32 and lower R23 ratios than the UV-bright ones.}
    \label{fig:r23_o32}
\end{figure}

In Figure~\ref{fig:r23_o32}, we show R23 vs O32 line ratios for our Deep- and Medium-tier samples of LAEs. These line ratios are widely used tracers of metallicity and ionization parameter respectively, with the former forming a two-valued relation with metallicity.
For comparison, we show $z<0.1$ galaxies from the SDSS MPA-JHU catalogues \citep{aih11}\footnote{\url{https://www.sdss3.org/dr10/spectro/galaxy_mpajhu.php}}, as well as non-Lyman-$\alpha$-emitting galaxies at $z>5.5$ from JADES Deep (\citealt{cam23}), and measurements from individual and stacked galaxies at $z\gtrsim5$ not selected on presence or otherwise of \lya\ \citep{mas23, nak23, san23, tan23}.
Our LAEs on average appear to be metal poor with high ionization parameters, lying away from the locus of typical star-forming galaxies at $z<0.1$ from SDSS towards high O32 and R23. Instead, they are more similar to what has been reported for the general galaxy population at $z>6$.

We do note that LAEs from our Deep Tier survey, which tend to have fainter UV magnitudes ($\lesssim-20.1$), show slightly higher O32 ratios and lower R23 ratios compared to LAEs found in the Medium Tier survey as well as other brighter LAEs at $z>6$. This is consistent with the finding in \citet{cam23} that $z\sim6$ galaxies from deep JADES observations show much higher O32, and lower R23 than those measured from stacks of CEERS galaxies \citep{san23}, which are typically brighter.
This is indicative of higher ionization parameters and lower chemical enrichment in fainter, less massive galaxies at $z>6$.

Overall we find that the parameter space on this plot occupied by LAEs at $z\gtrsim6$ is roughly the same as the general galaxy population at these redshifts. This suggests that the detection of \lya\ emission from a galaxy in the EoR may not necessarily depend on the chemical or ionization state that is in, but may be more driven by opportune sight-lines probing sufficiently ionized regions of the Universe.

We do not measure any presence of dust from Balmer decrements derived from \ha/\hb\ ratios for our LAEs (in agreement with \citealt{sandles23}), which suggests that such systems are relatively dust-free, which is also a prerequisite for the leakage of significant fractions of Lyman continuum photons from a galaxy into the IGM.

\subsection{Ionizing photon production}
\begin{figure*}
    \centering
    \includegraphics[width=0.49\linewidth]{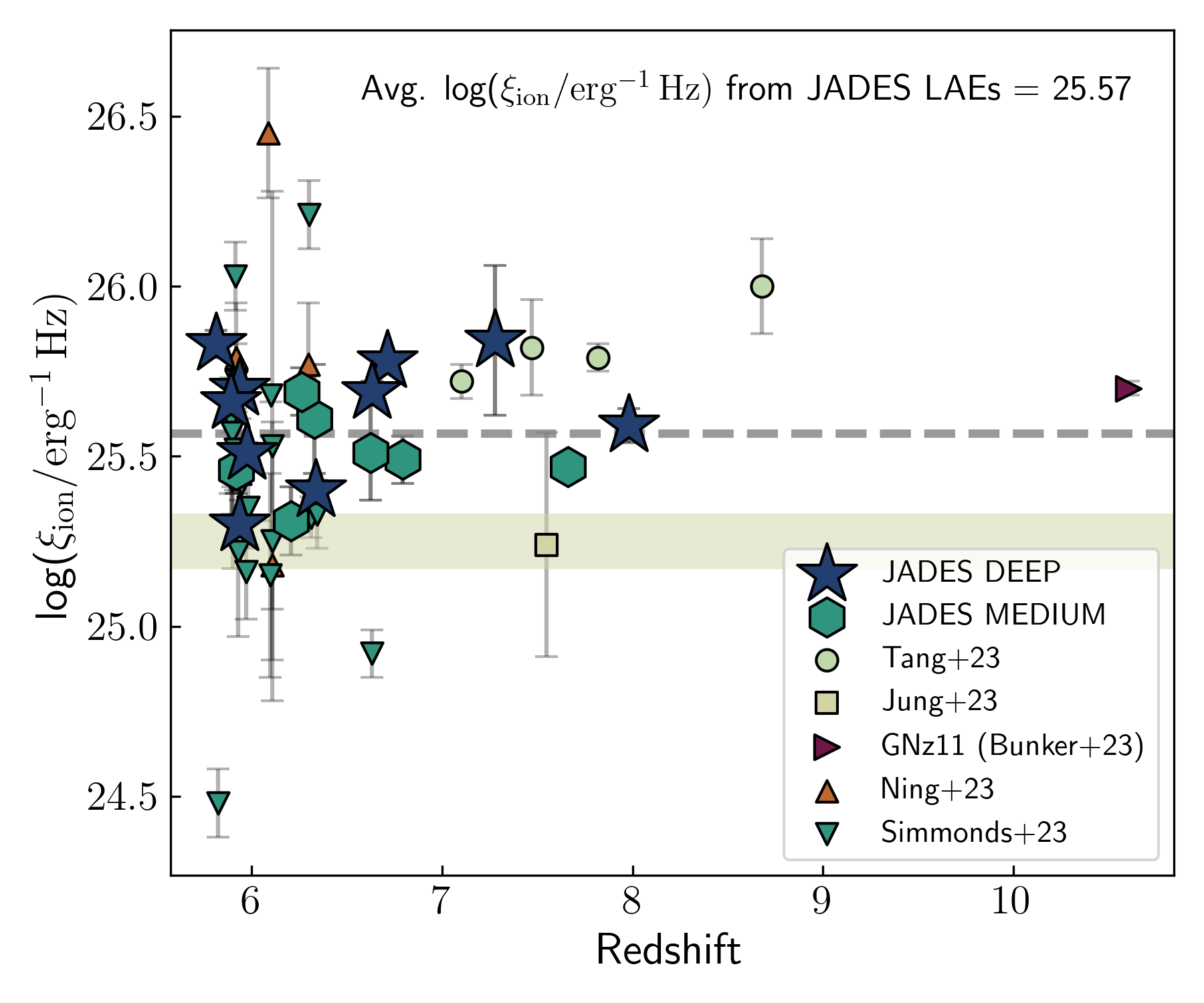}
    \includegraphics[width=0.49\linewidth]{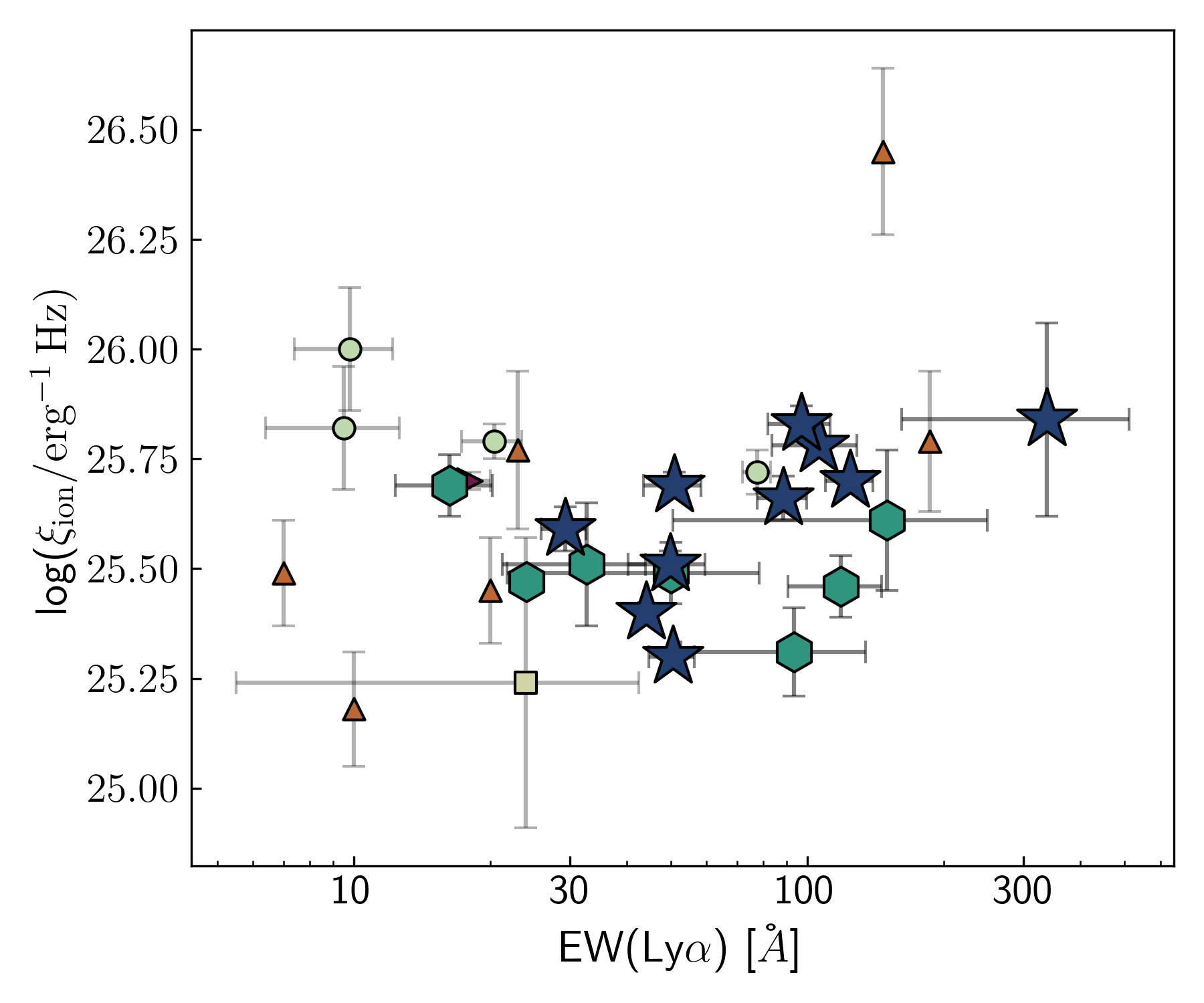}
    \caption{Dependence of the ionizing photon production efficiency (\xiion) from LAEs at $z\gtrsim 6$ on redshift (left) and EW(\lya) (right). The dashed line indicates the average \xiion\ value (log(\xiion/erg$^{-1}$\,Hz)\,$= 25.57$) measured across the new LAEs reported in this study. The shaded region marks the canonical value of log(\xiion/erg$^{-1}$\,Hz)\,$= 25.2-25.3$ from \citet{kuh12, rob13, rob15}. Overall, we do not find a significant evolution in the \xiion\ of LAEs across redshifts particularly at $z>6$, indicating that the ionizing properties do not seem to evolve strongly between $z=6-8.5$. There is also no strong correlation between \xiion\ and the strength of \lya\ emission across LAEs at $z>6$, which one would naively expect from simple Case B recombination in the absence of significant absorption/scattering of \lya\ photons. This lack of correlation indicates that the processes that control the escape of \lya\ photons may not necessarily be dependent on the processes that produce ionizing photons.}
    \label{fig:z_xiion}
\end{figure*}

The average ionizing photon production efficiency across our sample of faint LAEs is \xiion\,$=25.57$\,Hz\,erg$^{-1}$ shown as a dashed line in Figure \ref{fig:z_xiion}, which is $0.3-0.4$\,dex higher than the canonical value of $25.2-25.3$ typically assumed in reionization models \citep[e.g.][]{kuh12, rob13}. The higher \xiion\ values may also be indicative of elevated ionizing photon production due to non-thermal processes such as X-ray binary stars, whose impact is expected to increase with decreasing metallicities \citep[e.g.][]{sax21}. 

When comparing with other measurements for LAEs at the highest redshifts in the literature, we do not see significant evolution in \xiion\ as a function of redshift at $z>6$. Our measurements are higher than \xiion\,$=25.33$\,Hz\,erg$^{-1}$ that was reported for a sample of UV faint galaxies in the redshift range $3<z<7$ by \citet{prieto23}, where they reported a $\sim0.1$\,dex higher measurement for LAEs.

Interestingly, there does not seem to be any strong dependence of \xiion\ on the equivalent width of \lya\ emission either. Assuming Case B recombination and \fesc(\lya) of unity, \xiion\ may be expected to increase linearly with EW(\lya). The fact that there is no clear correlation between these two quantities across a wider sample of known LAEs at $z>6$ spanning orders of magnitude in brightness suggests that the mechanisms that are responsible for the production of ionizing photons in a galaxy are not the ones that also control the escape of \lya\ photons from the galaxy. In other words, the neutral gas and dust content, which preferentially affects the transmission of \lya\ photons, does not seem to closely depend on properties such as stellar metallicities or ages that control the production of ionizing photons. 

Combined with the lack of redshift evolution in \xiion, the picture that emerges is that the ionizing photon production is not closely linked to the strength of the emergent \lya\ line emission, as it is likely more dependent on the physical and chemical properties of star-forming regions that do not seem to evolve strongly between $z=6-8.5$. This may have important consequences for modelling the production and escape of ionizing photons from galaxies within the reionization epoch, which we revisit in Section \ref{sec:reionization}. The chemical enrichment and ionizing properties of LAEs in this study are given in Table~\ref{tab:derived}.

\begin{table*}
    \centering
    \caption{Chemical enrichment, ionization properties, and Lyman continuum escape fractions of \lya\ emitters presented in this study. The flux units are \flux\ (cgs).} 
    \begin{tabular}{l c c c c c c c}
    \toprule
    ID & $z_{\mathrm{spec}}$ & $F^{\mathrm{H}\beta}$ & $F^{\mathrm{H}\alpha}$ & $\log(\xi_{\mathrm{ion}}$/Hz\,erg$^{-1}$) & \oiii/\oii\ & (\oii+\oiii)/\hb\ & \fesc(LyC)  \\
    & & ($\times 10^{-19}$\,cgs) & ($\times 10^{-19}$\,cgs) & & (O32) & (R23) & \\
     \midrule
    \emph{Deep Tier} \\
  21842 & $7.982$ & $3.2 \pm 0.5$ & $-$ & $25.59^{+0.05}_{-0.05}$ & $11.5 \pm 1.6$ & $8.8 \pm 1.1$ & $0.08 \pm 0.01$ \\
  10013682 & $7.276$ & $0.8 \pm 0.2$ & $-$ & $25.66^{+0.11}_{-0.14}$ & $12.0 \pm 2.6$ & $9.5 \pm 1.7$ & $0.03 \pm 0.01$ \\
  4297 & $6.712$ & $5.3 \pm 0.6$ & $16.2 \pm 0.04$ & $25.78^{+0.02}_{-0.02}$ & $19.8 \pm 4.7$ & $7.5 \pm 1.4$ & $0.02 \pm 0.01$ \\
  16625 & $6.631$ & $5.9 \pm 0.3$ & $16.5 \pm 0.2$ & $25.69^{+0.03}_{-0.04}$ & $24.5 \pm 3.0$ & $4.9 \pm 0.5$ & $0.04 \pm 0.01$ \\
  18846 & $6.336$ & $11.4 \pm 0.2$ & $30.3 \pm 0.3$ & $25.32^{+0.01}_{-0.01}$ & $25.8 \pm 1.3$ & $6.3 \pm 0.2$ & $0.07 \pm 0.01$ \\
  19342 & $5.974$ & $4.1 \pm 0.2$ & $10.8 \pm 0.3$ & $25.42^{+0.01}_{-0.01}$ & $29.5 \pm 1.8$ & $6.4 \pm 0.4$ & $0.06 \pm 0.01$ \\
  9422 & $5.937$ & $18.4 \pm 0.1$ & $50.0 \pm 0.7$ & $25.65^{+0.01}_{-0.01}$ & $70.6 \pm 4.1$ & $7.7 \pm 0.3$ & $0.01 \pm 0.001$ \\
  6002 & $5.937$ & $2.8 \pm 0.1$ & $8.1 \pm 0.2$ & $25.19^{+0.02}_{-0.02}$ & $10.9 \pm 1.1$ & $7.6 \pm 0.6$ & $0.06 \pm 0.01$ \\
  19606 & $5.889$ & $4.6 \pm 0.4$ & $13.5 \pm 0.5$ & $25.51^{+0.04}_{-0.05}$ & $21.9 \pm 2.5$ & $7.5 \pm 0.7$ & $0.06 \pm 0.01$ \\
  10056849 & $5.814$ & $4.0 \pm 0.4$ & $10.9 \pm 0.3$ & $25.65^{+0.02}_{-0.02}$ & $17.22 \pm 1.3$ & $4.6 \pm 0.3$ & $0.03 \pm 0.01$ \\
  \emph{Medium Tier} \\
  12637 & $7.660$ & $13.9 \pm 0.5$ & $-$ & $25.45^{+0.02}_{-0.02}$ & $9.4 \pm 0.6$ & $9.7 \pm 0.5$ & $0.10 \pm 0.01$ \\
  15362 & $6.794$ & $4.3 \pm 0.4$ & $11.8 \pm 1.2$ & $25.49^{+0.07}_{-0.09}$ & $2.7 \pm 0.3$ & $6.2 \pm 0.6$ & $0.16 \pm 0.02$ \\
  13607 & $6.622$ & $3.6 \pm 0.5$ & $11.9 \pm 0.9$ & $25.51^{+0.14}_{-0.22}$ & $1.6 \pm 0.1$ & $11.5 \pm 1.1$ & $0.80 \pm 0.11$ \\
  14123 & $6.327$ & $5.6 \pm 0.9$ & $15.1 \pm 1.2$ & $25.61^{+0.16}_{-0.25}$ & $12.2 \pm 2.6$ & $8.9 \pm 2.4$ & $0.04 \pm 0.01$ \\
  58850 & $6.263$ & $26.0 \pm 1.7$ & $65.9 \pm 2.0$ & $25.83^{+0.09}_{-0.11}$ & $23.6 \pm 4.7$ & $11.2 \pm 1.6$ & $0.02 \pm 0.01$ \\
  17138 & $6.204$ & $3.6 \pm 3.0^\dagger$ & $10.4 \pm 1.9$ & $25.31^{+0.10}_{-0.13}$ & $1.8 \pm 0.3$ & $8.1 \pm 1.2$ & $0.38 \pm 0.10$ \\
  9365 & $5.917$ & $11.1 \pm 1.8$ & $30.8 \pm 1.3$ & $25.40^{+0.07}_{-0.08}$ & $12.3 \pm 1.7$ & $10.1 \pm 1.2$ & $0.14 \pm 0.02$ \\  
\bottomrule
 \end{tabular}

 $^\dagger$ The \hb\ line was affected by a possible cosmic ray in the spectrum, and therefore, \ha\ flux was used to estimate \hb\ flux assuming no dust.
    \label{tab:derived}
\end{table*}

\section{The escape of Lyman-alpha photons}
\label{sec:lya_escape}
\begin{figure}
    \centering
    \includegraphics[width=\linewidth]{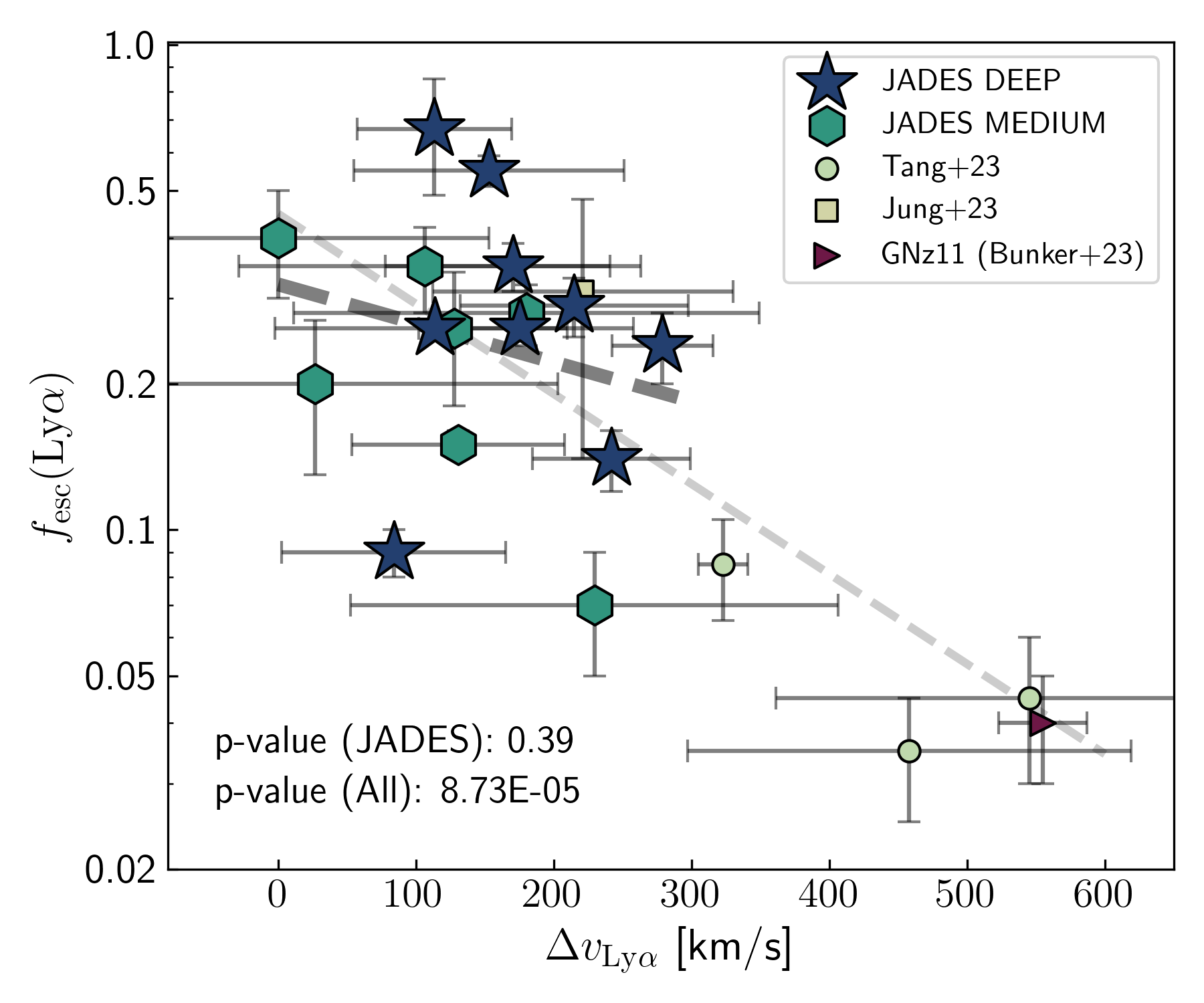}
    \caption{\lya\ velocity offset vs \fesc(\lya), where a strong anti-correlation between these quantities is observed both in the JADES-only sample (thick dashed line) as well as when looking at the full high redshift \lya\ emitter sample (thin dashed line). High \fesc(\lya) and small velocity offsets likely trace relatively dust and gas-free conditions, which neither leads to considerable resonant scattering of \lya\ photons as they travel along a sight line, nor does it attenuate \lya\ emission via absorption and scattering.}
    \label{fig:offset_escape}
\end{figure}
In this section we explore which galaxy property best traces the \lya\ velocity offset from the systemic redshift as well as the escape fraction of \lya\ photons, \fesc(\lya), which are widely regarded to be tracing escape channels for hydrogen ionizing LyC photons. The goal of this section is to determine the best tracer for LyC leakage when \lya\ emission may not be visible from galaxies in the reionization era.

\subsection{High Lyman-alpha escape fraction implies more Lyman-alpha photons escape closer to systemic velocity}
We begin by demonstrating that the escape fraction of \lya\ photons appears to be anti-correlated with the velocity offset of the \lya\ emission compared to the systemic redshift of a galaxy, as shown in Figure \ref{fig:offset_escape}. The anti-correlation is not as strong when considering only our magnitude limited JADES sample (p-value $=0.39$), but when expanding the dynamic range by including other known strong LAEs at $z>6$ with \lya\ velocity offset and escape fraction measurements, we recover an anti-correlation with a very high significance (p-value $=8.7\times 10^{-5}$). 

This anti-correlation can be explained using neutral gas column densities -- a low column density of neutral gas will lead to less resonant scattering of \lya\ photons out of the line of sight, thereby resulting in both a high observed \fesc(\lya) as well as low velocity offsets from systemic as has been predicted by theoretical models \citep{neufeld90, dijkstra06, verhamme06, laursen09}. Low neutral gas density environments are also though to be conducive to the escape of LyC photons from a galaxy (at least along the same line of sight as \lya). Therefore, both \lya\ velocity offsets and/or escape fractions can be important to ascertain the escape of ionizing photons from galaxies that drive reionization. 

In the following sections we explore correlations between various galaxy properties and each of \lya\ velocity offset and \fesc(\lya) to establish dependencies and/or observational biases that impact the \lya\ strength and line profile in LAEs at $z\gtrsim6$.

\subsection{Insights from Lyman-alpha velocity offsets} 
\label{sec:veloffsets}

\begin{figure*}
    \centering
    \includegraphics[width=0.49\linewidth]{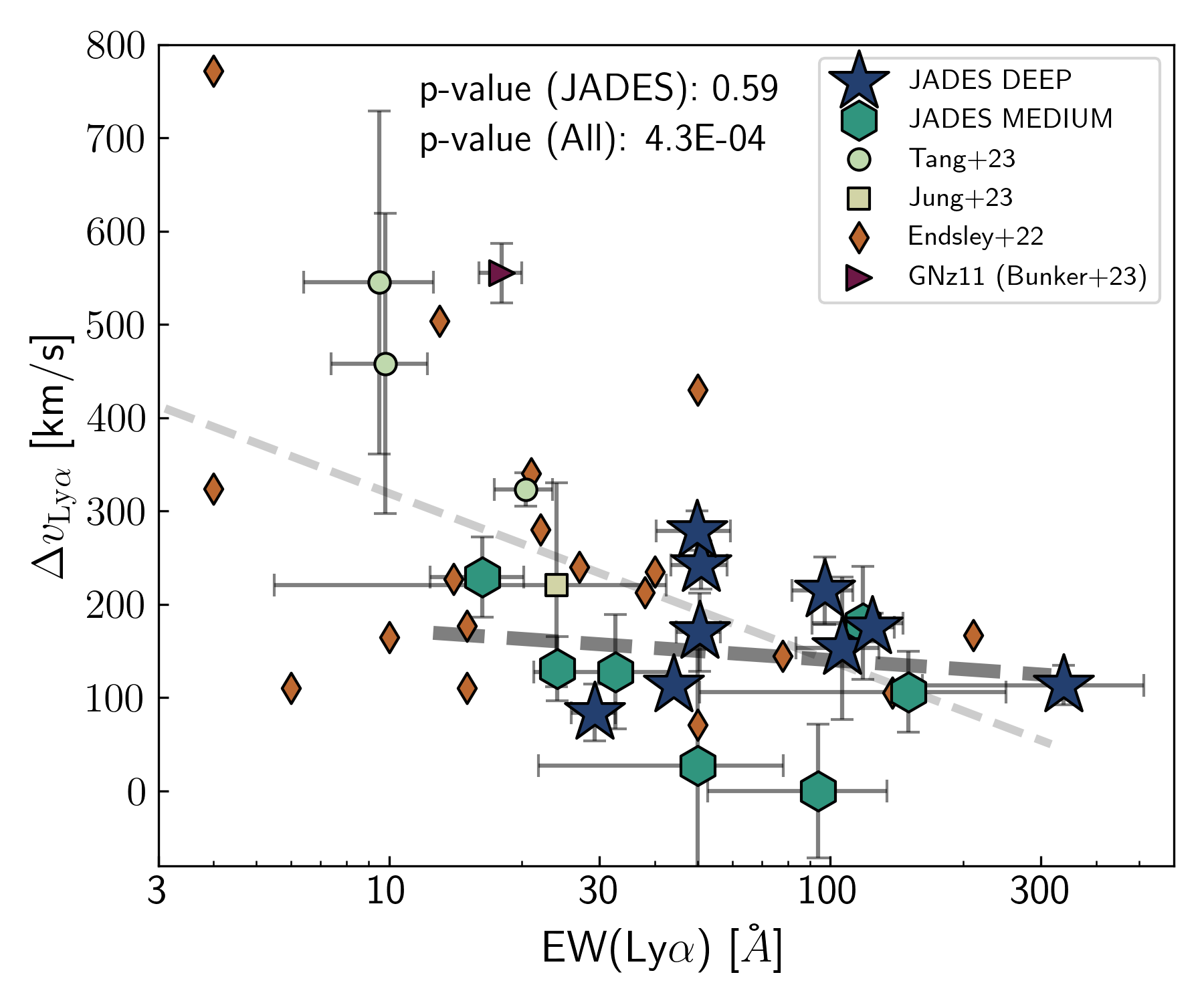}
    \includegraphics[width=0.49\linewidth]{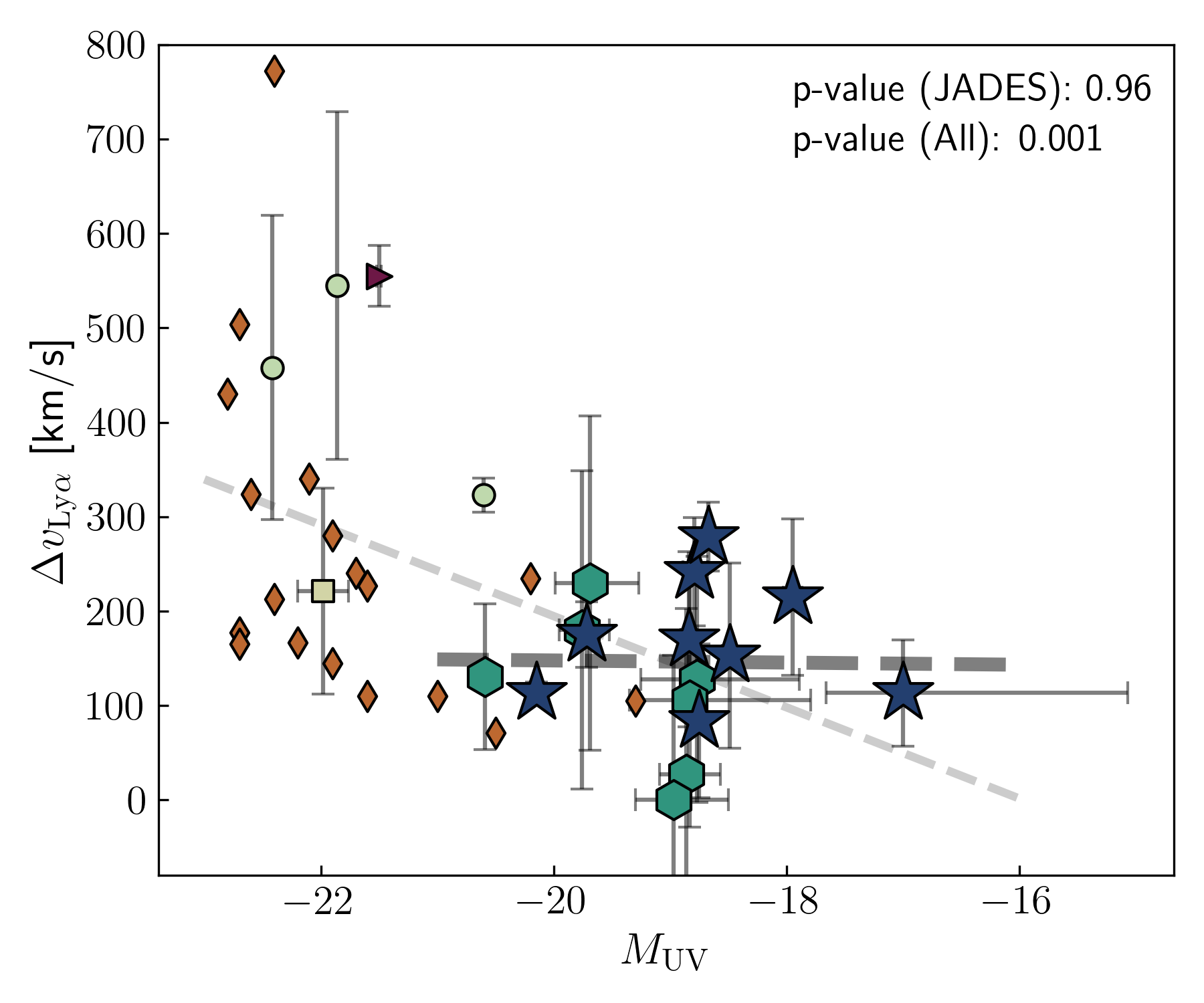}
    
    \includegraphics[width=0.49\linewidth]{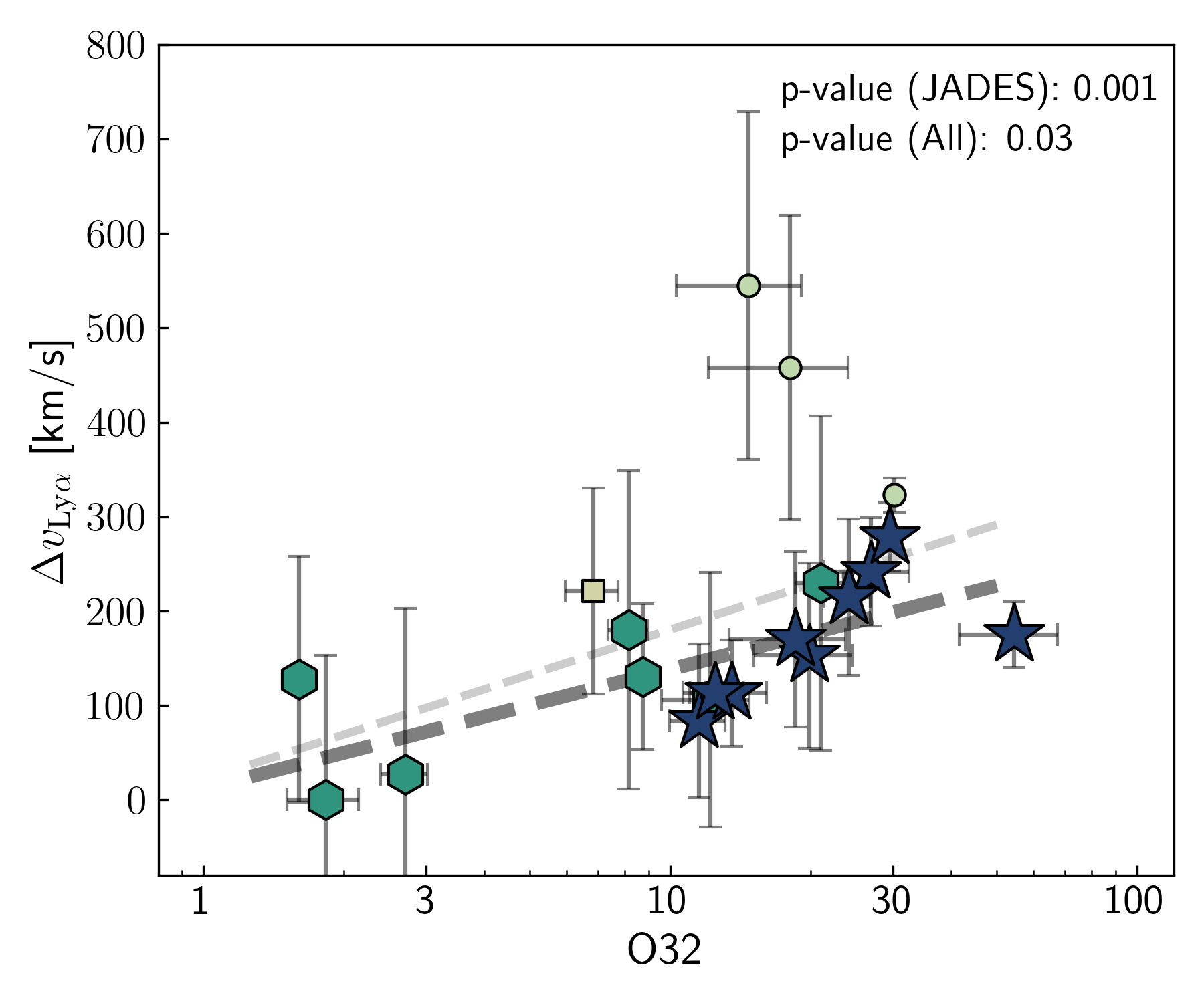}
    \includegraphics[width=0.49\linewidth]{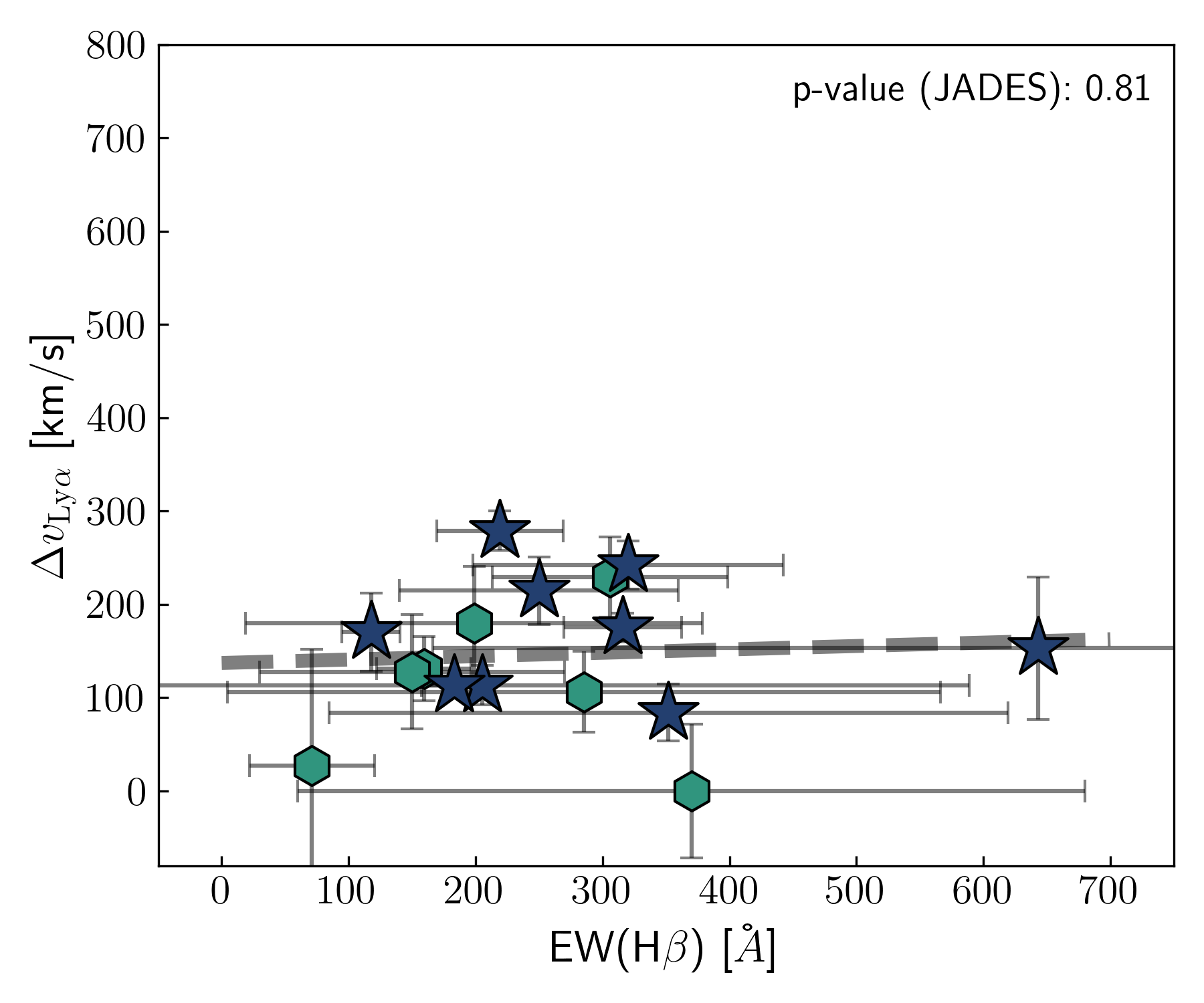}

    \caption{Dependence of the \lya\ velocity offset on EW(\lya) (top left), $M_{\rm{UV}}$ (top right), \oiii/\oii\ ratio (O32) (bottom left) and EW(\hb) (bottom right). In these plots, we show the correlations between quantities derived using only our JADES sample as the thicker, darker dashed line, and those derived using the extended LAE sample with the thinner, lighter dashed line, with p-values given for both. We find an anti-correlation between \lya\ velocity offset and EW(\lya), which is likely driven by the neutral gas content in the galaxy, whereby a larger reservoir of neutral gas both attenuates \lya\ flux close to the systemic velocity as well as moves the apparent peak of the emission line away from systemic velocity due to resonant scattering of \lya\ photons. \lya\ emission from UV-fainter LAEs also peaks closer to the systemic redshift, which is indicative of decreasing neutral gas content at fainter luminosities/galaxy masses. No strong correlations exist between \lya\ offset and O32 ratios or EW(\hb), interestingly with three low \lya\ velocity offset sources showing low ($<3$) O32 ratios.}
    \label{fig:veloffset}
\end{figure*}

In Figure \ref{fig:veloffset} we compare the \lya\ velocity offsets observed for our sample of galaxies with other observables that trace both the underlying stellar populations as well as the state of the ISM. We also include measurements of brighter LAEs in the EoR from the literature to increase the baseline of any trends that may become apparent.

When comparing \lya\ velocity offsets with EW(\lya), we find that our JADES sample shows a very weak anti-correlation (p-value $=0.59$), whereas the extended $z>6$ sample shows a strong anti-correlation (p-value = $4.3\times10^{-4}$), albeit with considerable scatter, as shown in Figure \ref{fig:veloffset} (top left). Such an anti-correlation has previously been reported in the literature across redshifts \citep[e.g.][]{izo21} and mainly stems from the resonant scattering of \lya\ photons by the neutral gas within the galaxies -- a higher velocity offset compared to the systemic redshift is indicative of more resonant scattering of the emergent \lya\ photons, which results in decreased \lya\ flux observed along the line-of-sight. Therefore, the same scattering mechanism is responsible for increased offset from systemic velocity as well as the reduction of EW(\lya) across galaxies. With high EW \lya\ emission that peaks close to the systemic redshift likely tracing low covering fractions of neutral gas, galaxies that exhibit such \lya\ profiles and strengths are also likely to be leaking significant amounts of LyC photons \citep{ver15, ver17}. 

For the sample of UV-faint LAEs probed by the JADES sample presented in this paper, we find the equivalent widths to be higher and the velocity offsets to be lower compared to UV-bright LAEs in the literature (Figure \ref{fig:veloffset}, top right), which may indicate that UV-fainter LAEs could be more likely to host conditions required for efficient \lya\ as well as LyC escape (see also \citealt{prieto23a}), which we subsequently explore in detail in the later sections. For our JADES sample, do not find any correlation between the \lya\ velocity offset and the UV magnitude (p-value $=0.96$). However, when looking at the extended LAE sample with a larger dynamic range in UV magnitude, we find that the \lya\ velocity offsets appear to be reducing at fainter UV magnitudes (p-value $=0.001$). There have been suggestions in the literature that the \lya\ velocity offset increases with stellar mass \citep[e.g.][]{erb14, hay23a}, which may explain the trends that we see with UV magnitude for the full sample.

Particularly within the context of ionized bubbles within which LAEs in the EoR must reside to be able to freely transmit \lya\ photons along the line of sight, smaller velocity offsets are also expected to trace large ionized bubble sizes \citep[see][for example]{mason20, sax23a}, which would lead to considerably less attenuation by the intervening IGM and therefore, higher transmission of \lya\ leading to high equivalent width measurements.

Next we compare the \lya\ offset with spectroscopic indicators of the ionization parameter (that is the ratio of ionizing photons to the hydrogen density) probed by indicators such as the O32 ratio (Figure \ref{fig:veloffset}, bottom left). Interestingly, we observe a mild positive correlation between O32 and \lya\ velocity offset, but there is considerable scatter on this relation and the relation becomes less significant when we also include LAEs from \citet{tan23}. A high O32 ratio has been proposed as an indicator for both high \lya\ and LyC escape \citep[e.g.][]{izo21} and we do find that our UV-faint LAEs with small velocity offsets on average show high O32 ratios. However, given a lack of strong correlation between O32 and \lya\ velocity offset we conclude that high O32 ratios are a necessary but not a sufficient condition for efficient ionizing photon escape \citep[see also][]{cho23}.

Looking at the dependence of \lya\ velocity offset on EW(\hb), which is a good tracer for star formation rates and consequently the production rates of Hydrogen ionizing photons, only for our JADES sample, we do not find any strong correlation (Figure \ref{fig:veloffset}, bottom right). EW(\hb) has also been proposed as a robust indicator of \fesc(LyC) \citep[e.g.][]{zac13, flu22b}. The requirement of a relatively high EW(\hb) is perhaps similar in nature to the requirement of high O32 from LyC leaking galaxies, necessary but insufficient on its own to enable efficient \lya/LyC escape.

\subsection{Insights from the Lyman-alpha escape fraction}
\label{sec:lya_fesc}
\begin{figure*}
    \centering
    \includegraphics[width=0.49\linewidth]{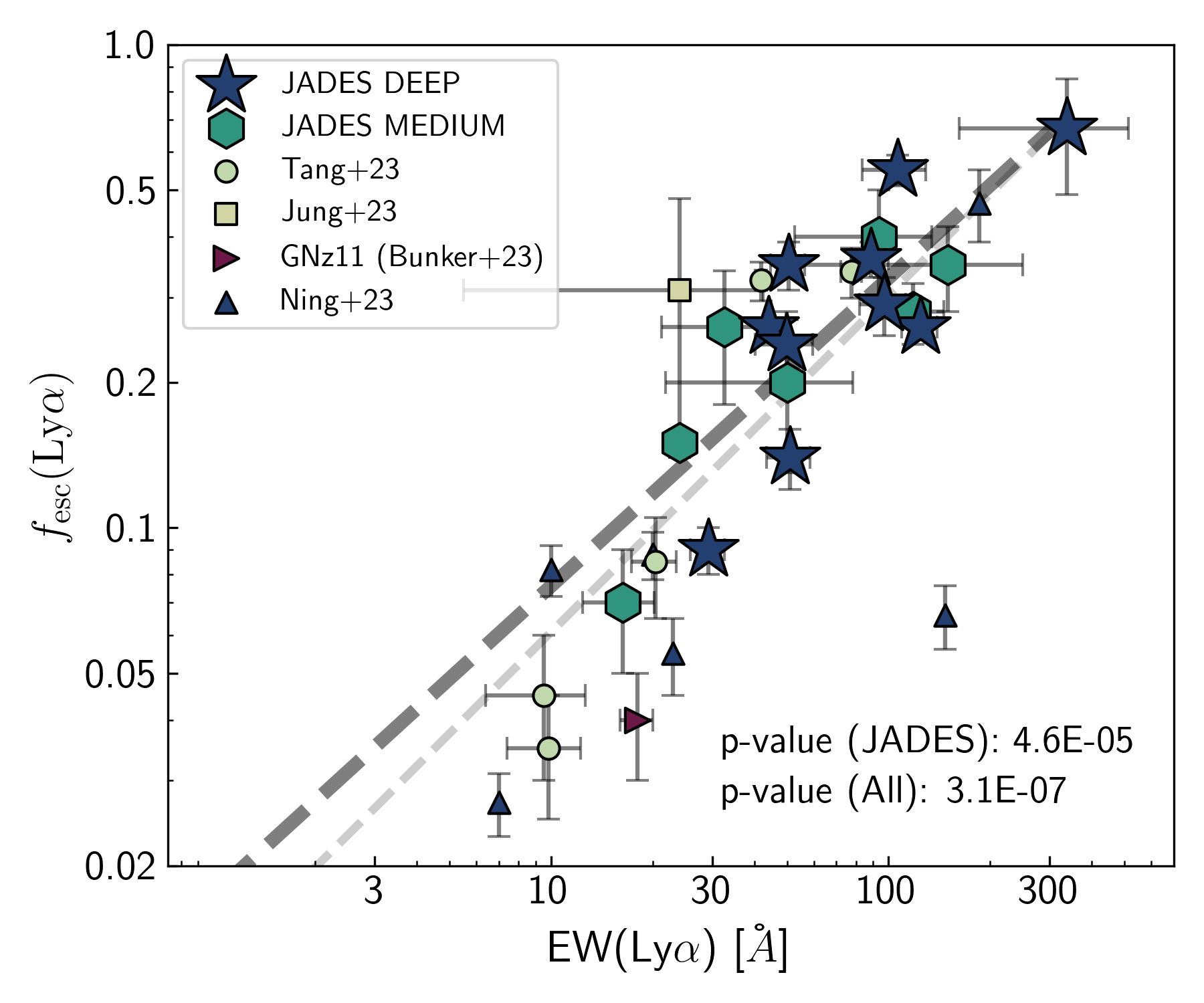}
    \includegraphics[width=0.49\linewidth]{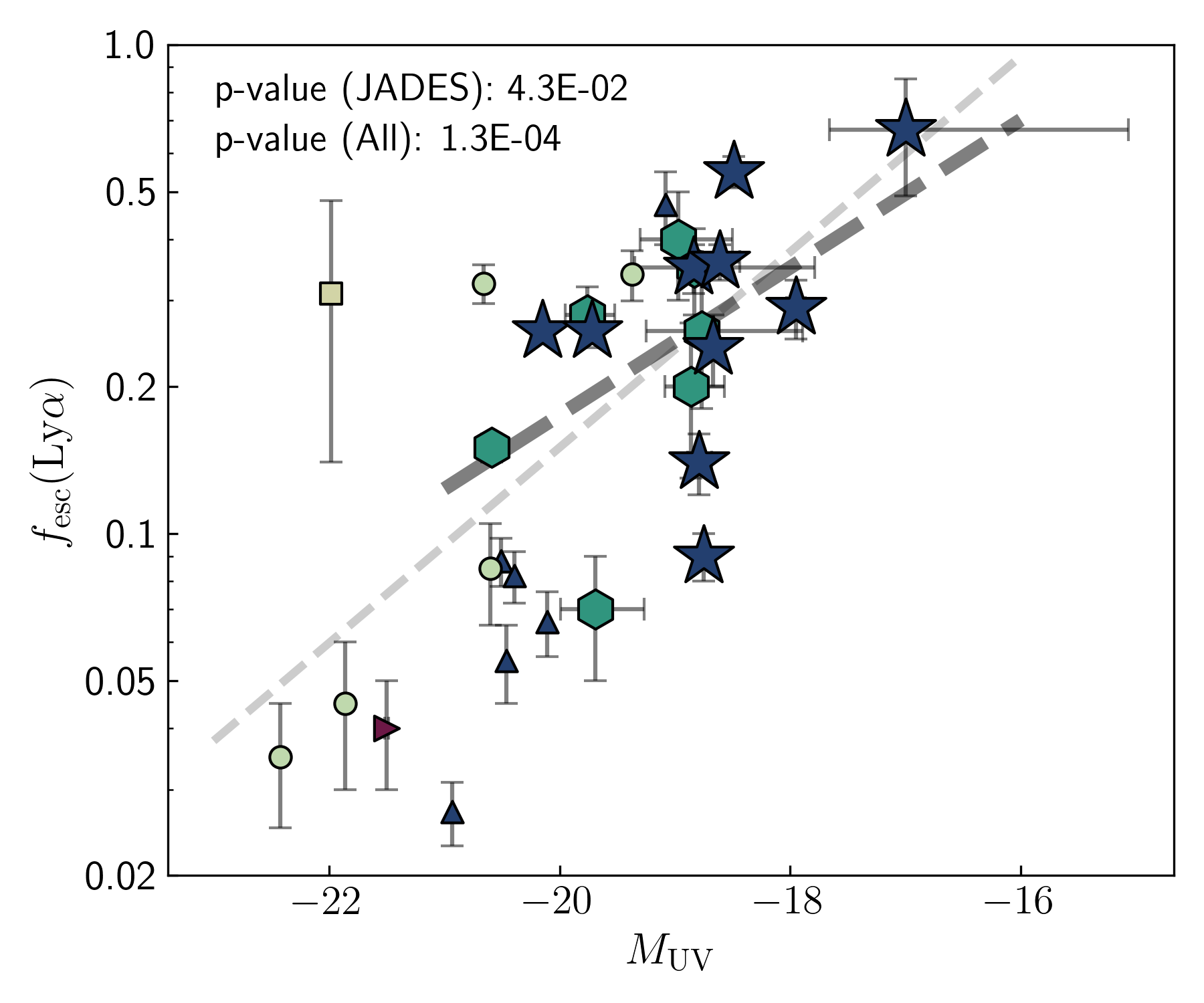}
    
    \includegraphics[width=0.49\linewidth]{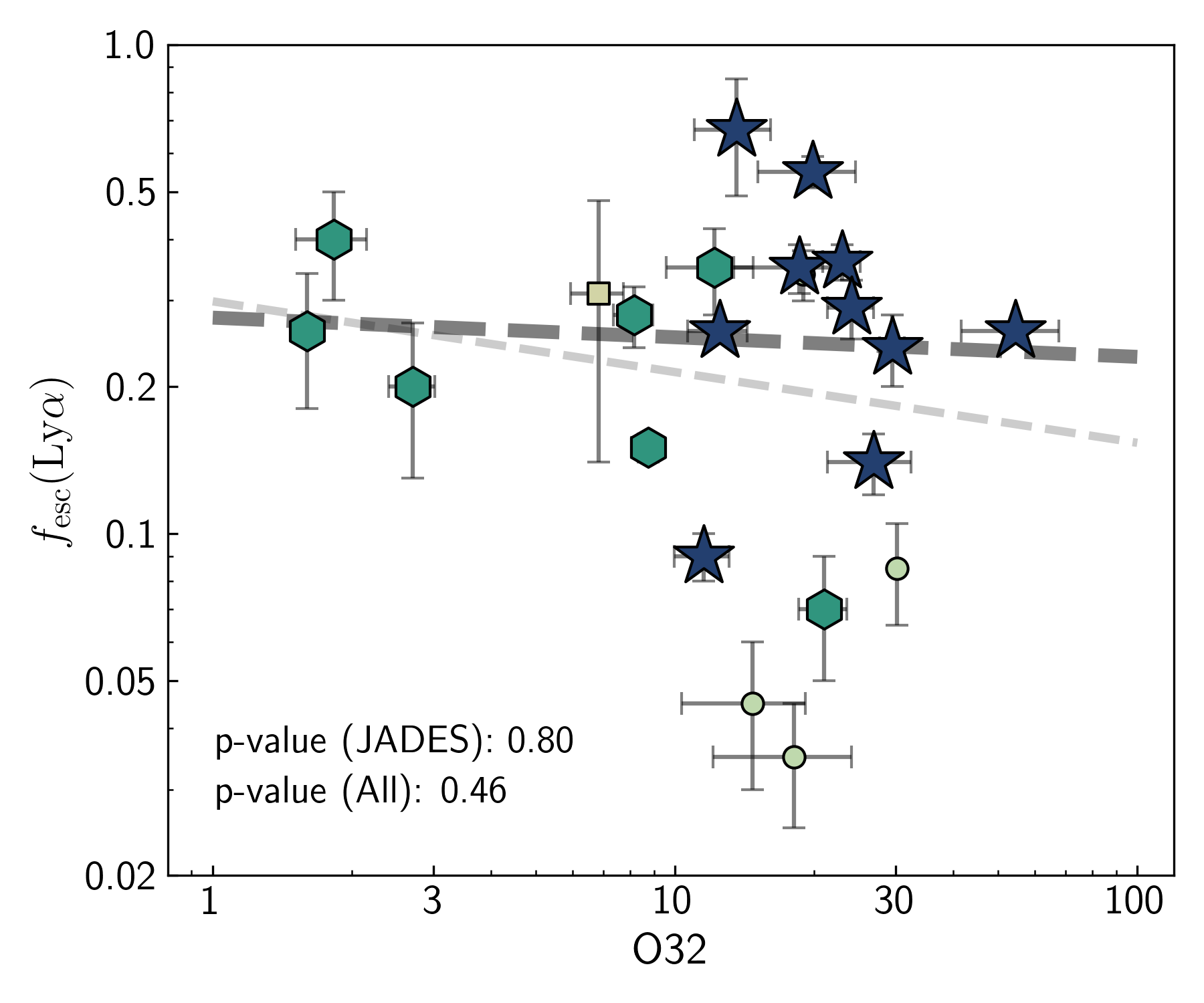}
    \includegraphics[width=0.49\linewidth]{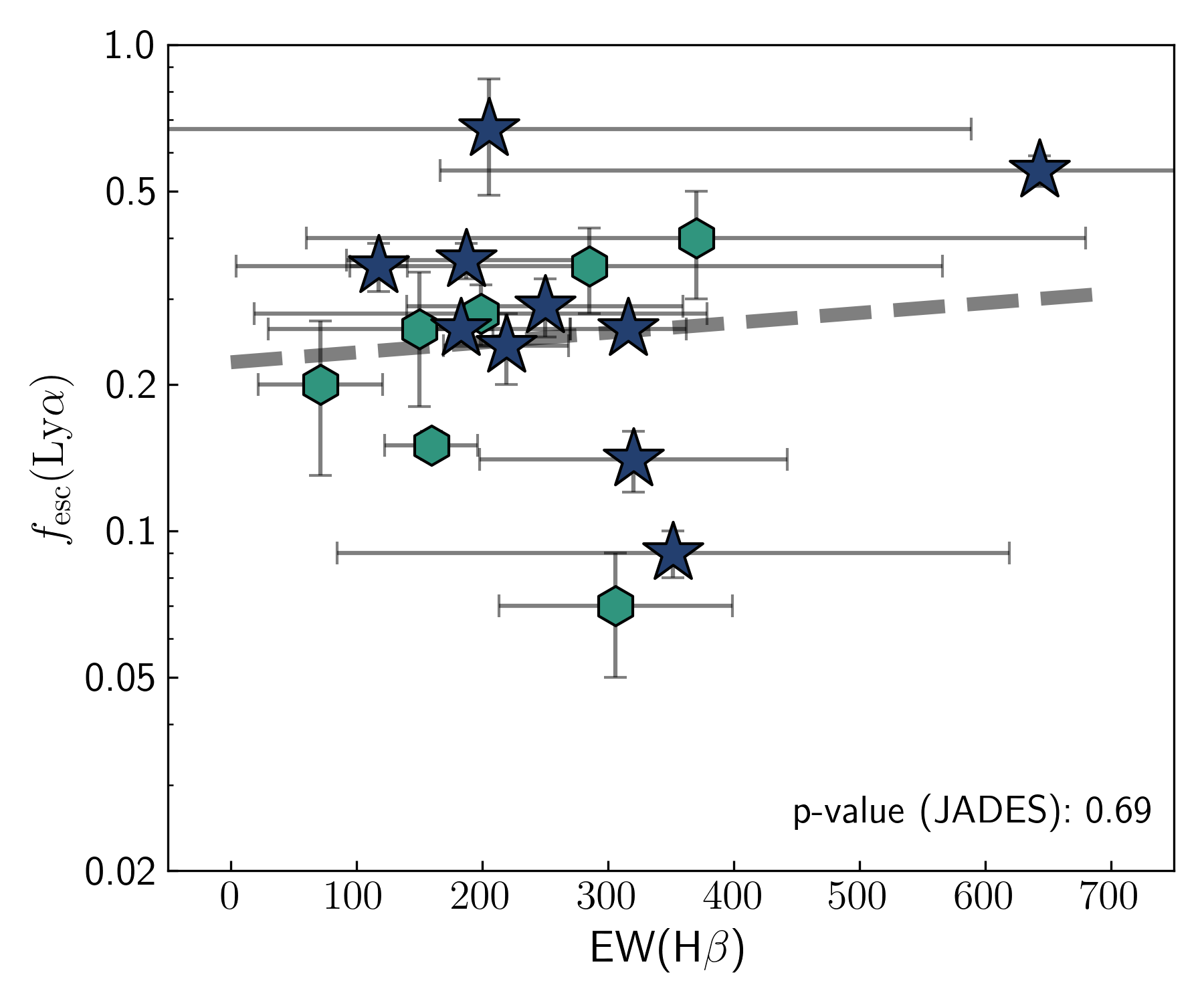}
    \caption{Dependence of the \lya\ escape fraction on EW(\lya) (top left), $M_{\rm{UV}}$ (top right), \oiii/\oii\ ratio (O32) (bottom left) and EW(\hb) (bottom right). The correlations follow the same convention as in Figure \ref{fig:veloffset}. Unsurprisingly, we find that \fesc(\lya) correlates strongly with EW(\lya). We also note that UV-faint galaxies show higher \fesc(\lya), although this could be attributed purely to the flux limited nature of our spectroscopic survey. We do not find strong correlations between \fesc(\lya) and O32 or \hb\ strength, although as previously noted we do find that our faint LAEs on average show high O32 ratios and \hb\ line strengths.}
    \label{fig:lya_fesc}
\end{figure*}

We now explore the dependence of \fesc(\lya) measured directly from the spectra with other galaxy properties and show the dependence of \fesc(\lya) on EW(\lya), $M_{\mathrm{UV}}$, O32 and EW(\hb) in Figure \ref{fig:lya_fesc}.

We find a strong correlation at high significance levels between \fesc(\lya) and EW(\lya) for both our JADES LAEs (p-value = $4.6\times10^{-5}$) as well as the extended sample (p-value = $3.1\times10^{-7}$) (Figure \ref{fig:lya_fesc}, top left) Since \fesc(\lya) is calculated using the observed ratio of \lya\ to \ha\ (or \hb) emission, this strong correlation suggests that the \ha\ (or \hb) line fluxes do not scale in proportion with the \lya\ escape fractions in LAEs with higher EW \lya\ emission. 

We also find that the observed \fesc(\lya) (just like EW(\lya)) increases consistently with decreasing UV magnitudes (Figure \ref{fig:lya_fesc}, top right), with strong correlations seen both for the JADES sample (p-value $=4.3\times10^{-2}$) and the full sample (p-value $=1.3\times10^{-2}$). This may be indicative of decreasing neutral gas covering fractions that potentially play a more important role in dictating the strength of the observed \lya\ emission than the intrinsic production of ionizing photons. 

However, We note that the lack of low \fesc(\lya) (or low EW(\lya)) detections from the faintest galaxies may also be a consequence of the flux limited nature of the spectroscopic data used in this study. It is also worth noting that the lack of high \fesc(\lya) observations from the UV-brightest galaxies is once again indicative of increasing neutral gas fractions in more luminous/massive systems, which likely attenuates and/or scatters the \lya\ flux along the line-of-sight.

Comparing with O32 and EW(\hb), we find that neither of these quantities correlates strongly with \fesc(\lya) across both the JADES-only sample as well as when including other LAEs (p-values $>0.4$). Several studies at lower redshifts have found that O32 correlates with \fesc(\lya) \citep[e.g.][]{yang17, flu22b}, whereas \citet{izotov20} reported no correlation between O32 and \fesc(\lya) for their sample of extreme \oiii\ emitters at low redshifts. We note that high O32 ratios and/or EW(\hb) are perhaps needed to have a higher chance of observing high \fesc(\lya) as was also noted by \citet{flu22b}, but there is considerable scatter in the O32 ratios that we measure for our LAEs, with some LAEs showing O32\,$<3$. Therefore, the O32 ratio may not be a good predictor of the expected \fesc(\lya) (and consequently LyC) from galaxies in the reionization era.

\subsection{Role of the IGM in attenuating Lyman-alpha emission at $z>6$}
\begin{figure*}
    \centering
    \includegraphics[width=\linewidth]{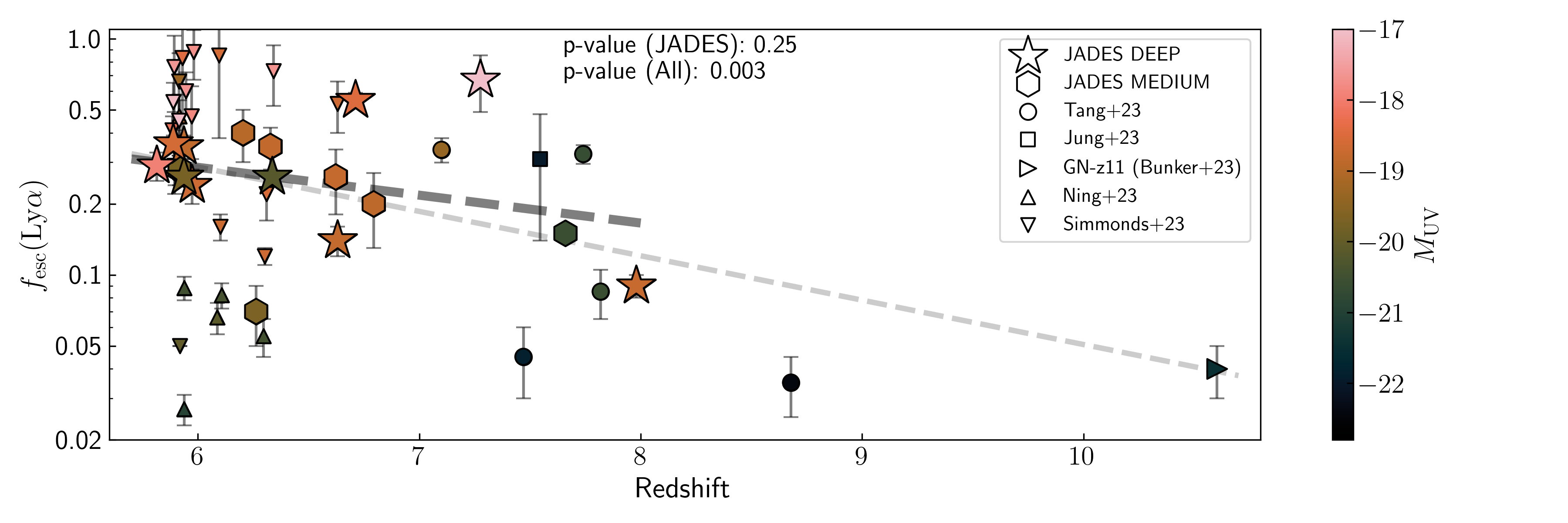}
    \caption{Evolution of \fesc(\lya) with redshift, colour-coded by the UV magnitude, with significance of the correlations following the convention from Figure \ref{fig:veloffset}. We see a gradual decrease of \fesc(\lya) at increasing redshifts, and at the same redshifts we see a decreasing \fesc(\lya) with increasing UV luminosity as was previously noted. The decrease of \fesc(\lya) with redshift could be driven by both selection biases as well as increasing IGM neutral fraction, and we explore the redshift evolution of only the UV-bright LAEs in Figure \ref{fig:fesc_redshift_bright}.}
    \label{fig:fesc_redshift}
\end{figure*}

\begin{figure*}
    \centering
    \includegraphics[width=\linewidth]{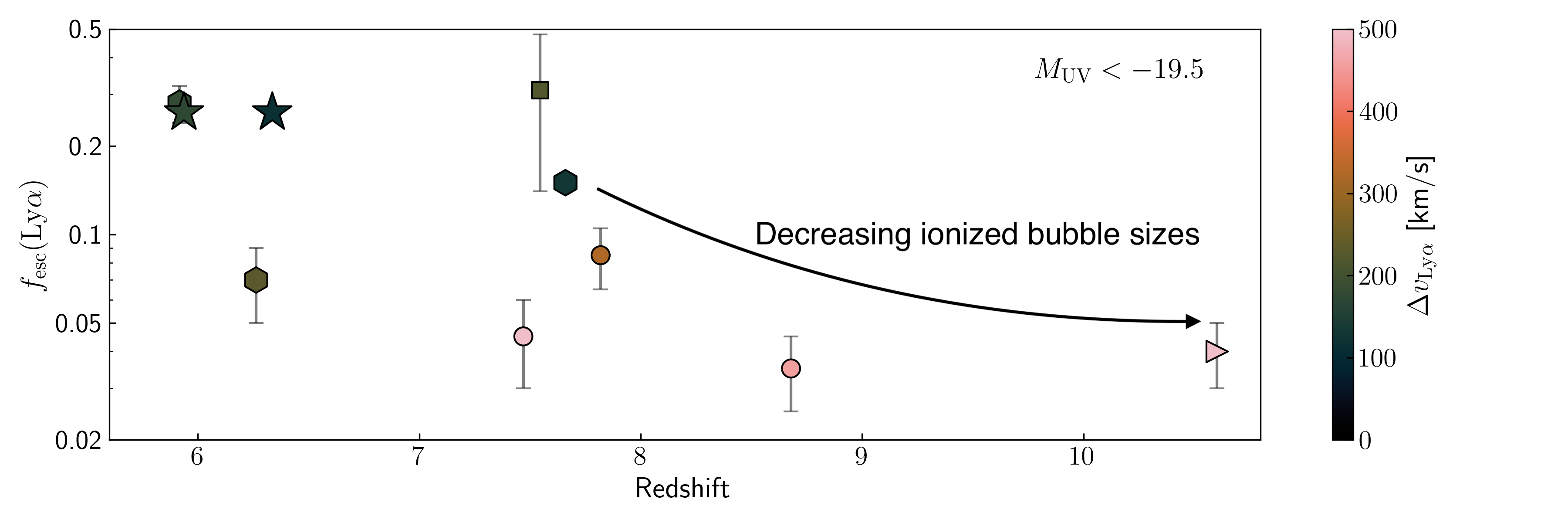}
    \caption{\fesc(\lya) as a function of redshift for galaxies brighter than $M_{\mathrm{UV}} < -19.5$, colour-coded by \lya\ velocity offsets. The symbols are the same as in Figure \ref{fig:fesc_redshift}. We find that the \lya\ escape fraction also seems to decrease with redshift for brighter galaxies, which is likely driven by the size of the ionized bubbles surrounding each galaxy indicated by the increasing $\Delta v_{\mathrm{Ly}\alpha}$ at the highest redshifts, indicating more neutral bubbles around the LAEs. However, increased neutral gas column densities at brighter UV magnitudes may also play a role in redshifting the emergent \lya\ emission.}
    \label{fig:fesc_redshift_bright}
\end{figure*}

Finally in this section, we look at the evolution of \fesc(\lya) with redshift, focusing particularly on $z\gtrsim6$ where the IGM is expected to play a dominant role in attenuating \lya\ emission, unless the LAEs live in large ionized bubbles. In Figure \ref{fig:fesc_redshift} we show \fesc(\lya) as a function of redshift, colour-coded by $M_{\mathrm{UV}}$ for our LAEs along with others known at $z\gtrsim5.8$. A decrease in the \fesc(\lya) is apparent: the decrease in \fesc(\lya) for only the JADES LAEs is less significant with p-value $=0.25$, however when including all the LAEs in the redshift of interest from the literature, the decrease is highly significant with p-value $=0.003$. Admittedly, the high significance may be driven by the UV magnitude limited nature of these surveys, where \lya\ detections from rarer galaxies at the highest redshifts is challenging purely from an observational point of view.

It is also interesting to note that at any given redshift, UV-fainter galaxies exhibit higher \fesc(\lya), as we had previously noted. The question that arises from this is, what is truly driving the decrease of \fesc(\lya) with redshift? Is the increasing neutrality of the IGM with redshift more dominant than the observational biases associated with being able to only observe UV-bright galaxies at high redshifts from flux limited studies?

To explore this effect, in Figure \ref{fig:fesc_redshift_bright} we show the \fesc(\lya) as a function of redshift for only the UV-brightest galaxies with $M_{\mathrm{UV}} < -19.5$ (arbitrarily chosen) to study the evolution in a more flux complete sample of LAEs across a large redshift baseline. This time around, we colour code the data points with \lya\ velocity offset, which is a good proxy for the size of the ionized bubble around the LAE. 

Figure \ref{fig:fesc_redshift_bright} clearly shows that for UV-bright galaxies at $z>6$, the decrease in \fesc(\lya) is accompanied by an increasing \lya\ velocity offset at the highest redshifts, indicating that the decline of EW(\lya) seen with redshift in UV-bright LAEs is likely driven by the reduction in the sizes of the ionized bubbles traced by the \lya\ velocity offset from systemic redshift. This demonstrates that increased IGM attenuation at the highest redshifts is playing an important role in the observed evolution of EW(\lya), at least within the UV-brightest sample, by attenuating the emergent \lya\ flux and the measured \fesc. The increasing neutral fraction of the IGM from our sample is also clearly seen in the companion paper by \citet{jon23}. 

However, we also note that very high stellar masses for some of the UV-brightest galaxies in the literature sample \citep[e.g.][]{end22c}, accompanied by high neutral gas densities and dust may also attenuate \lya\ emission close to the systemic redshift, leading to the same decrease in \fesc(\lya) and increase in the observed \lya\ velocity offset. This is similar to the degeneracies between the effect of the ISM vs the IGM in shaping \lya\ properties that was discussed earlier.

Therefore, a combination of selection effects as well as increasingly neutral IGM, which manifests itself as smaller ionized bubbles around LAEs at the highest redshifts play an important role in regulating \fesc(\lya). Estimates on the sizes of ionized regions around JADES LAEs have been presented in a companion paper \citep{witstok23} and offer a powerful probe of the spatial as well as temporal evolution of the IGM neutral fraction in this field.

The comparisons we have presented in this section demonstrate that to use \lya\ emission to infer significant LyC photon leakage from galaxies in the reionization era, both high \fesc(\lya) and low \lya\ velocity offsets compared to systemic redshift are required. We have found that the dependence of both quantities on other spectroscopic and photometric galaxy properties are filled with complexity and are impacted by observational biases (most importantly the flux limited nature of spectroscopic observations in a field). However, the detection of \lya\ emission from a $z>6$ galaxy is a powerful probe nonetheless at identifying LyC leakage, as has also been noted at lower redshifts \citep{ver17, izo21, sax22b}.

In the next section we attempt to move beyond a simple \fesc(\lya) and use all of the available spectroscopic and photometric indicators to estimate \fesc(LyC), which is the quantity that is needed to capture the contribution of galaxies to the reionization budget of the Universe at $z\gtrsim6$.

\section{Implications for LyC photon production, escape and reionization}
\label{sec:reionization}
\begin{figure*}
    \centering
    \includegraphics[width=0.49\linewidth]{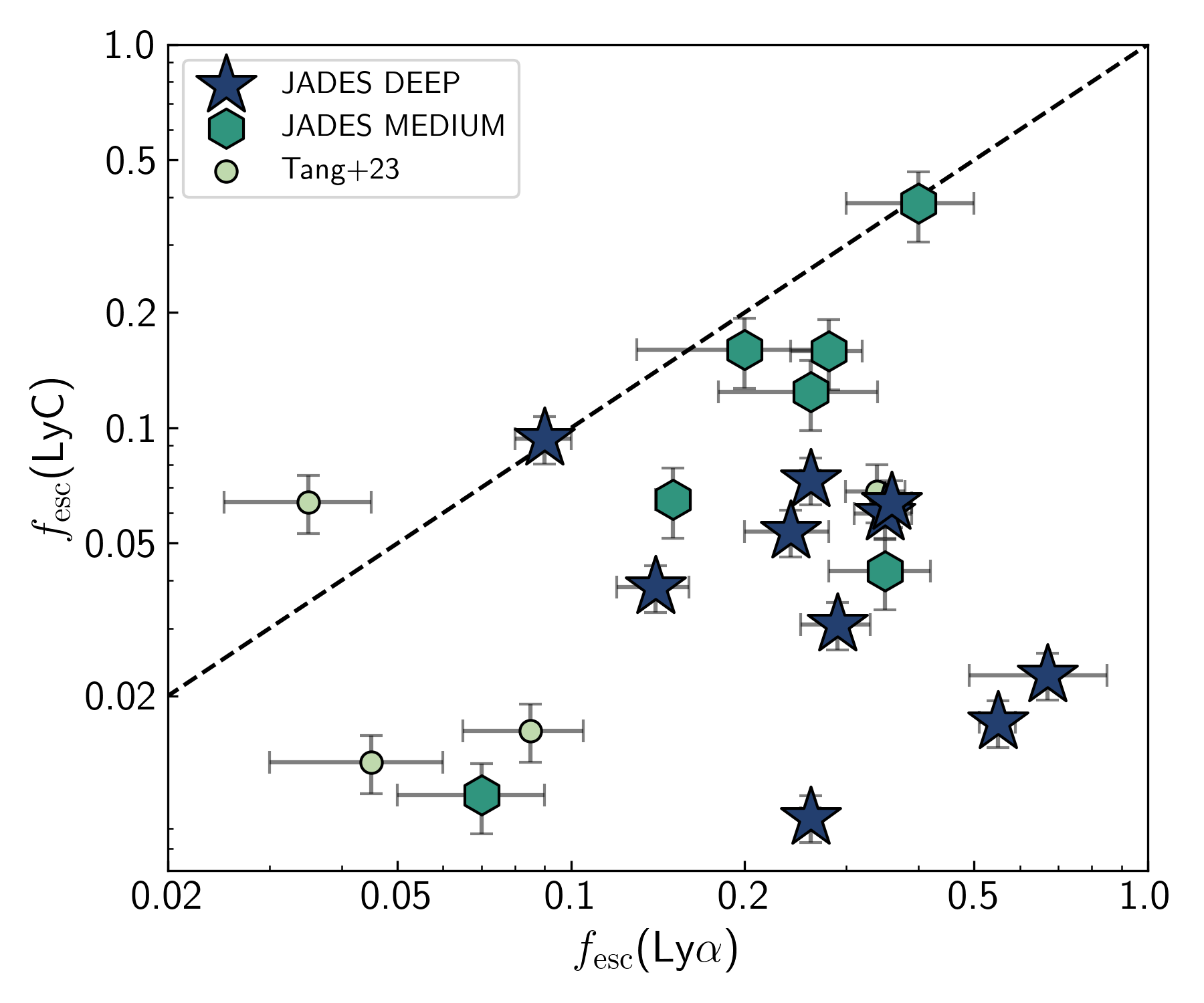}
    \includegraphics[width=0.49\linewidth]{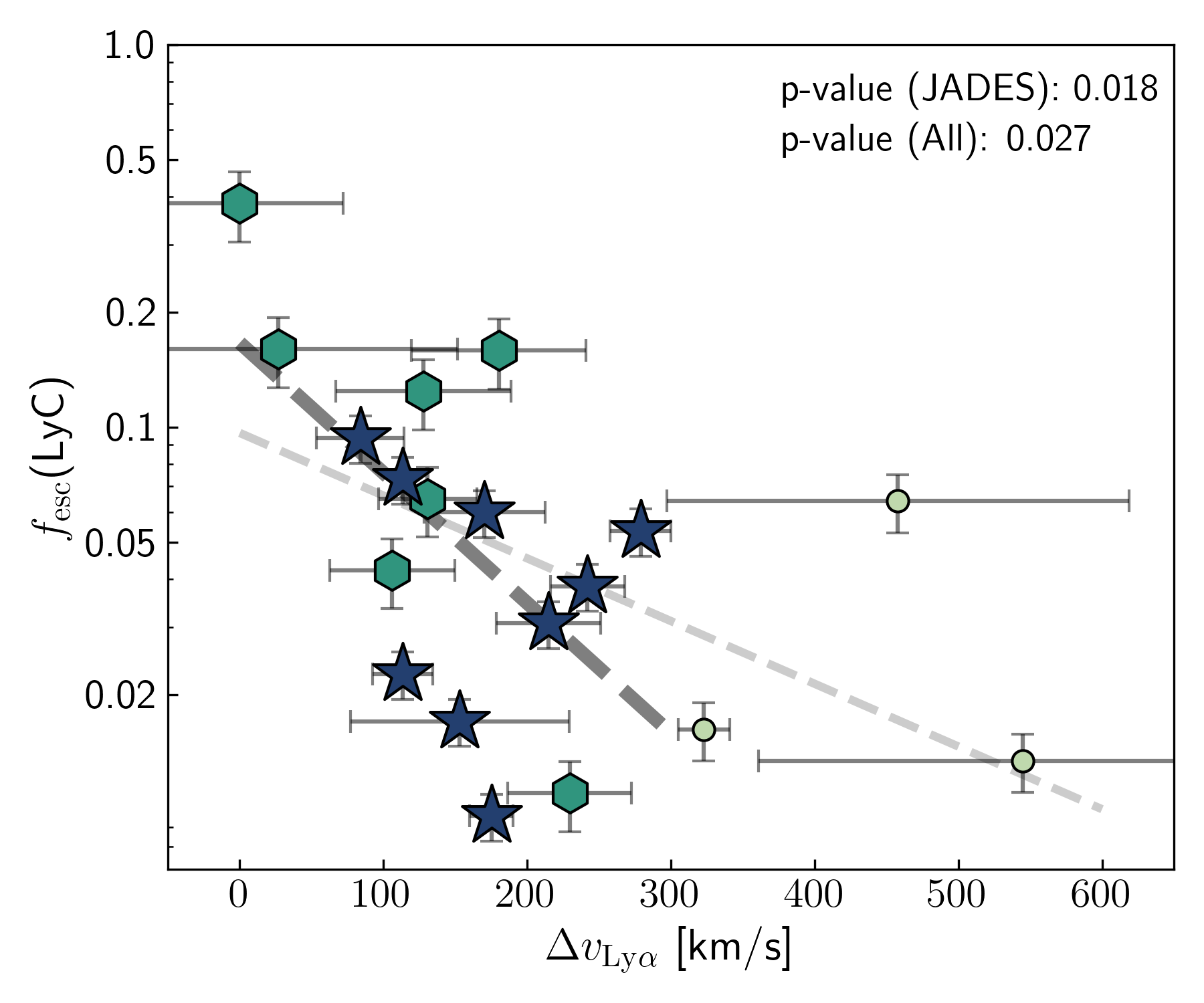}
    \caption{Comparison of the calculated \fesc(LyC) using the relation from \citet{cho23} with the observed \fesc(\lya) (left, with the one-to-one relation shown as the dashed line) and \lya\ velocity offset (right) for LAEs. With the exception of two LAEs from \citet{tan23}, the \fesc(LyC) we infer is always lower than \fesc(\lya), consistent with model predictions \citep[e.g.][]{maj22}. Interestingly, one of our faint LAEs has \fesc(LyC)\,$>0.2$. We find that qualitatively, \fesc(LyC) and \lya\ velocity offset anti-correlate, but a low \lya\ velocity offset does not necessarily guarantee high \fesc(LyC). Similar trends were reported for low redshift LyC leaking galaxies by \citet{izo21}.}
    \label{fig:lya_lyc}
\end{figure*}

Although the presence of strong \lya\ emission peaking close to the systemic velocity has been used to infer high LyC escape fractions \citep{ver15, izo21, nai22}, the physics that control the escape of LyC photons from star-forming galaxies are much more complicated \citep[e.g.][]{dij16, bar20, kat20, gar21, maj22, cho23}. The neutral gas content within a galaxy, in particular, can affect the \lya\ and LyC photons differently, which combined with the line-of-sight dependence of both \lya\ and LyC photon escape can often complicate the inference of LyC photon escape from \lya\ alone. For example, one of the most well-studied LyC leakers, \emph{Ion1} at $z\approx3.8$ \citep{van12, ji20}  actually does not show any \lya\ emission, which demonstrates the complex relationship between \lya\ and LyC photons.

Therefore, in this section we fold in other photometric and spectroscopic properties of our faint LAEs to make a more informed inference on the LyC escape fractions. Several observational studies as well as simulations have attempted to connect the leakage of LyC photons to spectroscopic properties. Some of the most exciting observational results linking LyC leakage to galaxy properties are being delivered by the Low-z Lyman Continuum Survey (LzLCS; \citealt{flu22a, flu22b}). State-of-the-art high-resolution cosmological simulations such as \textsc{sphinx}$^{20}$ are also now being used to study the dependence of LyC photon escape on galaxy properties \citep[e.g.][]{ros22, kat23a, cho23}, which offers much more control on the sample sizes and selection functions when attempting to use observations of galaxy properties to predict \fesc(LyC).

Using the \textsc{sphinx}$^{20}$ simulation, \citet{cho23} specifically focused on observables that trace conditions within galaxies that enable both the production as well as escape of LyC photons. Briefly, these conditions mainly require the galaxies to have (i) relatively high star formation rates (sSFR $>10^{-9}$\,yr$^{-1}$); (ii) stellar ages in the range $3.5-10$\,Myr, a time long enough for the first generation of supernovae to have cleared out channels in the ISM for LyC escape, while short enough that UV photons are still being produced in abundance by the stellar population, and (iii) low dust and neutral gas content. Using these criteria, \citet{cho23} report a six-parameter equation to predict the angle-averaged (and not sight-line dependent) \fesc(LyC) based on observed galaxy properties. These parameters include the UV slope, $\beta$, dust attenuation $E(B-V)$ (typically measured from the Balmer line decrement), \hb\ line luminosity, $M_{\mathrm{UV}}$, R23 and O32.

The predicted \fesc(LyC) from \citet{cho23} have been validated against comprehensive observational efforts to measure \fesc(LyC) from low redshift galaxies such as LzLCS \citep{flu22a, flu22b}. Each of the individual diagnostics for LyC leakage analysed by \citet{cho23} agree with what was reported for a handful of candidate LyC leaking galaxies by LzLCS. Further, dependence of \fesc(LyC) on one of the most promising indicators, the UV slope ($\beta$), was also found to agree with observed trends reported by \citet{chi22}. This validation lends credibility to the LyC leakage predictions and the multi-variate fits derived by \citet{cho23} to estimate \fesc(LyC) for galaxies with good spectroscopic measurements.

Therefore, we use the relationship between \fesc(LyC) and galaxy properties derived by \citet{cho23} to predict \fesc(LyC) for our sample of LAEs. The choice of using LyC leakage predictors from \citet{cho23} over other observational predictors has been made for two main reasons: As \citet{cho23} demonstrated, single diagnostic based LyC leakage predictors often suffer from scatter, where many ISM/dust/stellar conditions often being necessary but not sufficient for high LyC \fesc. These observations may also suffer from sample selection biases, which can be easily overcome when using data from simulations. Secondly, using a multi-variate \fesc(LyC) predictor helps encapsulate several competing physical processes within galaxies that eventually result in LyC leakage. With our JWST spectra, these multi-variate measurements are now possible, thereby maximizing the information that can be derived about LyC leakage from our galaxies.

All of the input parameters required to predict \fesc(LyC) have been observed for our faint LAEs from JADES, which makes predicting \fesc(LyC) using Equation (4) from \citet{cho23} relatively straightforward. We note that to predict the angle-averaged \fesc(LyC), the properties of the \lya\ emission line, which can often be highly sight-line dependent, are not taken into account by \citet{cho23} when estimating \fesc(LyC) \citep[c.f.][]{maj22}. The calculated \fesc(LyC) for our LAEs are given in Table \ref{tab:derived}.

In Figure \ref{fig:lya_lyc} we show \fesc(LyC) calculated using observed galaxy properties compared with \fesc(\lya) measured directly from the spectra (left) and the \lya\ velocity offset from the systemic (right) from our faint LAEs as well as those that we calculate using observed quantities from \citet{tan23}. We find that the predicted \fesc(LyC) remains below \fesc(\lya) for all our LAEs, but one LAE from \citet{tan23} have higher \fesc(LyC) compared to \fesc(\lya).

In general we do not find any strong correlation between \fesc(LyC) and \fesc(\lya). We do find that \fesc(LyC) significantly anti-correlates with \lya\ velocity offset for both our JADES LAEs as well as when brighter LAEs from \citet{tan23} are included (Figure \ref{fig:lya_lyc}, right), but we note that a low ($\lesssim 200$,\kms) \lya\ velocity offset does not guarantee a high \fesc(LyC) and that there is considerable scatter in the plot. From a sample of low redshift LyC leakers, \citet{izo21} also found that \fesc(LyC) tends to always be lower than \fesc(\lya), and that \fesc(LyC) weakly anti-correlates with \lya\ velocity offset from systemic, consistent with our findings.

We do, however, note the lack of any LAE with high velocity offsets showing high \fesc(LyC), which seems to suggest that low \lya\ velocity offsets are a necessary but not a sufficient condition to enable high LyC photon escape, mainly tracing the absence of high column density neutral gas, and that galaxies that show large \lya\ velocity offsets compared to systemic likely trace highly dense neutral gas conditions, which may not be conducive for significant LyC escape fractions. 

\begin{figure}
    \centering
    \includegraphics[width=\linewidth]{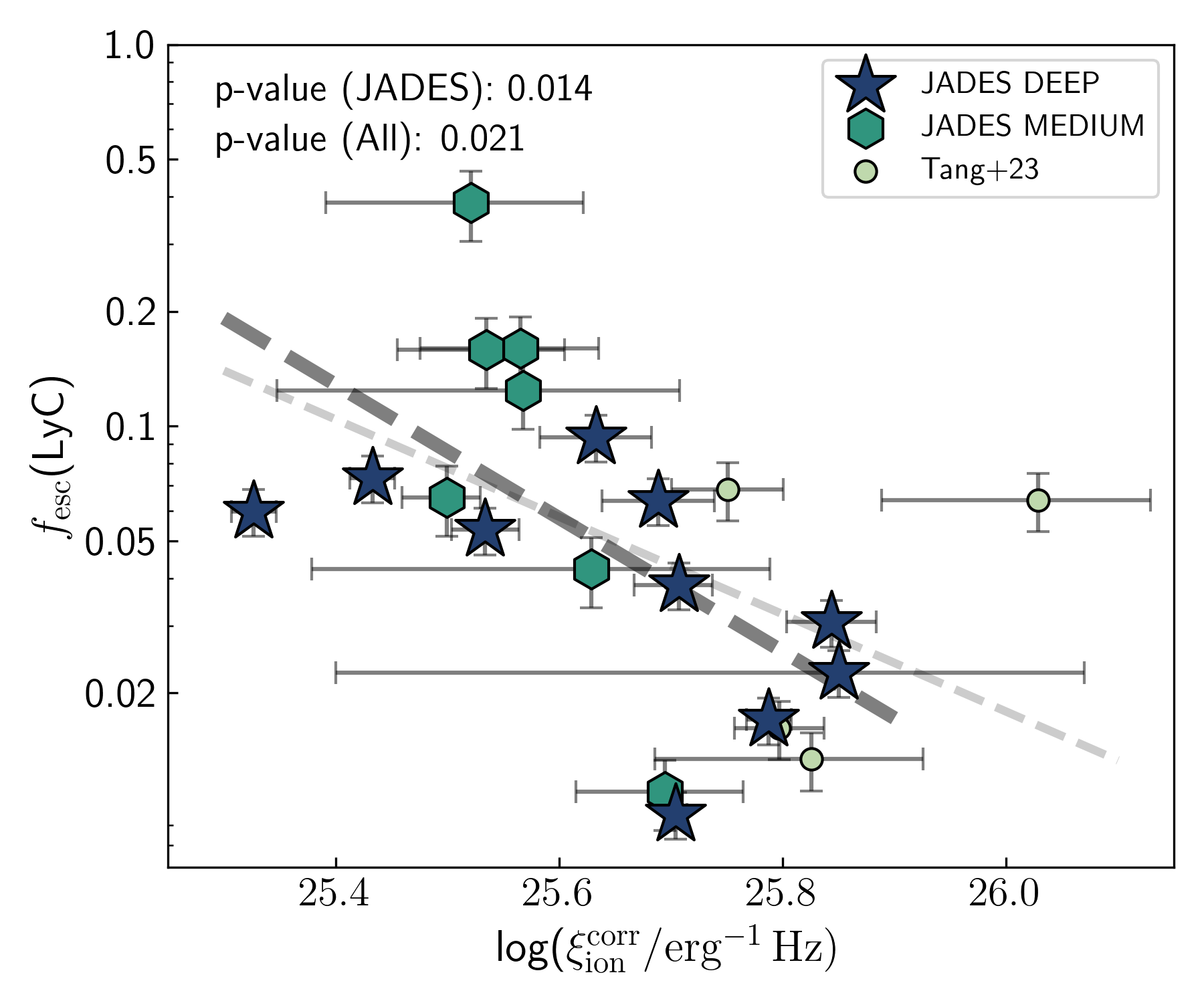}
    \caption{Corrected \xiion\ vs \fesc(LyC) calculated using the multi-parameter fit as described in the text. We find that LAEs that show higher \fesc(LyC) have lower measured \xiion\ compared to the sample average, which implies that a significant fraction of escaping ionizing photons will lead to decreased Balmer (and potentially nebular) line strengths at a given star formation rate \citep[e.g.][]{top22}. Additionally, this lack of correlation may also arise due to a time delay between the production and escape of ionizing photons from galaxies, whereby intense star formation activity that produces ionizing photons needs time to clear out channels to also facilitate LyC escape, as has been reported from several high-resolution simulations \citep[e.g.][]{bar20, kat20, cho23}.}
    \label{fig:lyc_xiion}
\end{figure}

Before assessing the co-dependence of ionizing photon production and escape from our faint LAEs, we note that when calculating \xiion\ for our LAEs using Equation (\ref{eq:xiion}) in Section \ref{sec:xiion} we assumed \fesc(LyC) to be zero. However, with \fesc(LyC) predictions for our LAEs, we now calculate \xiioncorr, which is corrected for the fraction of ionizing photons that escape out of the galaxy, thereby not contributing towards line emission. In this section going forward, we use the corrected \xiion\ value, \xiioncorr.

We now explore the dependence of \fesc(LyC) on the corrected ionizing photon production efficiencies in Figure \ref{fig:lyc_xiion}. We find that sources with the highest \fesc(LyC) do not necessarily show high values of \xiioncorr. Interestingly, there is am anti-correlation between the two quantities that seems significant based on the p-values, which may be not be entirely unexpected. When non-negligible fractions of ionizing photons begin escaping from the galaxy, there are fewer photons available to produce the Balmer line (as well as strong nebular line) emission \citep[e.g.][]{top22}. This would lead to low \xiioncorr\ values inferred when simply assuming Case-B recombination, which can consistently explain this observed mild anti-correlation. 

Another important effect that may be driving the scatter between \fesc(LyC) and \xiioncorr\ could be the expected time delays between significant production of ionizing photons and the emergence of escape channels that facilitate the escape of those photons. As noted in \citet{cho23} (but see also \citealt{bar20, kat20}), for a burst of star formation it is not up until $\sim3.5$\,Myr since the starburst is triggered that supernovae begin to clear channels in the ISM to allow significant LyC photon escape. Very early on in the starburst, there is a very high production rate of ionizing photons, but these photons are unable to escape out of the H\,\textsc{ii} regions. Therefore, the age of starburst and the time delay between the peak of ionizing photon production and the emergence of escape channels may lead to the observed scatter.

Finally, we explore the dependence of the product of the ionizing photon production efficiency and the ionizing photon escape fraction (\fesc(LyC)\,$\times$\,\xiioncorr), which is an important quantity needed to assess the contribution of individual star-forming galaxies to the reionization budget of the Universe, with UV magnitude, EW of \lya\ emission and its evolution with redshift. To overcome any potential sample selection biases, we restrict our analysis only to the JADES LAEs here, for which the selection functions and completeness is fairly well understood. The average ionizing photon output from our JADES LAEs is log(\fesc(LyC)\,$\times$\,\xiioncorr/erg$^{-1}$\,Hz) $= 24.49^{+0.29}_{-0.39}$. This measurement is in agreement with that reported by \citet{meyer19} for faint galaxy populations in the redshift range $4.5 < z < 6.2$. The correlations between \fesc(LyC)\,$\times$\,\xiioncorr\ and the above-mentioned quantities are shown in Figure \ref{fig:ionizing_output} and are discussed below. 
\begin{figure*}
    \centering
    \includegraphics[width=0.49\linewidth]{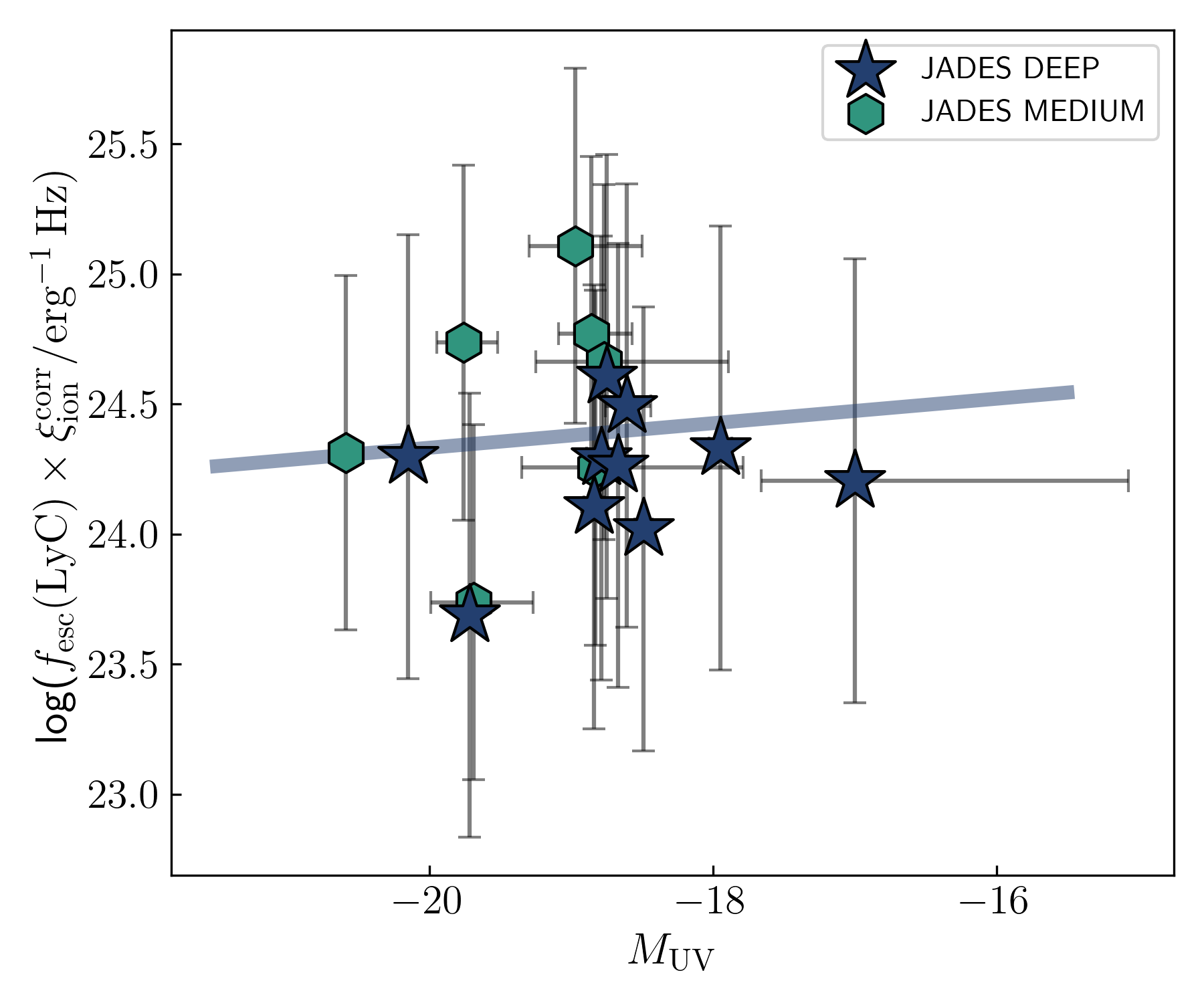}
    \includegraphics[width=0.49\linewidth]{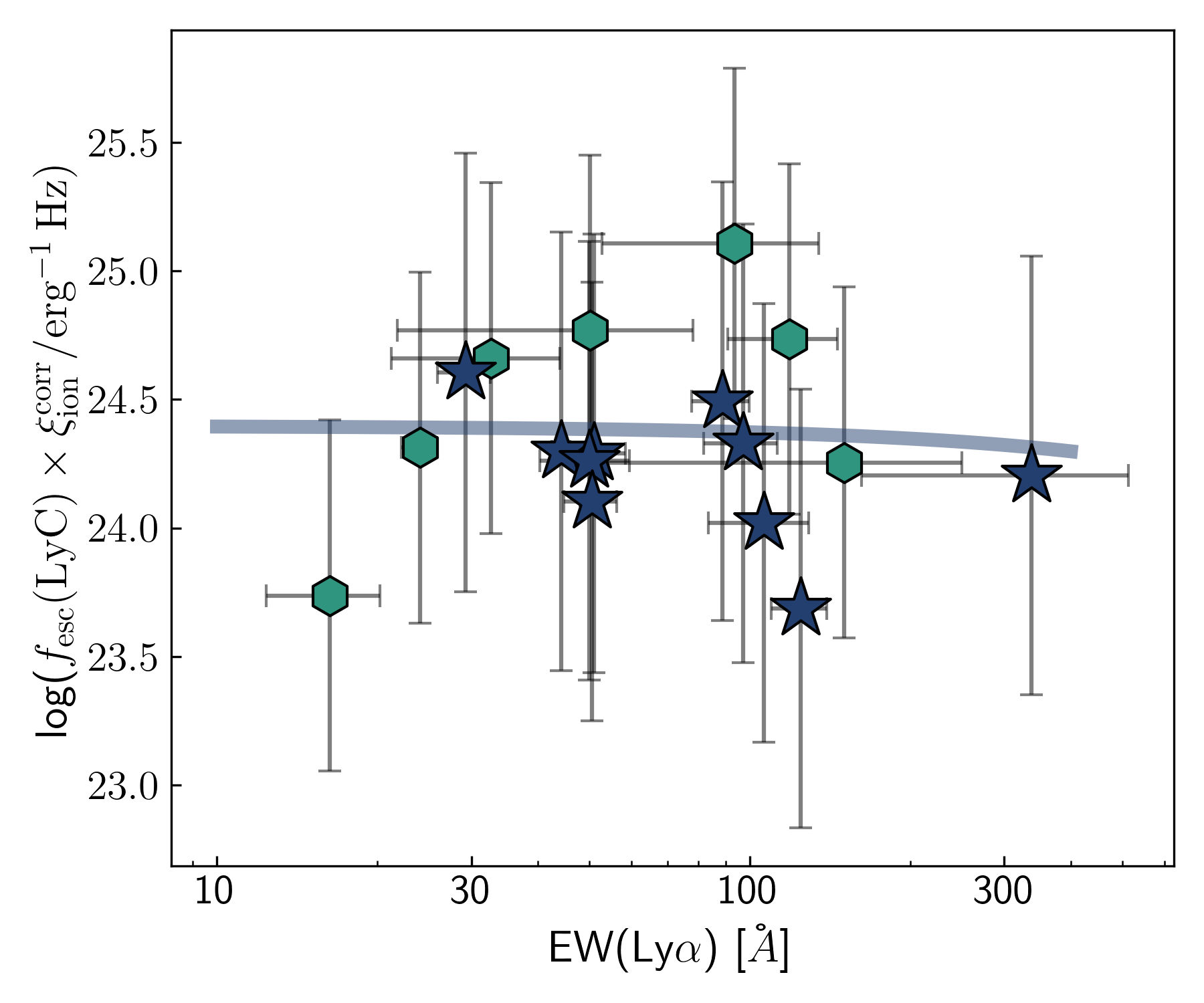}
    \includegraphics[width=\linewidth]{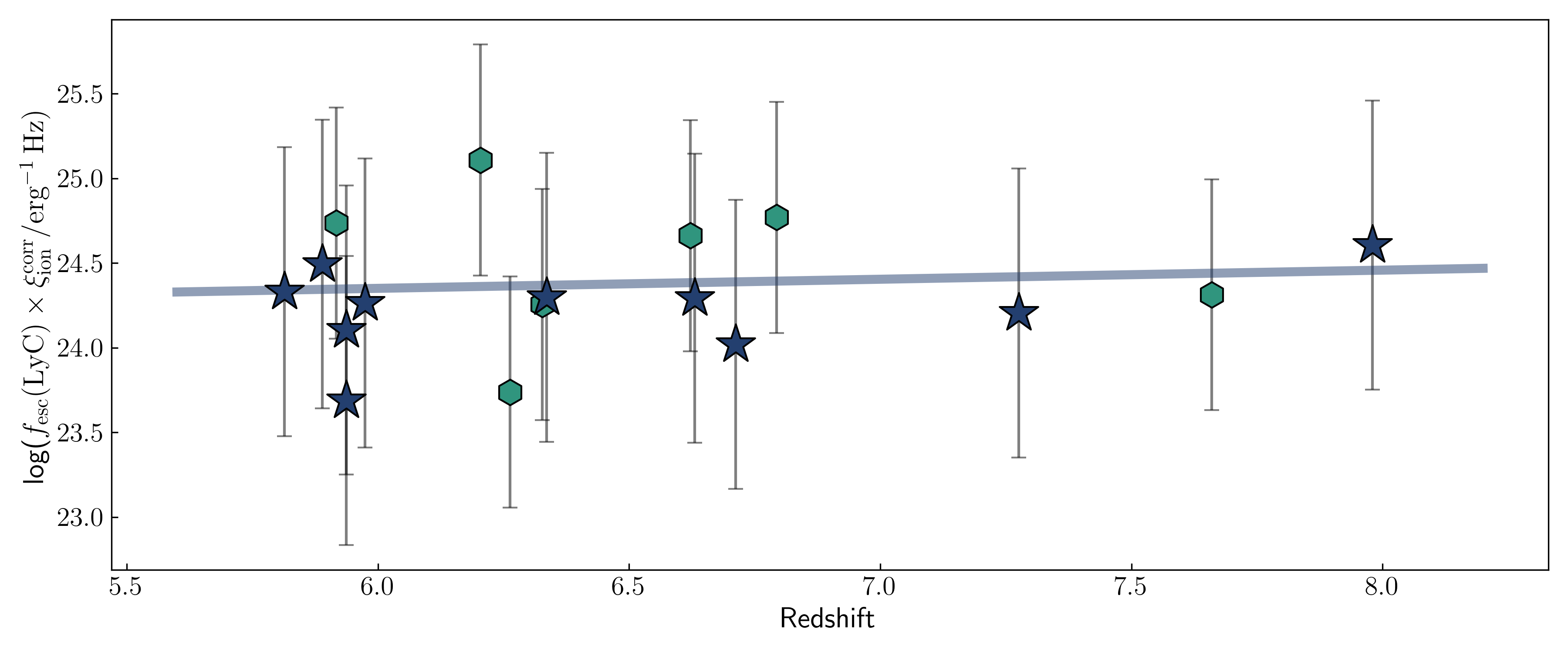}
    \caption{Dependence of the ionizing photon output from galaxies (that is \fesc(LyC)\,$\times$\,\xiion) as a function of UV magnitude (left) and \lya\ equivalent width (right). The best-fitting linear relations are shown as blue lines (with considerable scatter in the relation, which is not shown to preserve clarity). Overall, we do not find any strong dependence of the ionizing photon output on either the UV magnitude of the \lya\ strength. The ionizing photon output is key to quantifying the role of star-forming galaxies in contributing towards the reionization budget at $z>6$.}
    \label{fig:ionizing_output}
\end{figure*}

We begin by assessing the dependence of log(\fesc(LyC)\,$\times$\,\xiioncorr) on UV magnitude, finding that it increases very mildly with decreasing $M_\mathrm{UV}$ as shown in Figure \ref{fig:ionizing_output} (top left), following the linear relation:
\begin{equation}
    \log \left(\frac{f_\mathrm{esc} \times \xi_\mathrm{ion}^\mathrm{corr}}{\mathrm{erg}^{-1}\,\mathrm{Hz}} \right) = 0.047\,(\pm 0.014)\, M_{\mathrm{UV}} + 25.27\,(\pm 5.16).
\end{equation}
This lack of strong correlation for LAEs clearly disfavours enhanced ionizing photon output from UV-fainter LAEs, and may have important consequences for models of reionization.

We further find that the ionizing photon output from LAEs also remains roughly constant over a range of EW(\lya), best fit with the linear relation:
\begin{multline}
    \log \left(\frac{f_\mathrm{esc} \times \xi_\mathrm{ion}^\mathrm{corr}}{\mathrm{erg}^{-1}\,\mathrm{Hz}} \right) = -2.5\times10^{-4}\,(\pm 1.9\times10^{-6})\, \left(\frac{\mathrm{EW(\mathrm{Ly}\alpha)}}{\AA}\right) \\ + 24.39\,(\pm 0.02).
\end{multline}
The little to no dependence of log(\fesc\,$\times$\,\xiion) on the equivalent width of \lya\ emission is shown in Figure \ref{fig:ionizing_output} (top right; note that the x-axis is in log scale), implying that purely the observed \lya\ line strength is not a good independent indicator of the ionizing photon output from LAEs in the epoch of reionization. The added attenuation by the neutral IGM along the line of sight at $z>6$ and the inherent difficulty in disentangling the effects of the IGM and ISM/CGM at these redshifts likely contributes to the \lya\ line strength not being a robust tracer of the ionizing photon output.

Finally, we derive the dependence of the ionizing photon output of our LAEs with redshift, finding the best-fitting relation:
\begin{equation}
    \log \left(\frac{f_\mathrm{esc} \times \xi_\mathrm{ion}}{\mathrm{erg}^{-1}\,\mathrm{Hz}} \right) = 0.053\,(\pm 0.025)\, z + 24.027\,(\pm 1.050).
\end{equation}
We find that the ionizing photon output of LAEs shows a mild increase with redshift as shown in Figure \ref{fig:ionizing_output} (bottom). This is consistent with a picture whereby younger galaxies may be able to achieve higher ionizing photon production efficiencies, but the relatively mild evolution is also consistent with a picture whereby the production and escape of ionizing photons is dictated by physical processes that operate on much-shorter timescales (that is intense star formation or SNe activity), which cannot be captured via a strong trend with redshift.

The best-fitting relations of ionizing photon output with UV magnitude \lya\ EW and redshift can be used to estimate the total number of ionizing photons contributed by LAEs at a given redshift, depending on the space density of LAEs. Since the increasing neutrality of the IGM makes it impossible to obtain a complete \lya\ luminosity function at $z>6$, certain assumptions about the evolution of the \lya\ luminosity function may need to be made \citep[see][for example]{mat22}. 

Alternatively, if a good handle on the \lya\ emitter fraction (unaffected by the IGM attenuation) can be obtained at $z>6$ by extrapolating from fractions measured at lower redshifts \citep[e.g.][]{santos20}, then our best-fitting relations can also be used to estimate the total ionizing photon contribution towards the reionization budget from LAEs at $z>6$. Any dependence of the LAE fraction on UV magnitude must also be taken into account for this calculation \citep[see][]{jon23}.

\section{Conclusions}
\label{sec:conclusions}
In this study we have presented detailed properties of 17faint \lya\ emitting galaxies at $z>5.8$ from the \emph{JWST} Advanced Deep Extragalactic Survey (JADES) Deep and Medium Tier NIRSpec MSA surveys. These new \lya\ emitters, spanning absolute UV magnitudes of $-17.0$ to $-20.6$, are generally fainter compared to LAEs that were previously known in or near the epoch of reionization, opening up a new window into studying the properties of faint galaxies that reside within ionized bubbles in the reionization era.

Using measurements directly from the low resolution (R100) PRISM as well as medium resolution (R1000) grating spectra, we report the detection of other rest-frame optical emission lines such as \oii, \hb, \oiii\ and \ha. The detection of these lines enables a reliable measure of their spectroscopic redshift against which the velocity offsets of \lya\ emission can be accurately measured. In general, these LAEs have blue rest-UV spectral slopes ($-2.1$ to $-2.7$) and little to no dust measured from Balmer decrements.

Using rest-optical line ratios, we find that our LAEs appear to be metal poor with high ionization parameters, properties that are typical of \emph{JWST}-detected faint star-forming galaxies at $z>6$. These properties combined with steep UV slopes and no dust indicate that all of our LAEs are young, star-forming systems. We further measure the ionizing photon production efficiencies (\xiion) directly from Balmer line emission and find that our LAEs on average have log(\xiion/Hz\,erg$^{-1}) \approx 25.56$, which does not seem to evolve strongly with redshift. We also do not find a strong dependence of \xiion\ on the strength of \lya\ emission.

Using the \lya\ escape fraction (calculated using Balmer line emission) and the velocity offset of the peak of the \lya\ line compared to systemic redshift, we study the galaxy properties that govern the escape of \lya\ photons from reionization era galaxies. We note that the escape fraction of \lya\ photons is anti-correlated with its velocity offset from systemic, and is strongly correlated with \lya\ equivalent width, consistent with expectations from \lya\ emission models as well as high-resolution galaxy simulations employing radiative transfer to track the escape of \lya\ emission. 

We also find that LAEs that are fainter in the UV show higher \lya\ escape fractions, although this could be attributed to the flux limited nature of our spectroscopic surveys. We do not find strong correlations between \lya\ escape fraction or velocity offset with key ISM indicators such as \oiii/\oii\ ratios or \hb\ equivalent widths. We conclude that the escape of \lya\ emission is a complicated process, and may not necessarily depend strongly on the state of the ISM and stellar populations at any given time, especially at $z>6$ when the IGM attenuation also plays an important role.

We find a gradual decrease in \lya\ escape fractions with redshift, indicative of increasing IGM attenuation in diminishing \lya\ strengths at the highest redshifts. By making a UV cut to remove selection effects, we find that the \lya\ escape fraction still evolves weakly with redshift, but the escaping \lya\ emission is considerably more offset compared to systemic velocity at the highest redshifts, indicative of decreasing ionized fractions and sizes of the bubbles that must surround these LAEs.

Making use of several photometric and spectroscopic indicators for our LAEs, we then predict escape fractions of hydrogen ionizing Lyman continuum photons. We find that with the exception of one LAE in our sample, the LyC escape fraction is always lower than the \lya\ escape fraction, with no significant correlations between the two. We also do not find any significant correlation between LyC escape fraction and \xiion, which can likely explained by the reduced Balmer line emission in the presence of significant ionizing photon escape or by time delays between the production and escape of ionizing photons. 

By combining the production and escape of LyC photons (that is \fesc\,$\times$\,\xiion), we find that the quantity that is actually responsible for delivering ionizing photons from the galaxies to the IGM remains relatively consistent across UV magnitudes and EW(\lya), but increases gradually with redshift. Using these dependencies and assumptions about the \lya\ emitter fraction at any given redshift, more realistic models of reionization can be constructed.

Deeper and wider spectroscopic surveys in the future will help expand the samples of known LAEs in the reionization era. The availability of other spectroscopic indicators tracing the nature of stellar populations, ISM ionization and chemical conditions would be key to assess the role of \lya\ emitting galaxies in driving cosmic reionization, helping build more realistic models charting the reionization history of the Universe.

\begin{acknowledgements}
We would like to thank the referee for constructive comments that helped improve the quality of this manuscript. AS thanks Harley Katz, Richard Ellis, Jorryt Matthee and Anne Verhamme for insightful discussions. AS, AJB, GCJ, AJC and JC acknowledge funding from the “FirstGalaxies” Advanced Grant from the European Research Council (ERC) under the European Union’s Horizon 2020 research and innovation programme (Grant agreement No. 789056). JW, RM, MC, TJL, LS \& JS acknowledge support by the Science and Technology Facilities Council (STFC), ERC Advanced Grant 695671 "QUENCH". JW also acknowledges support from the Fondation MERAC. SA acknowledges support from the research project PID2021-127718NB- I00 of the Spanish Ministry of Science and Innovation/State Agency of Research (MICIN/AEI). RB acknowledges support from an STFC Ernest Rutherford Fellowship (ST/T003596/1). KB acknowledges support from the Australian Research Council Centre of Excellence for All Sky Astrophysics in 3 Dimensions (ASTRO 3D), through project number CE170100013. SC acknowledges support by European Union's HE ERC Starting Grant No. 101040227 - "WINGS". ECL acknowledges support of an STFC Webb Fellowship (ST/W001438/1). DJE is supported as a Simons Investigator and by JWST/NIRCam contract to the University of Arizona, NAS5- 02015. BDJ, BER and MR acknowledge support from the NIRCam Science Team contract to the University of Arizona, NAS5-02015. RS acknowledges support from a STFC Ernest Rutherford Fellowship (ST/S004831/1). The work of CCW is supported by NOIR-Lab, which is managed by the Association of Universities for Research in Astronomy (AURA) under a cooperative agreement with the National Science Foundation.  

This work is based on observations made with the NASA/ESA/CSA James Webb Space Telescope. The data were obtained from the Mikulski Archive for Space Telescopes at the Space Telescope Science Institute, which is operated by the Association of Universities for Research in Astronomy, Inc., under NASA contract NAS 5-03127 for JWST. These observations are associated with program \#1180 and \#1210.

\end{acknowledgements}

\bibliographystyle{aa}
\bibliography{GS_LAEs}

\begin{thebibliography}{144}
\expandafter\ifx\csname natexlab\endcsname\relax\def\natexlab#1{#1}\fi

\bibitem[{{Aihara} {et~al.}(2011){Aihara}, {Allende Prieto}, {An}, {Anderson},
  {Aubourg}, {Balbinot}, {Beers}, {Berlind}, {Bickerton}, {Bizyaev}, {Blanton},
  {Bochanski}, {Bolton}, {Bovy}, {Brandt}, {Brinkmann}, {Brown}, {Brownstein},
  {Busca}, {Campbell}, {Carr}, {Chen}, {Chiappini}, {Comparat}, {Connolly},
  {Cortes}, {Croft}, {Cuesta}, {da Costa}, {Davenport}, {Dawson}, {Dhital},
  {Ealet}, {Ebelke}, {Edmondson}, {Eisenstein}, {Escoffier}, {Esposito},
  {Evans}, {Fan}, {Femen{\'\i}a Castell{\'a}}, {Font-Ribera}, {Frinchaboy},
  {Ge}, {Gillespie}, {Gilmore}, {Gonz{\'a}lez Hern{\'a}ndez}, {Gott}, {Gould},
  {Grebel}, {Gunn}, {Hamilton}, {Harding}, {Harris}, {Hawley}, {Hearty}, {Ho},
  {Hogg}, {Holtzman}, {Honscheid}, {Inada}, {Ivans}, {Jiang}, {Johnson},
  {Jordan}, {Jordan}, {Kazin}, {Kirkby}, {Klaene}, {Knapp}, {Kneib},
  {Kochanek}, {Koesterke}, {Kollmeier}, {Kron}, {Lampeitl}, {Lang}, {Le Goff},
  {Lee}, {Lin}, {Long}, {Loomis}, {Lucatello}, {Lundgren}, {Lupton}, {Ma},
  {MacDonald}, {Mahadevan}, {Maia}, {Makler}, {Malanushenko}, {Malanushenko},
  {Mandelbaum}, {Maraston}, {Margala}, {Masters}, {McBride}, {McGehee},
  {McGreer}, {M{\'e}nard}, {Miralda-Escud{\'e}}, {Morrison}, {Mullally},
  {Muna}, {Munn}, {Murayama}, {Myers}, {Naugle}, {Neto}, {Nguyen}, {Nichol},
  {O'Connell}, {Ogando}, {Olmstead}, {Oravetz}, {Padmanabhan},
  {Palanque-Delabrouille}, {Pan}, {Pandey}, {P{\^a}ris}, {Percival},
  {Petitjean}, {Pfaffenberger}, {Pforr}, {Phleps}, {Pichon}, {Pieri}, {Prada},
  {Price-Whelan}, {Raddick}, {Ramos}, {Reyl{\'e}}, {Rich}, {Richards}, {Rix},
  {Robin}, {Rocha-Pinto}, {Rockosi}, {Roe}, {Rollinde}, {Ross}, {Ross},
  {Rossetto}, {S{\'a}nchez}, {Sayres}, {Schlegel}, {Schlesinger}, {Schmidt},
  {Schneider}, {Sheldon}, {Shu}, {Simmerer}, {Simmons}, {Sivarani}, {Snedden},
  {Sobeck}, {Steinmetz}, {Strauss}, {Szalay}, {Tanaka}, {Thakar}, {Thomas},
  {Tinker}, {Tofflemire}, {Tojeiro}, {Tremonti}, {Vandenberg}, {Vargas
  Maga{\~n}a}, {Verde}, {Vogt}, {Wake}, {Wang}, {Weaver}, {Weinberg}, {White},
  {White}, {Yanny}, {Yasuda}, {Yeche}, \& {Zehavi}}]{aih11}
{Aihara}, H., {Allende Prieto}, C., {An}, D., {et~al.} 2011, \apjs, 193, 29

\bibitem[{{Ajiki} {et~al.}(2003){Ajiki}, {Taniguchi}, {Fujita}, {Shioya},
  {Nagao}, {Murayama}, {Yamada}, {Umeda}, \& {Komiyama}}]{aji03}
{Ajiki}, M., {Taniguchi}, Y., {Fujita}, S.~S., {et~al.} 2003, \aj, 126, 2091

\bibitem[{{Arellano-C{\'o}rdova} {et~al.}(2022){Arellano-C{\'o}rdova}, {Berg},
  {Chisholm}, {Arrabal Haro}, {Dickinson}, {Finkelstein}, {Leclercq}, {Rogers},
  {Simons}, {Skillman}, {Trump}, \& {Kartaltepe}}]{arr22}
{Arellano-C{\'o}rdova}, K.~Z., {Berg}, D.~A., {Chisholm}, J., {et~al.} 2022,
  \apjl, 940, L23

\bibitem[{{Atek} {et~al.}(2008){Atek}, {Kunth}, {Hayes}, {{\"O}stlin}, \&
  {Mas-Hesse}}]{ate08}
{Atek}, H., {Kunth}, D., {Hayes}, M., {{\"O}stlin}, G., \& {Mas-Hesse}, J.~M.
  2008, \aap, 488, 491

\bibitem[{{Barrow} {et~al.}(2020){Barrow}, {Robertson}, {Ellis}, {Nakajima},
  {Saxena}, {Stark}, \& {Tang}}]{bar20}
{Barrow}, K. S.~S., {Robertson}, B.~E., {Ellis}, R.~S., {et~al.} 2020, \apjl,
  902, L39

\bibitem[{{B{\"o}ker} {et~al.}(2023){B{\"o}ker}, {Beck}, {Birkmann},
  {Giardino}, {Keyes}, {Kumari}, {Muzerolle}, {Rawle}, {Zeidler}, {Abul-Huda},
  {Alves de Oliveira}, {Arribas}, {Bechtold}, {Bhatawdekar}, {Bonaventura},
  {Bunker}, {Cameron}, {Carniani}, {Charlot}, {Curti}, {Espinoza}, {Ferruit},
  {Franx}, {Jakobsen}, {Karakla}, {L{\'o}pez-Caniego}, {L{\"u}tzgendorf},
  {Maiolino}, {Manjavacas}, {Marston}, {Moseley}, {Ogle}, {Perna},
  {Pe{\~n}a-Guerrero}, {Pirzkal}, {Plesha}, {Proffitt}, {Rauscher}, {Rix},
  {Rodr{\'\i}guez del Pino}, {Rustamkulov}, {Sabbi}, {Sing}, {Sirianni}, {te
  Plate}, {{\'U}beda}, {Wahlgren}, {Wislowski}, {Wu}, \& {Willott}}]{bok23}
{B{\"o}ker}, T., {Beck}, T.~L., {Birkmann}, S.~M., {et~al.} 2023, arXiv
  e-prints, arXiv:2301.13766

\bibitem[{{Bolan} {et~al.}(2022){Bolan}, {Lemaux}, {Mason}, {Brada{\v{c}}},
  {Treu}, {Strait}, {Pelliccia}, {Pentericci}, \& {Malkan}}]{bol22}
{Bolan}, P., {Lemaux}, B.~C., {Mason}, C., {et~al.} 2022, \mnras, 517, 3263

\bibitem[{{Bosman} {et~al.}(2022){Bosman}, {Davies}, {Becker}, {Keating},
  {Davies}, {Zhu}, {Eilers}, {D'Odorico}, {Bian}, {Bischetti}, {Cristiani},
  {Fan}, {Farina}, {Haehnelt}, {Hennawi}, {Kulkarni}, {Mesinger}, {Meyer},
  {Onoue}, {Pallottini}, {Qin}, {Ryan-Weber}, {Schindler}, {Walter}, {Wang}, \&
  {Yang}}]{bos22}
{Bosman}, S. E.~I., {Davies}, F.~B., {Becker}, G.~D., {et~al.} 2022, \mnras,
  514, 55

\bibitem[{{Bouwens} {et~al.}(2022){Bouwens}, {Smit}, {Schouws}, {Stefanon},
  {Bowler}, {Endsley}, {Gonzalez}, {Inami}, {Stark}, {Oesch}, {Hodge},
  {Aravena}, {da Cunha}, {Dayal}, {de Looze}, {Ferrara}, {Fudamoto},
  {Graziani}, {Li}, {Nanayakkara}, {Pallottini}, {Schneider}, {Sommovigo},
  {Topping}, {van der Werf}, {Algera}, {Barrufet}, {Hygate}, {Labb{\'e}},
  {Riechers}, \& {Witstok}}]{bou22b}
{Bouwens}, R.~J., {Smit}, R., {Schouws}, S., {et~al.} 2022, \apj, 931, 160

\bibitem[{{Bunker} {et~al.}(2023{\natexlab{a}}){Bunker}, {Cameron},
  {Curtis-Lake}, {Jakobsen}, {Carniani}, {Curti}, {Witstok}, {Maiolino},
  {D'Eugenio}, {Looser}, {Willott}, {Bonaventura}, {Hainline}, {Uebler},
  {Willmer}, {Saxena}, {Smit}, {Alberts}, {Arribas}, {Baker}, {Baum},
  {Bhatawdekar}, {Bowler}, {Boyett}, {Charlot}, {Chen}, {Chevallard},
  {Circosta}, {DeCoursey}, {de Graaff}, {Egami}, {Eisenstein}, {Endsley},
  {Ferruit}, {Giardino}, {Hausen}, {Helton}, {Hviding}, {Ji}, {Johnson},
  {Jones}, {Kumari}, {Laseter}, {Luetzgendorf}, {Maseda}, {Nelson}, {Parlanti},
  {Perna}, {Rawle}, {Rix}, {Rieke}, {Robertson}, {Rodriguez Del Pino},
  {Sandles}, {Scholtz}, {Sharpe}, {Skarbinski}, {Stark}, {Sun}, {Tacchella},
  {Topping}, {Villanueva}, {Wallace}, {Williams}, \& {Woodrum}}]{bun23b}
{Bunker}, A.~J., {Cameron}, A.~J., {Curtis-Lake}, E., {et~al.}
  2023{\natexlab{a}}, arXiv e-prints, arXiv:2306.02467

\bibitem[{{Bunker} {et~al.}(2023{\natexlab{b}}){Bunker}, {Saxena}, {Cameron},
  {Willott}, {Curtis-Lake}, {Jakobsen}, {Carniani}, {Smit}, {Maiolino},
  {Witstok}, {Curti}, {D'Eugenio}, {Jones}, {Ferruit}, {Arribas}, {Charlot},
  {Chevallard}, {Giardino}, {de Graaff}, {Looser}, {L{\"u}tzgendorf}, {Maseda},
  {Rawle}, {Rix}, {Del Pino}, {Alberts}, {Egami}, {Eisenstein}, {Endsley},
  {Hainline}, {Hausen}, {Johnson}, {Rieke}, {Rieke}, {Robertson}, {Shivaei},
  {Stark}, {Sun}, {Tacchella}, {Tang}, {Williams}, {Willmer}, {Baker}, {Baum},
  {Bhatawdekar}, {Bowler}, {Boyett}, {Chen}, {Circosta}, {Helton}, {Ji},
  {Kumari}, {Lyu}, {Nelson}, {Parlanti}, {Perna}, {Sandles}, {Scholtz},
  {Suess}, {Topping}, {{\"U}bler}, {Wallace}, \& {Whitler}}]{bun23}
{Bunker}, A.~J., {Saxena}, A., {Cameron}, A.~J., {et~al.} 2023{\natexlab{b}},
  \aap, 677, A88

\bibitem[{{Calzetti} {et~al.}(1994){Calzetti}, {Kinney}, \&
  {Storchi-Bergmann}}]{cal94}
{Calzetti}, D., {Kinney}, A.~L., \& {Storchi-Bergmann}, T. 1994, \apj, 429, 582

\bibitem[{{Cameron} {et~al.}(2023){Cameron}, {Saxena}, {Bunker}, {D'Eugenio},
  {Carniani}, {Maiolino}, {Curtis-Lake}, {Ferruit}, {Jakobsen}, {Arribas},
  {Bonaventura}, {Charlot}, {Chevallard}, {Curti}, {Looser}, {Maseda}, {Rawle},
  {Rodr{\'\i}guez Del Pino}, {Smit}, {{\"U}bler}, {Willott}, {Witstok},
  {Egami}, {Eisenstein}, {Johnson}, {Hainline}, {Rieke}, {Robertson}, {Stark},
  {Tacchella}, {Williams}, {Willmer}, {Bhatawdekar}, {Bowler}, {Boyett},
  {Circosta}, {Helton}, {Jones}, {Kumari}, {Ji}, {Nelson}, {Parlanti},
  {Sandles}, {Scholtz}, \& {Sun}}]{cam23}
{Cameron}, A.~J., {Saxena}, A., {Bunker}, A.~J., {et~al.} 2023, \aap, 677, A115

\bibitem[{{Carniani} {et~al.}(2017){Carniani}, {Maiolino}, {Pallottini},
  {Vallini}, {Pentericci}, {Ferrara}, {Castellano}, {Vanzella}, {Grazian},
  {Gallerani}, {Santini}, {Wagg}, \& {Fontana}}]{car17}
{Carniani}, S., {Maiolino}, R., {Pallottini}, A., {et~al.} 2017, \aap, 605, A42

\bibitem[{{Carniani} {et~al.}(2018){Carniani}, {Maiolino}, {Smit}, \&
  {Amor{\'\i}n}}]{carn18}
{Carniani}, S., {Maiolino}, R., {Smit}, R., \& {Amor{\'\i}n}, R. 2018, \apjl,
  854, L7

\bibitem[{{Caruana} {et~al.}(2012){Caruana}, {Bunker}, {Wilkins}, {Stanway},
  {Lacy}, {Jarvis}, {Lorenzoni}, \& {Hickey}}]{car12}
{Caruana}, J., {Bunker}, A.~J., {Wilkins}, S.~M., {et~al.} 2012, \mnras, 427,
  3055

\bibitem[{{Caruana} {et~al.}(2014){Caruana}, {Bunker}, {Wilkins}, {Stanway},
  {Lorenzoni}, {Jarvis}, \& {Ebert}}]{car14}
{Caruana}, J., {Bunker}, A.~J., {Wilkins}, S.~M., {et~al.} 2014, \mnras, 443,
  2831

\bibitem[{{Castellano} {et~al.}(2022){Castellano}, {Pentericci}, {Cupani},
  {Curtis-Lake}, {Vanzella}, {Amor{\'\i}n}, {Belfiori}, {Calabr{\`o}},
  {Carniani}, {Charlot}, {Chevallard}, {Dayal}, {Dickinson}, {Ferrara},
  {Fontana}, {Giallongo}, {Hutter}, {Merlin}, {Paris}, \& {Santini}}]{cas22}
{Castellano}, M., {Pentericci}, L., {Cupani}, G., {et~al.} 2022, \aap, 662,
  A115

\bibitem[{{Chisholm} {et~al.}(2022){Chisholm}, {Saldana-Lopez}, {Flury},
  {Schaerer}, {Jaskot}, {Amor{\'\i}n}, {Atek}, {Finkelstein}, {Fleming},
  {Ferguson}, {Fern{\'a}ndez}, {Giavalisco}, {Hayes}, {Heckman}, {Henry}, {Ji},
  {Marques-Chaves}, {Mauerhofer}, {McCandliss}, {Oey}, {{\"O}stlin},
  {Rutkowski}, {Scarlata}, {Thuan}, {Trebitsch}, {Wang}, {Worseck}, \&
  {Xu}}]{chi22}
{Chisholm}, J., {Saldana-Lopez}, A., {Flury}, S., {et~al.} 2022, \mnras, 517,
  5104

\bibitem[{{Choustikov} {et~al.}(2023){Choustikov}, {Katz}, {Saxena}, {Cameron},
  {Devriendt}, {Slyz}, {Rosdahl}, {Blaizot}, \& {Michel-Dansac}}]{cho23}
{Choustikov}, N., {Katz}, H., {Saxena}, A., {et~al.} 2023, arXiv e-prints,
  arXiv:2304.08526

\bibitem[{{Cuby} {et~al.}(2003){Cuby}, {Le F{\`e}vre}, {McCracken},
  {Cuillandre}, {Magnier}, \& {Meneux}}]{cub03}
{Cuby}, J.~G., {Le F{\`e}vre}, O., {McCracken}, H., {et~al.} 2003, \aap, 405,
  L19

\bibitem[{{Curti} {et~al.}(2023){Curti}, {D'Eugenio}, {Carniani}, {Maiolino},
  {Sandles}, {Witstok}, {Baker}, {Bennett}, {Piotrowska}, {Tacchella},
  {Charlot}, {Nakajima}, {Maheson}, {Mannucci}, {Amiri}, {Arribas}, {Belfiore},
  {Bonaventura}, {Bunker}, {Chevallard}, {Cresci}, {Curtis-Lake},
  {Hayden-Pawson}, {Jones}, {Kumari}, {Laseter}, {Looser}, {Marconi}, {Maseda},
  {Scholtz}, {Smit}, {{\"U}bler}, \& {Wallace}}]{cur23}
{Curti}, M., {D'Eugenio}, F., {Carniani}, S., {et~al.} 2023, \mnras, 518, 425

\bibitem[{{Curtis-Lake} {et~al.}(2023){Curtis-Lake}, {Carniani}, {Cameron},
  {Charlot}, {Jakobsen}, {Maiolino}, {Bunker}, {Witstok}, {Smit}, {Chevallard},
  {Willott}, {Ferruit}, {Arribas}, {Bonaventura}, {Curti}, {D'Eugenio},
  {Franx}, {Giardino}, {Looser}, {L{\"u}tzgendorf}, {Maseda}, {Rawle}, {Rix},
  {Rodr{\'\i}guez del Pino}, {{\"U}bler}, {Sirianni}, {Dressler}, {Egami},
  {Eisenstein}, {Endsley}, {Hainline}, {Hausen}, {Johnson}, {Rieke},
  {Robertson}, {Shivaei}, {Stark}, {Tacchella}, {Williams}, {Willmer},
  {Bhatawdekar}, {Bowler}, {Boyett}, {Chen}, {de Graaff}, {Helton}, {Hviding},
  {Jones}, {Kumari}, {Lyu}, {Nelson}, {Perna}, {Sandles}, {Saxena}, {Suess},
  {Sun}, {Topping}, {Wallace}, \& {Whitler}}]{curtis23}
{Curtis-Lake}, E., {Carniani}, S., {Cameron}, A., {et~al.} 2023, Nature
  Astronomy, 7, 622

\bibitem[{{Dayal} \& {Ferrara}(2018)}]{day18}
{Dayal}, P. \& {Ferrara}, A. 2018, \physrep, 780, 1

\bibitem[{{Dijkstra}(2014)}]{dij16}
{Dijkstra}, M. 2014, \pasa, 31, e040

\bibitem[{{Dijkstra} {et~al.}(2006){Dijkstra}, {Haiman}, \&
  {Spaans}}]{dijkstra06}
{Dijkstra}, M., {Haiman}, Z., \& {Spaans}, M. 2006, \apj, 649, 14

\bibitem[{{Eisenstein} {et~al.}(2023){Eisenstein}, {Willott}, {Alberts},
  {Arribas}, {Bonaventura}, {Bunker}, {Cameron}, {Carniani}, {Charlot},
  {Curtis-Lake}, {D'Eugenio}, {Endsley}, {Ferruit}, {Giardino}, {Hainline},
  {Hausen}, {Jakobsen}, {Johnson}, {Maiolino}, {Rieke}, {Rieke}, {Rix},
  {Robertson}, {Stark}, {Tacchella}, {Williams}, {Willmer}, {Baker}, {Baum},
  {Bhatawdekar}, {Boyett}, {Chen}, {Chevallard}, {Circosta}, {Curti},
  {Danhaive}, {DeCoursey}, {de Graaff}, {Dressler}, {Egami}, {Helton},
  {Hviding}, {Ji}, {Jones}, {Kumari}, {L{\"u}tzgendorf}, {Laseter}, {Looser},
  {Lyu}, {Maseda}, {Nelson}, {Parlanti}, {Perna}, {Pusk{\'a}s}, {Rawle},
  {Rodr{\'\i}guez Del Pino}, {Sandles}, {Saxena}, {Scholtz}, {Sharpe},
  {Shivaei}, {Silcock}, {Simmonds}, {Skarbinski}, {Smit}, {Stone}, {Suess},
  {Sun}, {Tang}, {Topping}, {{\"U}bler}, {Villanueva}, {Wallace}, {Whitler},
  {Witstok}, \& {Woodrum}}]{eis23}
{Eisenstein}, D.~J., {Willott}, C., {Alberts}, S., {et~al.} 2023, arXiv
  e-prints, arXiv:2306.02465

\bibitem[{{Endsley} {et~al.}(2022){Endsley}, {Stark}, {Bouwens}, {Schouws},
  {Smit}, {Stefanon}, {Inami}, {Bowler}, {Oesch}, {Gonzalez}, {Aravena}, {da
  Cunha}, {Dayal}, {Ferrara}, {Graziani}, {Nanayakkara}, {Pallottini},
  {Schneider}, {Sommovigo}, {Topping}, {van der Werf}, \& {Hutter}}]{end22c}
{Endsley}, R., {Stark}, D.~P., {Bouwens}, R.~J., {et~al.} 2022, \mnras, 517,
  5642

\bibitem[{{Endsley} {et~al.}(2023){Endsley}, {Stark}, {Whitler}, {Topping},
  {Chen}, {Plat}, {Chisholm}, \& {Charlot}}]{end23b}
{Endsley}, R., {Stark}, D.~P., {Whitler}, L., {et~al.} 2023, \mnras, 524, 2312

\bibitem[{{Erb} {et~al.}(2014){Erb}, {Steidel}, {Trainor}, {Bogosavljevi{\'c}},
  {Shapley}, {Nestor}, {Kulas}, {Law}, {Strom}, {Rudie}, {Reddy}, {Pettini},
  {Konidaris}, {Mace}, {Matthews}, \& {McLean}}]{erb14}
{Erb}, D.~K., {Steidel}, C.~C., {Trainor}, R.~F., {et~al.} 2014, \apj, 795, 33

\bibitem[{{Fan} {et~al.}(2006){Fan}, {Carilli}, \& {Keating}}]{fan06}
{Fan}, X., {Carilli}, C.~L., \& {Keating}, B. 2006, \araa, 44, 415

\bibitem[{{Feltre} {et~al.}(2020){Feltre}, {Maseda}, {Bacon}, {Pradeep},
  {Leclercq}, {Kusakabe}, {Wisotzki}, {Hashimoto}, {Schmidt}, {Blaizot},
  {Brinchmann}, {Boogaard}, {Cantalupo}, {Carton}, {Inami}, {Kollatschny},
  {Marino}, {Matthee}, {Nanayakkara}, {Richard}, {Schaye}, {Tresse}, {Urrutia},
  {Verhamme}, \& {Weilbacher}}]{fel20}
{Feltre}, A., {Maseda}, M.~V., {Bacon}, R., {et~al.} 2020, \aap, 641, A118

\bibitem[{{Ferruit} {et~al.}(2022){Ferruit}, {Jakobsen}, {Giardino}, {Rawle},
  {Alves de Oliveira}, {Arribas}, {Beck}, {Birkmann}, {B{\"o}ker}, {Bunker},
  {Charlot}, {de Marchi}, {Franx}, {Henry}, {Karakla}, {Kassin}, {Kumari},
  {L{\'o}pez-Caniego}, {L{\"u}tzgendorf}, {Maiolino}, {Manjavacas}, {Marston},
  {Moseley}, {Muzerolle}, {Pirzkal}, {Rauscher}, {Rix}, {Sabbi}, {Sirianni},
  {te Plate}, {Valenti}, {Willott}, \& {Zeidler}}]{fer22}
{Ferruit}, P., {Jakobsen}, P., {Giardino}, G., {et~al.} 2022, \aap, 661, A81

\bibitem[{{Fletcher} {et~al.}(2019){Fletcher}, {Tang}, {Robertson}, {Nakajima},
  {Ellis}, {Stark}, \& {Inoue}}]{fle19}
{Fletcher}, T.~J., {Tang}, M., {Robertson}, B.~E., {et~al.} 2019, \apj, 878, 87

\bibitem[{{Flury} {et~al.}(2022{\natexlab{a}}){Flury}, {Jaskot}, {Ferguson},
  {Worseck}, {Makan}, {Chisholm}, {Saldana-Lopez}, {Schaerer}, {McCandliss},
  {Wang}, {Ford}, {Heckman}, {Ji}, {Giavalisco}, {Amorin}, {Atek}, {Blaizot},
  {Borthakur}, {Carr}, {Castellano}, {Cristiani}, {De Barros}, {Dickinson},
  {Finkelstein}, {Fleming}, {Fontanot}, {Garel}, {Grazian}, {Hayes}, {Henry},
  {Mauerhofer}, {Micheva}, {Oey}, {Ostlin}, {Papovich}, {Pentericci},
  {Ravindranath}, {Rosdahl}, {Rutkowski}, {Santini}, {Scarlata}, {Teplitz},
  {Thuan}, {Trebitsch}, {Vanzella}, {Verhamme}, \& {Xu}}]{flu22a}
{Flury}, S.~R., {Jaskot}, A.~E., {Ferguson}, H.~C., {et~al.}
  2022{\natexlab{a}}, \apjs, 260, 1

\bibitem[{{Flury} {et~al.}(2022{\natexlab{b}}){Flury}, {Jaskot}, {Ferguson},
  {Worseck}, {Makan}, {Chisholm}, {Saldana-Lopez}, {Schaerer}, {McCandliss},
  {Xu}, {Wang}, {Oey}, {Ford}, {Heckman}, {Ji}, {Giavalisco}, {Amor{\'\i}n},
  {Atek}, {Blaizot}, {Borthakur}, {Carr}, {Castellano}, {De Barros},
  {Dickinson}, {Finkelstein}, {Fleming}, {Fontanot}, {Garel}, {Grazian},
  {Hayes}, {Henry}, {Mauerhofer}, {Micheva}, {Ostlin}, {Papovich},
  {Pentericci}, {Ravindranath}, {Rosdahl}, {Rutkowski}, {Santini}, {Scarlata},
  {Teplitz}, {Thuan}, {Trebitsch}, {Vanzella}, \& {Verhamme}}]{flu22b}
{Flury}, S.~R., {Jaskot}, A.~E., {Ferguson}, H.~C., {et~al.}
  2022{\natexlab{b}}, \apj, 930, 126

\bibitem[{{Fujimoto} {et~al.}(2023){Fujimoto}, {Arrabal Haro}, {Dickinson},
  {Finkelstein}, {Kartaltepe}, {Larson}, {Burgarella}, {Bagley}, {Behroozi},
  {Chworowsky}, {Hirschmann}, {Trump}, {Wilkins}, {Yung}, {Koekemoer},
  {Papovich}, {Pirzkal}, {Ferguson}, {Fontana}, {Grogin}, {Grazian}, {Kewley},
  {Kocevski}, {Lotz}, {Pentericci}, {Ravindranath}, {Somerville}, {Wilkins},
  {Amor{\'\i}n}, {Backhaus}, {Calabr{\`o}}, {Casey}, {Cooper}, {Fern{\'a}ndez},
  {Franco}, {Giavalisco}, {Hathi}, {Harish}, {Hutchison}, {Iyer}, {Jung},
  {Lucas}, \& {Zavala}}]{fuj23}
{Fujimoto}, S., {Arrabal Haro}, P., {Dickinson}, M., {et~al.} 2023, \apjl, 949,
  L25

\bibitem[{{Fuller} {et~al.}(2020){Fuller}, {Lemaux}, {Brada{\v{c}}}, {Hoag},
  {Schmidt}, {Huang}, {Strait}, {Mason}, {Treu}, {Pentericci}, {Trenti},
  {Henry}, \& {Malkan}}]{ful20}
{Fuller}, S., {Lemaux}, B.~C., {Brada{\v{c}}}, M., {et~al.} 2020, \apj, 896,
  156

\bibitem[{{Furlanetto} {et~al.}(2006){Furlanetto}, {Zaldarriaga}, \&
  {Hernquist}}]{fur06}
{Furlanetto}, S.~R., {Zaldarriaga}, M., \& {Hernquist}, L. 2006, \mnras, 365,
  1012

\bibitem[{{Furusawa} {et~al.}(2016){Furusawa}, {Kashikawa}, {Kobayashi},
  {Dunlop}, {Shimasaku}, {Takata}, {Sekiguchi}, {Naito}, {Furusawa}, {Ouchi},
  {Nakata}, {Yasuda}, {Okura}, {Taniguchi}, {Yamada}, {Kajisawa}, {Fynbo}, \&
  {Le F{\`e}vre}}]{fur16}
{Furusawa}, H., {Kashikawa}, N., {Kobayashi}, M. A.~R., {et~al.} 2016, \apj,
  822, 46

\bibitem[{{Garel} {et~al.}(2021){Garel}, {Blaizot}, {Rosdahl}, {Michel-Dansac},
  {Haehnelt}, {Katz}, {Kimm}, \& {Verhamme}}]{gar21}
{Garel}, T., {Blaizot}, J., {Rosdahl}, J., {et~al.} 2021, \mnras, 504, 1902

\bibitem[{{Giavalisco} {et~al.}(2004){Giavalisco}, {Ferguson}, {Koekemoer},
  {Dickinson}, {Alexander}, {Bauer}, {Bergeron}, {Biagetti}, {Brandt},
  {Casertano}, {Cesarsky}, {Chatzichristou}, {Conselice}, {Cristiani}, {Da
  Costa}, {Dahlen}, {de Mello}, {Eisenhardt}, {Erben}, {Fall}, {Fassnacht},
  {Fosbury}, {Fruchter}, {Gardner}, {Grogin}, {Hook}, {Hornschemeier}, {Idzi},
  {Jogee}, {Kretchmer}, {Laidler}, {Lee}, {Livio}, {Lucas}, {Madau},
  {Mobasher}, {Moustakas}, {Nonino}, {Padovani}, {Papovich}, {Park},
  {Ravindranath}, {Renzini}, {Richardson}, {Riess}, {Rosati}, {Schirmer},
  {Schreier}, {Somerville}, {Spinrad}, {Stern}, {Stiavelli}, {Strolger},
  {Urry}, {Vandame}, {Williams}, \& {Wolf}}]{gia04}
{Giavalisco}, M., {Ferguson}, H.~C., {Koekemoer}, A.~M., {et~al.} 2004, \apjl,
  600, L93

\bibitem[{{Gordon} {et~al.}(2003){Gordon}, {Clayton}, {Misselt}, {Landolt}, \&
  {Wolff}}]{gor03}
{Gordon}, K.~D., {Clayton}, G.~C., {Misselt}, K.~A., {Landolt}, A.~U., \&
  {Wolff}, M.~J. 2003, \apj, 594, 279

\bibitem[{{Haiman}(2002)}]{hai02}
{Haiman}, Z. 2002, \apjl, 576, L1

\bibitem[{{Harikane} {et~al.}(2018){Harikane}, {Ouchi}, {Shibuya}, {Kojima},
  {Zhang}, {Itoh}, {Ono}, {Higuchi}, {Inoue}, {Chevallard}, {Capak}, {Nagao},
  {Onodera}, {Faisst}, {Martin}, {Rauch}, {Bruzual}, {Charlot}, {Davidzon},
  {Fujimoto}, {Hilmi}, {Ilbert}, {Lee}, {Matsuoka}, {Silverman}, \&
  {Toft}}]{har18}
{Harikane}, Y., {Ouchi}, M., {Shibuya}, T., {et~al.} 2018, \apj, 859, 84

\bibitem[{{Hashimoto} {et~al.}(2019){Hashimoto}, {Inoue}, {Mawatari}, {Tamura},
  {Matsuo}, {Furusawa}, {Harikane}, {Shibuya}, {Knudsen}, {Kohno}, {Ono},
  {Zackrisson}, {Okamoto}, {Kashikawa}, {Oesch}, {Ouchi}, {Ota}, {Shimizu},
  {Taniguchi}, {Umehata}, \& {Watson}}]{has19}
{Hashimoto}, T., {Inoue}, A.~K., {Mawatari}, K., {et~al.} 2019, \pasj, 71, 71

\bibitem[{{Hayes} {et~al.}(2013){Hayes}, {{\"O}stlin}, {Schaerer}, {Verhamme},
  {Mas-Hesse}, {Adamo}, {Atek}, {Cannon}, {Duval}, {Guaita}, {Herenz}, {Kunth},
  {Laursen}, {Melinder}, {Orlitov{\'a}}, {Ot{\'\i}-Floranes}, \&
  {Sandberg}}]{hay13}
{Hayes}, M., {{\"O}stlin}, G., {Schaerer}, D., {et~al.} 2013, \apjl, 765, L27

\bibitem[{{Hayes} {et~al.}(2023){Hayes}, {Runnholm}, {Scarlata}, {Gronke}, \&
  {Rivera-Thorsen}}]{hay23a}
{Hayes}, M.~J., {Runnholm}, A., {Scarlata}, C., {Gronke}, M., \&
  {Rivera-Thorsen}, T.~E. 2023, \mnras, 520, 5903

\bibitem[{{Hayes} \& {Scarlata}(2023)}]{hay23b}
{Hayes}, M.~J. \& {Scarlata}, C. 2023, \apjl, 954, L14

\bibitem[{{Hoag} {et~al.}(2019){Hoag}, {Brada{\v{c}}}, {Huang}, {Mason},
  {Treu}, {Schmidt}, {Trenti}, {Strait}, {Lemaux}, {Finney}, \&
  {Paddock}}]{hoa19}
{Hoag}, A., {Brada{\v{c}}}, M., {Huang}, K., {et~al.} 2019, \apj, 878, 12

\bibitem[{{Inoue} {et~al.}(2014){Inoue}, {Shimizu}, {Iwata}, \&
  {Tanaka}}]{ino14}
{Inoue}, A.~K., {Shimizu}, I., {Iwata}, I., \& {Tanaka}, M. 2014, \mnras, 442,
  1805

\bibitem[{{Izotov} {et~al.}(2020){Izotov}, {Schaerer}, {Worseck}, {Verhamme},
  {Guseva}, {Thuan}, {Orlitov{\'a}}, \& {Fricke}}]{izotov20}
{Izotov}, Y.~I., {Schaerer}, D., {Worseck}, G., {et~al.} 2020, \mnras, 491, 468

\bibitem[{{Izotov} {et~al.}(2021){Izotov}, {Worseck}, {Schaerer}, {Guseva},
  {Chisholm}, {Thuan}, {Fricke}, \& {Verhamme}}]{izo21}
{Izotov}, Y.~I., {Worseck}, G., {Schaerer}, D., {et~al.} 2021, \mnras, 503,
  1734

\bibitem[{{Jakobsen} {et~al.}(2022){Jakobsen}, {Ferruit}, {Alves de Oliveira},
  {Arribas}, {Bagnasco}, {Barho}, {Beck}, {Birkmann}, {B{\"o}ker}, {Bunker},
  {Charlot}, {de Jong}, {de Marchi}, {Ehrenwinkler}, {Falcolini}, {Fels},
  {Franx}, {Franz}, {Funke}, {Giardino}, {Gnata}, {Holota}, {Honnen}, {Jensen},
  {Jentsch}, {Johnson}, {Jollet}, {Karl}, {Kling}, {K{\"o}hler}, {Kolm},
  {Kumari}, {Lander}, {Lemke}, {L{\'o}pez-Caniego}, {L{\"u}tzgendorf},
  {Maiolino}, {Manjavacas}, {Marston}, {Maschmann}, {Maurer}, {Messerschmidt},
  {Moseley}, {Mosner}, {Mott}, {Muzerolle}, {Pirzkal}, {Pittet}, {Plitzke},
  {Posselt}, {Rapp}, {Rauscher}, {Rawle}, {Rix}, {R{\"o}del}, {Rumler},
  {Sabbi}, {Salvignol}, {Schmid}, {Sirianni}, {Smith}, {Strada}, {te Plate},
  {Valenti}, {Wettemann}, {Wiehe}, {Wiesmayer}, {Willott}, {Wright}, {Zeidler},
  \& {Zincke}}]{jak22}
{Jakobsen}, P., {Ferruit}, P., {Alves de Oliveira}, C., {et~al.} 2022, \aap,
  661, A80

\bibitem[{{Ji} {et~al.}(2020){Ji}, {Giavalisco}, {Vanzella}, {Siana},
  {Pentericci}, {Jaskot}, {Liu}, {Nonino}, {Ferguson}, {Castellano},
  {Mannucci}, {Schaerer}, {Fynbo}, {Papovich}, {Carnall}, {Amorin}, {Simons},
  {Hathi}, {Cullen}, \& {McLeod}}]{ji20}
{Ji}, Z., {Giavalisco}, M., {Vanzella}, E., {et~al.} 2020, \apj, 888, 109

\bibitem[{{Jones} {et~al.}(2023){Jones}, {Bunker}, {Saxena}, {Witstok},
  {Stark}, {Arribas}, {Baker}, {Bhatawdekar}, {Bowler}, {Boyett}, {Cameron},
  {Carniani}, {Charlot}, {Chevallard}, {Curti}, {Curtis-Lake}, {Eisenstein},
  {Hainline}, {Hausen}, {Ji}, {Johnson}, {Kumari}, {Looser}, {Maiolino},
  {Maseda}, {Parlanti}, {Rix}, {Robertson}, {Sandles}, {Scholtz}, {Smit},
  {Tacchella}, {Ubler}, {Williams}, \& {Willott}}]{jon23}
{Jones}, G.~C., {Bunker}, A.~J., {Saxena}, A., {et~al.} 2023, arXiv e-prints,
  arXiv:2306.02471

\bibitem[{{Jung} {et~al.}(2023){Jung}, {Finkelstein}, {Arrabal Haro},
  {Dickinson}, {Ferguson}, {Hutchison}, {Kartaltepe}, {Larson}, {Simons},
  {Papovich}, {Park}, {Pentericci}, {Trump}, {Amorin}, {Backhaus}, {Casey},
  {Cheng}, {Cleri}, {Cooper}, {Cooper}, {Gardner}, {Gawiser}, {Grazian},
  {Hathi}, {Hirschmann}, {Koekemoer}, {Lucas}, {Mobasher}, {Ravindranath},
  {Straughn}, {Yung}, \& {de la Vega}}]{jun23}
{Jung}, I., {Finkelstein}, S.~L., {Arrabal Haro}, P., {et~al.} 2023, arXiv
  e-prints, arXiv:2304.05385

\bibitem[{{Katz} {et~al.}(2023{\natexlab{a}}){Katz}, {Saxena}, {Cameron},
  {Carniani}, {Bunker}, {Arribas}, {Bhatawdekar}, {Bowler}, {Boyett}, {Cresci},
  {Curtis-Lake}, {D'Eugenio}, {Kumari}, {Looser}, {Maiolino}, {{\"U}bler},
  {Willott}, \& {Witstok}}]{kat23b}
{Katz}, H., {Saxena}, A., {Cameron}, A.~J., {et~al.} 2023{\natexlab{a}},
  \mnras, 518, 592

\bibitem[{{Katz} {et~al.}(2023{\natexlab{b}}){Katz}, {Saxena}, {Rosdahl},
  {Kimm}, {Blaizot}, {Garel}, {Michel-Dansac}, {Haehnelt}, {Ellis},
  {Penterrici}, {Devriendt}, \& {Slyz}}]{kat23a}
{Katz}, H., {Saxena}, A., {Rosdahl}, J., {et~al.} 2023{\natexlab{b}}, \mnras,
  518, 270

\bibitem[{{Katz} {et~al.}(2020){Katz}, {{\v{D}}urov{\v{c}}{\'\i}kov{\'a}},
  {Kimm}, {Rosdahl}, {Blaizot}, {Haehnelt}, {Devriendt}, {Slyz}, {Ellis}, \&
  {Laporte}}]{kat20}
{Katz}, H., {{\v{D}}urov{\v{c}}{\'\i}kov{\'a}}, D., {Kimm}, T., {et~al.} 2020,
  \mnras, 498, 164

\bibitem[{{Keating} {et~al.}(2020){Keating}, {Weinberger}, {Kulkarni},
  {Haehnelt}, {Chardin}, \& {Aubert}}]{kea20}
{Keating}, L.~C., {Weinberger}, L.~H., {Kulkarni}, G., {et~al.} 2020, \mnras,
  491, 1736

\bibitem[{{Kerutt} {et~al.}(2022){Kerutt}, {Wisotzki}, {Verhamme}, {Schmidt},
  {Leclercq}, {Herenz}, {Urrutia}, {Garel}, {Hashimoto}, {Maseda}, {Matthee},
  {Kusakabe}, {Schaye}, {Richard}, {Guiderdoni}, {Mauerhofer}, {Nanayakkara},
  \& {Vitte}}]{ker22}
{Kerutt}, J., {Wisotzki}, L., {Verhamme}, A., {et~al.} 2022, \aap, 659, A183

\bibitem[{{Knudsen} {et~al.}(2016){Knudsen}, {Richard}, {Kneib}, {Jauzac},
  {Cl{\'e}ment}, {Drouart}, {Egami}, \& {Lindroos}}]{knu16}
{Knudsen}, K.~K., {Richard}, J., {Kneib}, J.-P., {et~al.} 2016, \mnras, 462, L6

\bibitem[{{Kuhlen} \& {Faucher-Gigu{\`e}re}(2012)}]{kuh12}
{Kuhlen}, M. \& {Faucher-Gigu{\`e}re}, C.-A. 2012, \mnras, 423, 862

\bibitem[{{Kusakabe} {et~al.}(2020){Kusakabe}, {Blaizot}, {Garel}, {Verhamme},
  {Bacon}, {Richard}, {Hashimoto}, {Inami}, {Conseil}, {Guiderdoni}, {Drake},
  {Christian Herenz}, {Schaye}, {Oesch}, {Matthee}, {Anna Marino}, {Borello
  Schmidt}, {Pell{\'o}}, {Maseda}, {Leclercq}, {Kerutt}, \& {Mahler}}]{kus20}
{Kusakabe}, H., {Blaizot}, J., {Garel}, T., {et~al.} 2020, \aap, 638, A12

\bibitem[{{Laporte} {et~al.}(2017){Laporte}, {Nakajima}, {Ellis}, {Zitrin},
  {Stark}, {Mainali}, \& {Roberts-Borsani}}]{lap17}
{Laporte}, N., {Nakajima}, K., {Ellis}, R.~S., {et~al.} 2017, \apj, 851, 40

\bibitem[{{Laursen} {et~al.}(2009){Laursen}, {Sommer-Larsen}, \&
  {Andersen}}]{laursen09}
{Laursen}, P., {Sommer-Larsen}, J., \& {Andersen}, A.~C. 2009, \apj, 704, 1640

\bibitem[{{Luridiana} {et~al.}(2015){Luridiana}, {Morisset}, \& {Shaw}}]{pyneb}
{Luridiana}, V., {Morisset}, C., \& {Shaw}, R.~A. 2015, \aap, 573, A42

\bibitem[{{Mainali} {et~al.}(2017){Mainali}, {Kollmeier}, {Stark}, {Simcoe},
  {Walth}, {Newman}, \& {Miller}}]{mai17}
{Mainali}, R., {Kollmeier}, J.~A., {Stark}, D.~P., {et~al.} 2017, \apjl, 836,
  L14

\bibitem[{{Maiolino} {et~al.}(2015){Maiolino}, {Carniani}, {Fontana},
  {Vallini}, {Pentericci}, {Ferrara}, {Vanzella}, {Grazian}, {Gallerani},
  {Castellano}, {Cristiani}, {Brammer}, {Santini}, {Wagg}, \&
  {Williams}}]{mai15}
{Maiolino}, R., {Carniani}, S., {Fontana}, A., {et~al.} 2015, \mnras, 452, 54

\bibitem[{{Maji} {et~al.}(2022){Maji}, {Verhamme}, {Rosdahl}, {Garel},
  {Blaizot}, {Mauerhofer}, {Pittavino}, {Victoria Feser}, {Chuniaud}, {Kimm},
  {Katz}, \& {Haehnelt}}]{maj22}
{Maji}, M., {Verhamme}, A., {Rosdahl}, J., {et~al.} 2022, \aap, 663, A66

\bibitem[{{Malhotra} \& {Rhoads}(2006)}]{mal06}
{Malhotra}, S. \& {Rhoads}, J.~E. 2006, \apjl, 647, L95

\bibitem[{{Mascia} {et~al.}(2023){Mascia}, {Pentericci}, {Calabr{\`o}}, {Treu},
  {Santini}, {Yang}, {Napolitano}, {Roberts-Borsani}, {Bergamini}, {Grillo},
  {Rosati}, {Vulcani}, {Castellano}, {Boyett}, {Fontana}, {Glazebrook},
  {Henry}, {Mason}, {Merlin}, {Morishita}, {Nanayakkara}, {Paris}, {Roy},
  {Williams}, {Wang}, {Brammer}, {Brada{\v{c}}}, {Chen}, {Kelly}, {Koekemoer},
  {Trenti}, \& {Windhorst}}]{mas23}
{Mascia}, S., {Pentericci}, L., {Calabr{\`o}}, A., {et~al.} 2023, \aap, 672,
  A155

\bibitem[{{Maseda} {et~al.}(2020){Maseda}, {Bacon}, {Lam}, {Matthee},
  {Brinchmann}, {Schaye}, {Labbe}, {Schmidt}, {Boogaard}, {Bouwens},
  {Cantalupo}, {Franx}, {Hashimoto}, {Inami}, {Kusakabe}, {Mahler},
  {Nanayakkara}, {Richard}, \& {Wisotzki}}]{mas20}
{Maseda}, M.~V., {Bacon}, R., {Lam}, D., {et~al.} 2020, \mnras, 493, 5120

\bibitem[{{Mason} \& {Gronke}(2020)}]{mason20}
{Mason}, C.~A. \& {Gronke}, M. 2020, \mnras, 499, 1395

\bibitem[{{Mason} {et~al.}(2018){Mason}, {Treu}, {Dijkstra}, {Mesinger},
  {Trenti}, {Pentericci}, {de Barros}, \& {Vanzella}}]{mas18}
{Mason}, C.~A., {Treu}, T., {Dijkstra}, M., {et~al.} 2018, \apj, 856, 2

\bibitem[{{Matthee} {et~al.}(2022){Matthee}, {Naidu}, {Pezzulli}, {Gronke},
  {Sobral}, {Oesch}, {Hayes}, {Erb}, {Schaerer}, {Amor{\'\i}n}, {Tacchella},
  {Paulino-Afonso}, {Llerena}, {Calhau}, \& {R{\"o}ttgering}}]{mat22}
{Matthee}, J., {Naidu}, R.~P., {Pezzulli}, G., {et~al.} 2022, \mnras, 512, 5960

\bibitem[{{Matthee} {et~al.}(2017{\natexlab{a}}){Matthee}, {Sobral}, {Best},
  {Smail}, {Bian}, {Darvish}, {R{\"o}ttgering}, \& {Fan}}]{mat17b}
{Matthee}, J., {Sobral}, D., {Best}, P., {et~al.} 2017{\natexlab{a}}, \mnras,
  471, 629

\bibitem[{{Matthee} {et~al.}(2019){Matthee}, {Sobral}, {Boogaard},
  {R{\"o}ttgering}, {Vallini}, {Ferrara}, {Paulino-Afonso}, {Boone},
  {Schaerer}, \& {Mobasher}}]{mat19}
{Matthee}, J., {Sobral}, D., {Boogaard}, L.~A., {et~al.} 2019, \apj, 881, 124

\bibitem[{{Matthee} {et~al.}(2017{\natexlab{b}}){Matthee}, {Sobral}, {Boone},
  {R{\"o}ttgering}, {Schaerer}, {Girard}, {Pallottini}, {Vallini}, {Ferrara},
  \& {Darvish}}]{mat17}
{Matthee}, J., {Sobral}, D., {Boone}, F., {et~al.} 2017{\natexlab{b}}, \apj,
  851, 145

\bibitem[{{Matthee} {et~al.}(2020){Matthee}, {Sobral}, {Gronke}, {Pezzulli},
  {Cantalupo}, {R{\"o}ttgering}, {Darvish}, \& {Santos}}]{mat20}
{Matthee}, J., {Sobral}, D., {Gronke}, M., {et~al.} 2020, \mnras, 492, 1778

\bibitem[{{Meyer} {et~al.}(2019){Meyer}, {Bosman}, {Kakiichi}, \&
  {Ellis}}]{meyer19}
{Meyer}, R.~A., {Bosman}, S. E.~I., {Kakiichi}, K., \& {Ellis}, R.~S. 2019,
  \mnras, 483, 19

\bibitem[{{Miralda-Escud{\'e}}(1998)}]{mir98}
{Miralda-Escud{\'e}}, J. 1998, \apj, 501, 15

\bibitem[{{Naidu} {et~al.}(2018){Naidu}, {Forrest}, {Oesch}, {Tran}, \&
  {Holden}}]{nai18}
{Naidu}, R.~P., {Forrest}, B., {Oesch}, P.~A., {Tran}, K.-V.~H., \& {Holden},
  B.~P. 2018, \mnras, 478, 791

\bibitem[{{Naidu} {et~al.}(2022){Naidu}, {Matthee}, {Oesch}, {Conroy},
  {Sobral}, {Pezzulli}, {Hayes}, {Erb}, {Amor{\'\i}n}, {Gronke}, {Schaerer},
  {Tacchella}, {Kerutt}, {Paulino-Afonso}, {Calhau}, {Llerena}, \&
  {R{\"o}ttgering}}]{nai22}
{Naidu}, R.~P., {Matthee}, J., {Oesch}, P.~A., {et~al.} 2022, \mnras, 510, 4582

\bibitem[{{Nakajima} {et~al.}(2020){Nakajima}, {Ellis}, {Robertson}, {Tang}, \&
  {Stark}}]{nak20}
{Nakajima}, K., {Ellis}, R.~S., {Robertson}, B.~E., {Tang}, M., \& {Stark},
  D.~P. 2020, \apj, 889, 161

\bibitem[{{Nakajima} {et~al.}(2023){Nakajima}, {Ouchi}, {Isobe}, {Harikane},
  {Zhang}, {Ono}, {Umeda}, \& {Oguri}}]{nak23}
{Nakajima}, K., {Ouchi}, M., {Isobe}, Y., {et~al.} 2023, \apjs, 269, 33

\bibitem[{{Neufeld}(1990)}]{neufeld90}
{Neufeld}, D.~A. 1990, \apj, 350, 216

\bibitem[{{Ning} {et~al.}(2023){Ning}, {Cai}, {Jiang}, {Lin}, {Fu}, \&
  {Spinoso}}]{nin23}
{Ning}, Y., {Cai}, Z., {Jiang}, L., {et~al.} 2023, \apjl, 944, L1

\bibitem[{{Oesch} {et~al.}(2015){Oesch}, {van Dokkum}, {Illingworth},
  {Bouwens}, {Momcheva}, {Holden}, {Roberts-Borsani}, {Smit}, {Franx},
  {Labb{\'e}}, {Gonz{\'a}lez}, \& {Magee}}]{oes15}
{Oesch}, P.~A., {van Dokkum}, P.~G., {Illingworth}, G.~D., {et~al.} 2015,
  \apjl, 804, L30

\bibitem[{{Oke} \& {Gunn}(1983)}]{oke83}
{Oke}, J.~B. \& {Gunn}, J.~E. 1983, \apj, 266, 713

\bibitem[{{Orlitov{\'a}} {et~al.}(2018){Orlitov{\'a}}, {Verhamme}, {Henry},
  {Scarlata}, {Jaskot}, {Oey}, \& {Schaerer}}]{orl18}
{Orlitov{\'a}}, I., {Verhamme}, A., {Henry}, A., {et~al.} 2018, \aap, 616, A60

\bibitem[{{Osterbrock}(1989)}]{ost89}
{Osterbrock}, D.~E. 1989, {Astrophysics of gaseous nebulae and active galactic
  nuclei}

\bibitem[{{Ouchi} {et~al.}(2013){Ouchi}, {Ellis}, {Ono}, {Nakanishi}, {Kohno},
  {Momose}, {Kurono}, {Ashby}, {Shimasaku}, {Willner}, {Fazio}, {Tamura}, \&
  {Iono}}]{ouc13}
{Ouchi}, M., {Ellis}, R., {Ono}, Y., {et~al.} 2013, \apj, 778, 102

\bibitem[{{Pahl} {et~al.}(2021){Pahl}, {Shapley}, {Steidel}, {Chen}, \&
  {Reddy}}]{pah21}
{Pahl}, A.~J., {Shapley}, A., {Steidel}, C.~C., {Chen}, Y., \& {Reddy}, N.~A.
  2021, \mnras, 505, 2447

\bibitem[{{Pentericci} {et~al.}(2016){Pentericci}, {Carniani}, {Castellano},
  {Fontana}, {Maiolino}, {Guaita}, {Vanzella}, {Grazian}, {Santini}, {Yan},
  {Cristiani}, {Conselice}, {Giavalisco}, {Hathi}, \& {Koekemoer}}]{pen16}
{Pentericci}, L., {Carniani}, S., {Castellano}, M., {et~al.} 2016, \apjl, 829,
  L11

\bibitem[{{Pentericci} {et~al.}(2011){Pentericci}, {Fontana}, {Vanzella},
  {Castellano}, {Grazian}, {Dijkstra}, {Boutsia}, {Cristiani}, {Dickinson},
  {Giallongo}, {Giavalisco}, {Maiolino}, {Moorwood}, {Paris}, \&
  {Santini}}]{pen11}
{Pentericci}, L., {Fontana}, A., {Vanzella}, E., {et~al.} 2011, \apj, 743, 132

\bibitem[{{Pentericci} {et~al.}(2018{\natexlab{a}}){Pentericci}, {McLure},
  {Garilli}, {Cucciati}, {Franzetti}, {Iovino}, {Amorin}, {Bolzonella},
  {Bongiorno}, \& {Carnall}}]{pen18}
{Pentericci}, L., {McLure}, R.~J., {Garilli}, B., {et~al.} 2018{\natexlab{a}},
  \aap, 616, A174

\bibitem[{{Pentericci} {et~al.}(2018{\natexlab{b}}){Pentericci}, {Vanzella},
  {Castellano}, {Fontana}, {De Barros}, {Grazian}, {Marchi}, {Bradac},
  {Conselice}, {Cristiani}, {Dickinson}, {Finkelstein}, {Giallongo}, {Guaita},
  {Koekemoer}, {Maiolino}, {Santini}, \& {Tilvi}}]{pen18b}
{Pentericci}, L., {Vanzella}, E., {Castellano}, M., {et~al.}
  2018{\natexlab{b}}, \aap, 619, A147

\bibitem[{{Planck Collaboration} {et~al.}(2020){Planck Collaboration},
  {Aghanim}, {Akrami}, {Ashdown}, {Aumont}, {Baccigalupi}, {Ballardini},
  {Banday}, {Barreiro}, {Bartolo}, {Basak}, {Battye}, {Benabed}, {Bernard},
  {Bersanelli}, {Bielewicz}, {Bock}, {Bond}, {Borrill}, {Bouchet}, {Boulanger},
  {Bucher}, {Burigana}, {Butler}, {Calabrese}, {Cardoso}, {Carron},
  {Challinor}, {Chiang}, {Chluba}, {Colombo}, {Combet}, {Contreras}, {Crill},
  {Cuttaia}, {de Bernardis}, {de Zotti}, {Delabrouille}, {Delouis}, {Di
  Valentino}, {Diego}, {Dor{\'e}}, {Douspis}, {Ducout}, {Dupac}, {Dusini},
  {Efstathiou}, {Elsner}, {En{\ss}lin}, {Eriksen}, {Fantaye}, {Farhang},
  {Fergusson}, {Fernandez-Cobos}, {Finelli}, {Forastieri}, {Frailis},
  {Fraisse}, {Franceschi}, {Frolov}, {Galeotta}, {Galli}, {Ganga},
  {G{\'e}nova-Santos}, {Gerbino}, {Ghosh}, {Gonz{\'a}lez-Nuevo}, {G{\'o}rski},
  {Gratton}, {Gruppuso}, {Gudmundsson}, {Hamann}, {Handley}, {Hansen},
  {Herranz}, {Hildebrandt}, {Hivon}, {Huang}, {Jaffe}, {Jones}, {Karakci},
  {Keih{\"a}nen}, {Keskitalo}, {Kiiveri}, {Kim}, {Kisner}, {Knox},
  {Krachmalnicoff}, {Kunz}, {Kurki-Suonio}, {Lagache}, {Lamarre}, {Lasenby},
  {Lattanzi}, {Lawrence}, {Le Jeune}, {Lemos}, {Lesgourgues}, {Levrier},
  {Lewis}, {Liguori}, {Lilje}, {Lilley}, {Lindholm}, {L{\'o}pez-Caniego},
  {Lubin}, {Ma}, {Mac{\'\i}as-P{\'e}rez}, {Maggio}, {Maino}, {Mandolesi},
  {Mangilli}, {Marcos-Caballero}, {Maris}, {Martin}, {Martinelli},
  {Mart{\'\i}nez-Gonz{\'a}lez}, {Matarrese}, {Mauri}, {McEwen}, {Meinhold},
  {Melchiorri}, {Mennella}, {Migliaccio}, {Millea}, {Mitra},
  {Miville-Desch{\^e}nes}, {Molinari}, {Montier}, {Morgante}, {Moss}, {Natoli},
  {N{\o}rgaard-Nielsen}, {Pagano}, {Paoletti}, {Partridge}, {Patanchon},
  {Peiris}, {Perrotta}, {Pettorino}, {Piacentini}, {Polastri}, {Polenta},
  {Puget}, {Rachen}, {Reinecke}, {Remazeilles}, {Renzi}, {Rocha}, {Rosset},
  {Roudier}, {Rubi{\~n}o-Mart{\'\i}n}, {Ruiz-Granados}, {Salvati}, {Sandri},
  {Savelainen}, {Scott}, {Shellard}, {Sirignano}, {Sirri}, {Spencer},
  {Sunyaev}, {Suur-Uski}, {Tauber}, {Tavagnacco}, {Tenti}, {Toffolatti},
  {Tomasi}, {Trombetti}, {Valenziano}, {Valiviita}, {Van Tent}, {Vibert},
  {Vielva}, {Villa}, {Vittorio}, {Wandelt}, {Wehus}, {White}, {White},
  {Zacchei}, \& {Zonca}}]{planck}
{Planck Collaboration}, {Aghanim}, N., {Akrami}, Y., {et~al.} 2020, \aap, 641,
  A6

\bibitem[{{Prieto-Lyon} {et~al.}(2023{\natexlab{a}}){Prieto-Lyon}, {Mason},
  {Mascia}, {Merlin}, {Roy}, {Henry}, {Roberts-Borsani}, {Morishita}, {Wang},
  {Boyett}, {Bolan}, {Bradac}, {Castellano}, {Mercurio}, {Nanayakkara},
  {Paris}, {Pentericci}, {Scarlata}, {Trenti}, {Treu}, \&
  {Vanzella}}]{prieto23a}
{Prieto-Lyon}, G., {Mason}, C., {Mascia}, S., {et~al.} 2023{\natexlab{a}},
  \apj, 956, 136

\bibitem[{{Prieto-Lyon} {et~al.}(2023{\natexlab{b}}){Prieto-Lyon}, {Strait},
  {Mason}, {Brammer}, {Caminha}, {Mercurio}, {Acebron}, {Bergamini}, {Grillo},
  {Rosati}, {Vanzella}, {Castellano}, {Merlin}, {Paris}, {Boyett},
  {Calabr{\`o}}, {Morishita}, {Mascia}, {Pentericci}, {Roberts-Borsani}, {Roy},
  {Treu}, \& {Vulcani}}]{prieto23}
{Prieto-Lyon}, G., {Strait}, V., {Mason}, C.~A., {et~al.} 2023{\natexlab{b}},
  \aap, 672, A186

\bibitem[{{Rieke} {et~al.}(2023{\natexlab{a}}){Rieke}, {Kelly}, {Misselt},
  {Stansberry}, {Boyer}, {Beatty}, {Egami}, {Florian}, {Greene}, {Hainline},
  {Leisenring}, {Roellig}, {Schlawin}, {Sun}, {Tinnin}, {Williams}, {Willmer},
  {Wilson}, {Clark}, {Rohrbach}, {Brooks}, {Canipe}, {Correnti}, {DiFelice},
  {Gennaro}, {Girard}, {Hartig}, {Hilbert}, {Koekemoer}, {Nikolov}, {Pirzkal},
  {Rest}, {Robberto}, {Sunnquist}, {Telfer}, {Wu}, {Ferry}, {Lewis}, {Baum},
  {Beichman}, {Doyon}, {Dressler}, {Eisenstein}, {Ferrarese}, {Hodapp},
  {Horner}, {Jaffe}, {Johnstone}, {Krist}, {Martin}, {McCarthy}, {Meyer},
  {Rieke}, {Trauger}, \& {Young}}]{rie22}
{Rieke}, M.~J., {Kelly}, D.~M., {Misselt}, K., {et~al.} 2023{\natexlab{a}},
  \pasp, 135, 028001

\bibitem[{{Rieke} {et~al.}(2023{\natexlab{b}}){Rieke}, {Robertson},
  {Tacchella}, {Hainline}, {Johnson}, {Hausen}, {Ji}, {Willmer}, {Eisenstein},
  {Pusk{\'a}s}, {Alberts}, {Arribas}, {Baker}, {Baum}, {Bhatawdekar},
  {Bonaventura}, {Boyett}, {Bunker}, {Cameron}, {Carniani}, {Charlot},
  {Chevallard}, {Chen}, {Curti}, {Curtis-Lake}, {Danhaive}, {DeCoursey},
  {Dressler}, {Egami}, {Endsley}, {Helton}, {Hviding}, {Kumari}, {Looser},
  {Lyu}, {Maiolino}, {Maseda}, {Nelson}, {Rieke}, {Rix}, {Sandles}, {Saxena},
  {Sharpe}, {Shivaei}, {Skarbinski}, {Smit}, {Stark}, {Stone}, {Suess}, {Sun},
  {Topping}, {{\"U}bler}, {Villanueva}, {Wallace}, {Williams}, {Willott},
  {Whitler}, {Witstok}, \& {Woodrum}}]{rie23}
{Rieke}, M.~J., {Robertson}, B., {Tacchella}, S., {et~al.} 2023{\natexlab{b}},
  \apjs, 269, 16

\bibitem[{{Roberts-Borsani} {et~al.}(2016){Roberts-Borsani}, {Bouwens},
  {Oesch}, {Labbe}, {Smit}, {Illingworth}, {van Dokkum}, {Holden}, {Gonzalez},
  {Stefanon}, {Holwerda}, \& {Wilkins}}]{rob16}
{Roberts-Borsani}, G.~W., {Bouwens}, R.~J., {Oesch}, P.~A., {et~al.} 2016,
  \apj, 823, 143

\bibitem[{{Robertson}(2022)}]{rob22a}
{Robertson}, B.~E. 2022, \araa, 60, 121

\bibitem[{{Robertson} {et~al.}(2015){Robertson}, {Ellis}, {Furlanetto}, \&
  {Dunlop}}]{rob15}
{Robertson}, B.~E., {Ellis}, R.~S., {Furlanetto}, S.~R., \& {Dunlop}, J.~S.
  2015, \apjl, 802, L19

\bibitem[{{Robertson} {et~al.}(2013){Robertson}, {Furlanetto}, {Schneider},
  {Charlot}, {Ellis}, {Stark}, {McLure}, {Dunlop}, {Koekemoer}, {Schenker},
  {Ouchi}, {Ono}, {Curtis-Lake}, {Rogers}, {Bowler}, \& {Cirasuolo}}]{rob13}
{Robertson}, B.~E., {Furlanetto}, S.~R., {Schneider}, E., {et~al.} 2013, \apj,
  768, 71

\bibitem[{{Robertson} {et~al.}(2023){Robertson}, {Tacchella}, {Johnson},
  {Hainline}, {Whitler}, {Eisenstein}, {Endsley}, {Rieke}, {Stark}, {Alberts},
  {Dressler}, {Egami}, {Hausen}, {Rieke}, {Shivaei}, {Williams}, {Willmer},
  {Arribas}, {Bonaventura}, {Bunker}, {Cameron}, {Carniani}, {Charlot},
  {Chevallard}, {Curti}, {Curtis-Lake}, {D'Eugenio}, {Jakobsen}, {Looser},
  {L{\"u}tzgendorf}, {Maiolino}, {Maseda}, {Rawle}, {Rix}, {Smit}, {{\"U}bler},
  {Willott}, {Witstok}, {Baum}, {Bhatawdekar}, {Boyett}, {Chen}, {de Graaff},
  {Florian}, {Helton}, {Hviding}, {Ji}, {Kumari}, {Lyu}, {Nelson}, {Sandles},
  {Saxena}, {Suess}, {Sun}, {Topping}, \& {Wallace}}]{rob23}
{Robertson}, B.~E., {Tacchella}, S., {Johnson}, B.~D., {et~al.} 2023, Nature
  Astronomy, 7, 611

\bibitem[{{Rosdahl} {et~al.}(2022){Rosdahl}, {Blaizot}, {Katz}, {Kimm},
  {Garel}, {Haehnelt}, {Keating}, {Martin-Alvarez}, {Michel-Dansac}, \&
  {Ocvirk}}]{ros22}
{Rosdahl}, J., {Blaizot}, J., {Katz}, H., {et~al.} 2022, \mnras, 515, 2386

\bibitem[{{Roy} {et~al.}(2023){Roy}, {Henry}, {Treu}, {Jones}, {Prieto-Lyon},
  {Mason}, {Heckman}, {Nanayakkara}, {Pentericci}, {Mascia}, {Brada{\v{c}}},
  {Vanzella}, {Scarlata}, {Boyett}, {Trenti}, \& {Wang}}]{roy23}
{Roy}, N., {Henry}, A., {Treu}, T., {et~al.} 2023, \apjl, 952, L14

\bibitem[{{Sanders} {et~al.}(2023){Sanders}, {Shapley}, {Topping}, {Reddy}, \&
  {Brammer}}]{san23}
{Sanders}, R.~L., {Shapley}, A.~E., {Topping}, M.~W., {Reddy}, N.~A., \&
  {Brammer}, G.~B. 2023, \apj, 955, 54

\bibitem[{{Sandles} {et~al.}(2023){Sandles}, {D'Eugenio}, {Maiolino}, {Looser},
  {Arribas}, {Baker}, {Bonaventura}, {Bunker}, {Cameron}, {Carniani},
  {Charlot}, {Chevallard}, {Curti}, {Curtis-Lake}, {de Graaff}, {Eisenstein},
  {Hainline}, {Ji}, {Johnson}, {Jones}, {Kumari}, {Nelson}, {Perna}, {Rawle},
  {Rix}, {Robertson}, {Rodriguez Del Pino}, {Scholtz}, {Shivaei}, {Smit},
  {Sun}, {Tacchella}, {Uebler}, {Williams}, {Willott}, \&
  {Witstok}}]{sandles23}
{Sandles}, L., {D'Eugenio}, F., {Maiolino}, R., {et~al.} 2023, arXiv e-prints,
  arXiv:2306.03931

\bibitem[{{Santos} {et~al.}(2020){Santos}, {Sobral}, {Matthee}, {Calhau}, {da
  Cunha}, {Ribeiro}, {Paulino-Afonso}, {Arrabal Haro}, \&
  {Butterworth}}]{santos20}
{Santos}, S., {Sobral}, D., {Matthee}, J., {et~al.} 2020, \mnras, 493, 141

\bibitem[{{Saxena} {et~al.}(2022{\natexlab{a}}){Saxena}, {Cryer}, {Ellis},
  {Pentericci}, {Calabr{\`o}}, {Mascia}, {Saldana-Lopez}, {Schaerer}, {Katz},
  {Llerena}, \& {Amor{\'\i}n}}]{sax22b}
{Saxena}, A., {Cryer}, E., {Ellis}, R.~S., {et~al.} 2022{\natexlab{a}}, \mnras,
  517, 1098

\bibitem[{{Saxena} {et~al.}(2021){Saxena}, {Ellis}, {F{\"o}rster},
  {Calabr{\`o}}, {Pentericci}, {Carnall}, {Castellano}, {Cullen}, {Fontana},
  {Franco}, {Fynbo}, {Gargiulo}, {Garilli}, {Hathi}, {McLeod}, {Amor{\'\i}n},
  \& {Zamorani}}]{sax21}
{Saxena}, A., {Ellis}, R.~S., {F{\"o}rster}, P.~U., {et~al.} 2021, \mnras, 505,
  4798

\bibitem[{{Saxena} {et~al.}(2022{\natexlab{b}}){Saxena}, {Pentericci}, {Ellis},
  {Guaita}, {Calabr{\`o}}, {Schaerer}, {Vanzella}, {Amor{\'\i}n}, {Bolzonella},
  {Castellano}, {Fontanot}, {Hathi}, {Hibon}, {Llerena}, {Mannucci},
  {Saldana-Lopez}, {Talia}, \& {Zamorani}}]{sax22a}
{Saxena}, A., {Pentericci}, L., {Ellis}, R.~S., {et~al.} 2022{\natexlab{b}},
  \mnras, 511, 120

\bibitem[{{Saxena} {et~al.}(2023){Saxena}, {Robertson}, {Bunker}, {Endsley},
  {Cameron}, {Charlot}, {Simmonds}, {Tacchella}, {Witstok}, {Willott},
  {Carniani}, {Curtis-Lake}, {Ferruit}, {Jakobsen}, {Arribas}, {Chevallard},
  {Curti}, {D'Eugenio}, {De Graaff}, {Jones}, {Looser}, {Maseda}, {Rawle},
  {Rix}, {Del Pino}, {Smit}, {{\"U}bler}, {Eisenstein}, {Hainline}, {Hausen},
  {Johnson}, {Rieke}, {Williams}, {Willmer}, {Baker}, {Bhatawdekar}, {Bowler},
  {Boyett}, {Chen}, {Egami}, {Ji}, {Kumari}, {Nelson}, {Perna}, {Sandles},
  {Scholtz}, \& {Shivaei}}]{sax23a}
{Saxena}, A., {Robertson}, B.~E., {Bunker}, A.~J., {et~al.} 2023, \aap, 678,
  A68

\bibitem[{{Shivaei} {et~al.}(2020){Shivaei}, {Reddy}, {Rieke}, {Shapley},
  {Kriek}, {Battisti}, {Mobasher}, {Sanders}, {Fetherolf}, {Azadi}, {Coil},
  {Freeman}, {de Groot}, {Leung}, {Price}, {Siana}, \& {Zick}}]{shi20}
{Shivaei}, I., {Reddy}, N., {Rieke}, G., {et~al.} 2020, \apj, 899, 117

\bibitem[{{Simmonds} {et~al.}(2024){Simmonds}, {Tacchella}, {Hainline},
  {Johnson}, {McClymont}, {Robertson}, {Saxena}, {Sun}, {Witten}, {Baker},
  {Bhatawdekar}, {Boyett}, {Bunker}, {Charlot}, {Curtis-Lake}, {Egami},
  {Eisenstein}, {Hausen}, {Maiolino}, {Maseda}, {Scholtz}, {Williams},
  {Willott}, \& {Witstok}}]{sim24}
{Simmonds}, C., {Tacchella}, S., {Hainline}, K., {et~al.} 2024, \mnras, 527,
  6139

\bibitem[{{Simmonds} {et~al.}(2023){Simmonds}, {Tacchella}, {Maseda},
  {Williams}, {Baker}, {Witten}, {Johnson}, {Robertson}, {Saxena}, {Sun},
  {Witstok}, {Bhatawdekar}, {Boyett}, {Bunker}, {Charlot}, {Curtis-Lake},
  {Egami}, {Eisenstein}, {Ji}, {Maiolino}, {Sandles}, {Smit}, {{\"U}bler}, \&
  {Willott}}]{sim23a}
{Simmonds}, C., {Tacchella}, S., {Maseda}, M., {et~al.} 2023, \mnras, 523, 5468

\bibitem[{{Smith} {et~al.}(2019){Smith}, {Ma}, {Bromm}, {Finkelstein},
  {Hopkins}, {Faucher-Gigu{\`e}re}, \& {Kere{\v{s}}}}]{smi19}
{Smith}, A., {Ma}, X., {Bromm}, V., {et~al.} 2019, \mnras, 484, 39

\bibitem[{{Sobral} {et~al.}(2015){Sobral}, {Matthee}, {Darvish}, {Schaerer},
  {Mobasher}, {R{\"o}ttgering}, {Santos}, \& {Hemmati}}]{sob15}
{Sobral}, D., {Matthee}, J., {Darvish}, B., {et~al.} 2015, \apj, 808, 139

\bibitem[{{Stark}(2016)}]{star16}
{Stark}, D.~P. 2016, \araa, 54, 761

\bibitem[{{Stark} {et~al.}(2017){Stark}, {Ellis}, {Charlot}, {Chevallard},
  {Tang}, {Belli}, {Zitrin}, {Mainali}, {Gutkin}, {Vidal-Garc{\'\i}a},
  {Bouwens}, \& {Oesch}}]{sta17}
{Stark}, D.~P., {Ellis}, R.~S., {Charlot}, S., {et~al.} 2017, \mnras, 464, 469

\bibitem[{{Stark} {et~al.}(2015){Stark}, {Richard}, {Charlot}, {Cl{\'e}ment},
  {Ellis}, {Siana}, {Robertson}, {Schenker}, {Gutkin}, \& {Wofford}}]{stark15a}
{Stark}, D.~P., {Richard}, J., {Charlot}, S., {et~al.} 2015, \mnras, 450, 1846

\bibitem[{{Steidel} {et~al.}(1996){Steidel}, {Giavalisco}, {Dickinson}, \&
  {Adelberger}}]{ste96}
{Steidel}, C.~C., {Giavalisco}, M., {Dickinson}, M., \& {Adelberger}, K.~L.
  1996, \aj, 112, 352

\bibitem[{{Sun} {et~al.}(2023){Sun}, {Egami}, {Pirzkal}, {Rieke}, {Baum},
  {Boyer}, {Boyett}, {Bunker}, {Cameron}, {Curti}, {Eisenstein}, {Gennaro},
  {Greene}, {Jaffe}, {Kelly}, {Koekemoer}, {Kumari}, {Maiolino}, {Maseda},
  {Perna}, {Rest}, {Robertson}, {Schlawin}, {Smit}, {Stansberry}, {Sunnquist},
  {Tacchella}, {Williams}, \& {Willmer}}]{sun23}
{Sun}, F., {Egami}, E., {Pirzkal}, N., {et~al.} 2023, \apj, 953, 53

\bibitem[{{Tacchella} {et~al.}(2023){Tacchella}, {Johnson}, {Robertson},
  {Carniani}, {D'Eugenio}, {Kumari}, {Maiolino}, {Nelson}, {Suess},
  {{\"U}bler}, {Williams}, {Adebusola}, {Alberts}, {Arribas}, {Bhatawdekar},
  {Bonaventura}, {Bowler}, {Bunker}, {Cameron}, {Curti}, {Egami}, {Eisenstein},
  {Frye}, {Hainline}, {Helton}, {Ji}, {Looser}, {Lyu}, {Perna}, {Rawle},
  {Rieke}, {Rieke}, {Saxena}, {Sandles}, {Shivaei}, {Simmonds}, {Sun},
  {Willmer}, {Willott}, \& {Witstok}}]{tac23}
{Tacchella}, S., {Johnson}, B.~D., {Robertson}, B.~E., {et~al.} 2023, \mnras,
  522, 6236

\bibitem[{{Tang} {et~al.}(2023){Tang}, {Stark}, {Chen}, {Mason}, {Topping},
  {Endsley}, {Senchyna}, {Plat}, {Lu}, {Whitler}, {Robertson}, \&
  {Charlot}}]{tan23}
{Tang}, M., {Stark}, D.~P., {Chen}, Z., {et~al.} 2023, \mnras, 526, 1657

\bibitem[{{Topping} {et~al.}(2022){Topping}, {Stark}, {Endsley}, {Plat},
  {Whitler}, {Chen}, \& {Charlot}}]{top22}
{Topping}, M.~W., {Stark}, D.~P., {Endsley}, R., {et~al.} 2022, \apj, 941, 153

\bibitem[{{Trapp} {et~al.}(2023){Trapp}, {Furlanetto}, \& {Davies}}]{tra23}
{Trapp}, A.~C., {Furlanetto}, S.~R., \& {Davies}, F.~B. 2023, \mnras, 524, 5891

\bibitem[{{Trump} {et~al.}(2023){Trump}, {Arrabal Haro}, {Simons}, {Backhaus},
  {Amor{\'\i}n}, {Dickinson}, {Fern{\'a}ndez}, {Papovich}, {Nicholls},
  {Kewley}, {Brunker}, {Salzer}, {Wilkins}, {Almaini}, {Bagley}, {Berg},
  {Bhatawdekar}, {Bisigello}, {Buat}, {Burgarella}, {Calabr{\`o}}, {Casey},
  {Ciesla}, {Cleri}, {Cole}, {Cooper}, {Cooray}, {Costantin}, {Croton},
  {Ferguson}, {Finkelstein}, {Fujimoto}, {Gardner}, {Gawiser}, {Giavalisco},
  {Grazian}, {Grogin}, {Hathi}, {Hirschmann}, {Holwerda}, {Huertas-Company},
  {Hutchison}, {Jogee}, {Juneau}, {Jung}, {Kartaltepe}, {Kirkpatrick},
  {Kocevski}, {Koekemoer}, {Lotz}, {Lucas}, {Magnelli}, {Matharu},
  {P{\'e}rez-Gonz{\'a}lez}, {Pirzkal}, {Rafelski}, {Rose}, {Seill{\'e}},
  {Somerville}, {Straughn}, {Tacchella}, {Vanderhoof}, {Weiner}, {Wuyts},
  {Yung}, \& {Zavala}}]{tru23}
{Trump}, J.~R., {Arrabal Haro}, P., {Simons}, R.~C., {et~al.} 2023, \apj, 945,
  35

\bibitem[{{Vanzella} {et~al.}(2012){Vanzella}, {Guo}, {Giavalisco}, {Grazian},
  {Castellano}, {Cristiani}, {Dickinson}, {Fontana}, {Nonino}, {Giallongo},
  {Pentericci}, {Galametz}, {Faber}, {Ferguson}, {Grogin}, {Koekemoer},
  {Newman}, \& {Siana}}]{van12}
{Vanzella}, E., {Guo}, Y., {Giavalisco}, M., {et~al.} 2012, \apj, 751, 70

\bibitem[{{Vanzella} {et~al.}(2011){Vanzella}, {Pentericci}, {Fontana},
  {Grazian}, {Castellano}, {Boutsia}, {Cristiani}, {Dickinson}, {Gallozzi},
  {Giallongo}, {Giavalisco}, {Maiolino}, {Moorwood}, {Paris}, \&
  {Santini}}]{van11}
{Vanzella}, E., {Pentericci}, L., {Fontana}, A., {et~al.} 2011, \apjl, 730, L35

\bibitem[{{Verhamme} {et~al.}(2015){Verhamme}, {Orlitov{\'a}}, {Schaerer}, \&
  {Hayes}}]{ver15}
{Verhamme}, A., {Orlitov{\'a}}, I., {Schaerer}, D., \& {Hayes}, M. 2015, \aap,
  578, A7

\bibitem[{{Verhamme} {et~al.}(2017){Verhamme}, {Orlitov{\'a}}, {Schaerer},
  {Izotov}, {Worseck}, {Thuan}, \& {Guseva}}]{ver17}
{Verhamme}, A., {Orlitov{\'a}}, I., {Schaerer}, D., {et~al.} 2017, \aap, 597,
  A13

\bibitem[{{Verhamme} {et~al.}(2006){Verhamme}, {Schaerer}, \&
  {Maselli}}]{verhamme06}
{Verhamme}, A., {Schaerer}, D., \& {Maselli}, A. 2006, \aap, 460, 397

\bibitem[{{Weinberger} {et~al.}(2019){Weinberger}, {Haehnelt}, \&
  {Kulkarni}}]{wei19}
{Weinberger}, L.~H., {Haehnelt}, M.~G., \& {Kulkarni}, G. 2019, \mnras, 485,
  1350

\bibitem[{{Willott} {et~al.}(2015){Willott}, {Carilli}, {Wagg}, \&
  {Wang}}]{wil15}
{Willott}, C.~J., {Carilli}, C.~L., {Wagg}, J., \& {Wang}, R. 2015, \apj, 807,
  180

\bibitem[{{Witstok} {et~al.}(2023){Witstok}, {Smit}, {Saxena}, {Jones},
  {Helton}, {Sun}, {Maiolino}, {Kumari}, {Stark}, {Bunker}, {Arribas}, {Baker},
  {Bhatawdekar}, {Boyett}, {Cameron}, {Carniani}, {Charlot}, {Chevallard},
  {Curti}, {Curtis-Lake}, {Eisenstein}, {Endsley}, {Hainline}, {Ji}, {Johnson},
  {Looser}, {Nelson}, {Perna}, {Rix}, {Robertson}, {Sandles}, {Scholtz},
  {Simmonds}, {Tacchella}, {{\"U}bler}, {Williams}, {Willmer}, \&
  {Willott}}]{witstok23}
{Witstok}, J., {Smit}, R., {Saxena}, A., {et~al.} 2023, arXiv e-prints,
  arXiv:2306.04627

\bibitem[{{Witten} {et~al.}(2023){Witten}, {Laporte}, {Martin-Alvarez},
  {Sijacki}, {Yuan}, {Haehnelt}, {Baker}, {Dunlop}, {Ellis}, {Grogin},
  {Illingworth}, {Katz}, {Koekemoer}, {Magee}, {Maiolino}, {McClymont},
  {P{\'e}rez-Gonz{\'a}lez}, {Puskas}, {Roberts-Borsani}, {Santini}, \&
  {Simmonds}}]{wit23}
{Witten}, C., {Laporte}, N., {Martin-Alvarez}, S., {et~al.} 2023, arXiv
  e-prints, arXiv:2303.16225

\bibitem[{{Yang} {et~al.}(2017){Yang}, {Malhotra}, {Gronke}, {Rhoads},
  {Leitherer}, {Wofford}, {Jiang}, {Dijkstra}, {Tilvi}, \& {Wang}}]{yang17}
{Yang}, H., {Malhotra}, S., {Gronke}, M., {et~al.} 2017, \apj, 844, 171

\bibitem[{{Zackrisson} {et~al.}(2013){Zackrisson}, {Inoue}, \&
  {Jensen}}]{zac13}
{Zackrisson}, E., {Inoue}, A.~K., \& {Jensen}, H. 2013, \apj, 777, 39

\end{thebibliography}

\begin{appendix}

\section{1D spectra of LAEs}
\label{app:a}
\noindent\begin{minipage}{\textwidth}
\centering
    \includegraphics[width=\linewidth]{figs/spec_1d/021842_1D_combined.png}
    \includegraphics[width=\linewidth]{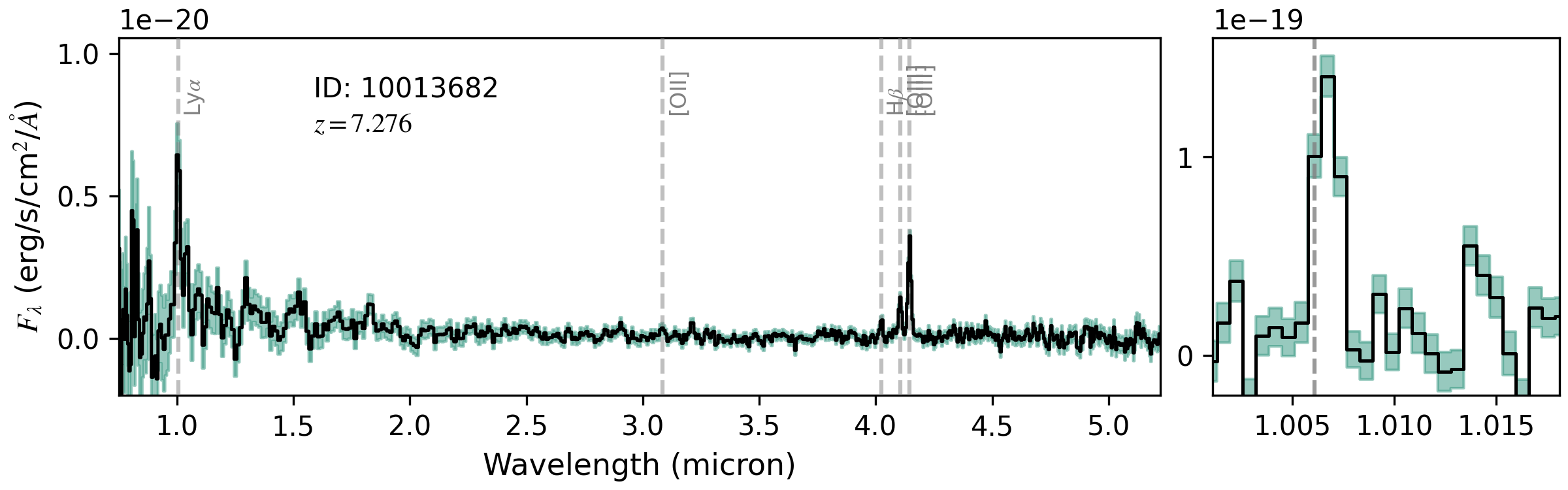}
    \includegraphics[width=\linewidth]{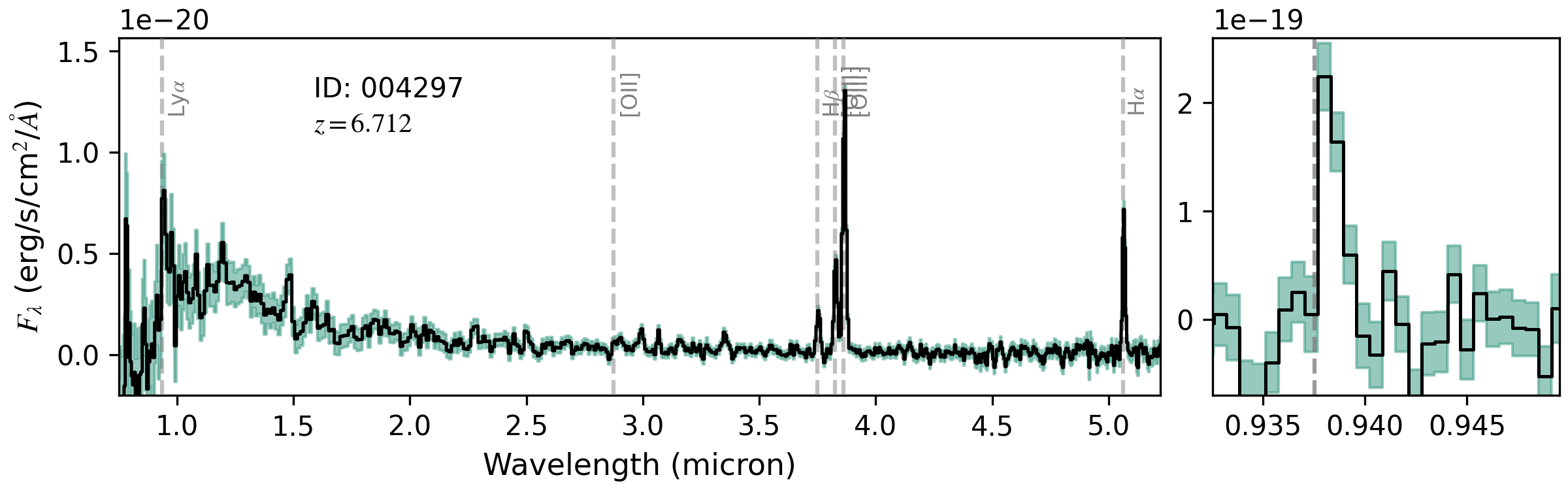}
    \includegraphics[width=\linewidth]{figs/spec_1d/016625_1D_combined.png}
    \captionof{figure}{1D spectra of LAEs from PRISM (R100) with a zoom-in on \lya\ emission from G140M (R1000).}
    \label{fig:1dspectra}
\end{minipage}

\begin{figure*}
    \centering 
    \ContinuedFloat
    \includegraphics[width=\linewidth]{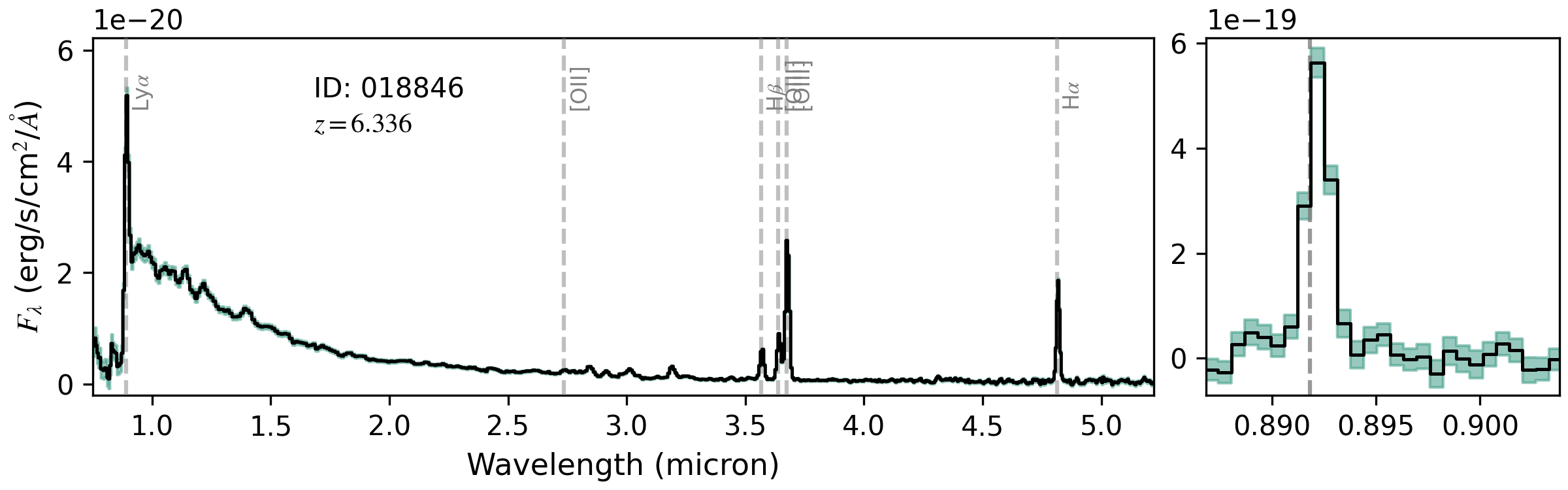}
    \includegraphics[width=\linewidth]{figs/spec_1d/019342_1D_combined.png}
    \includegraphics[width=\linewidth]{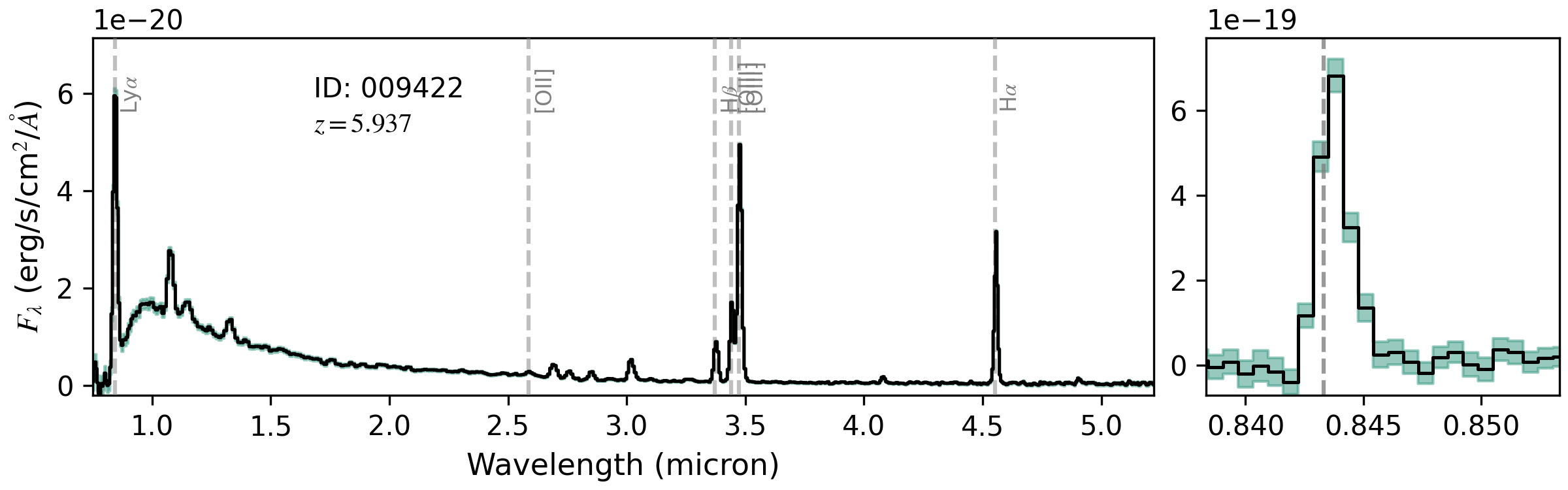}
    \includegraphics[width=\linewidth]{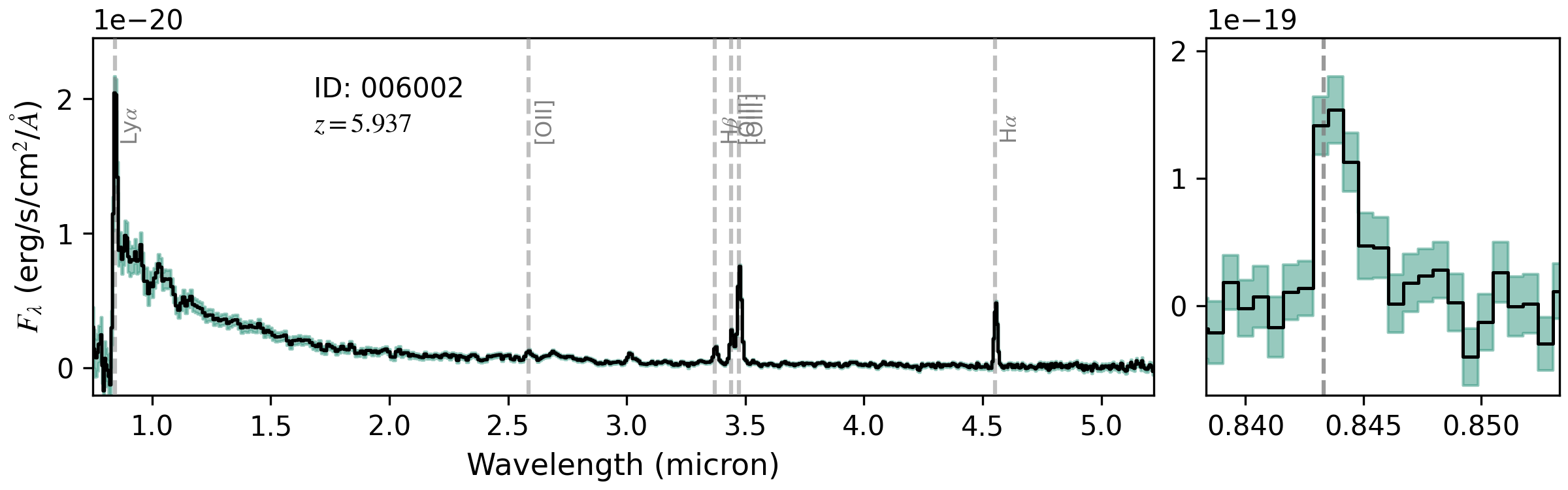}
    \caption[]{continued.}
\end{figure*}

\begin{figure*}
    \centering 
    \ContinuedFloat
    \includegraphics[width=\linewidth]{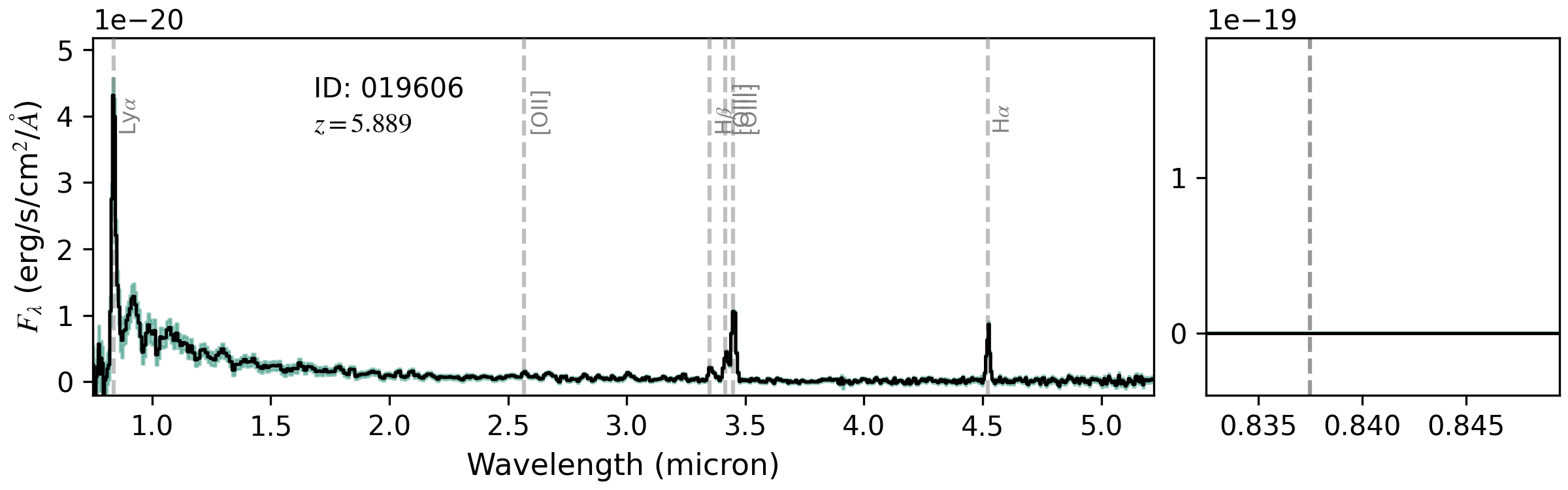}
    \includegraphics[width=\linewidth]{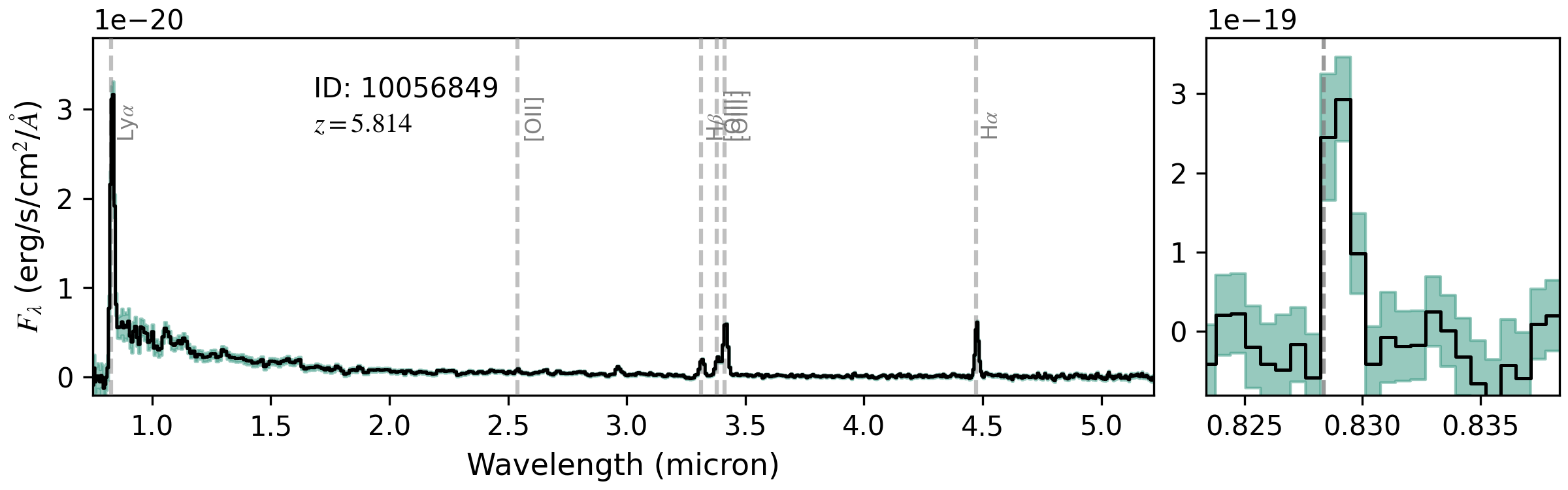}
    \includegraphics[width=\linewidth]{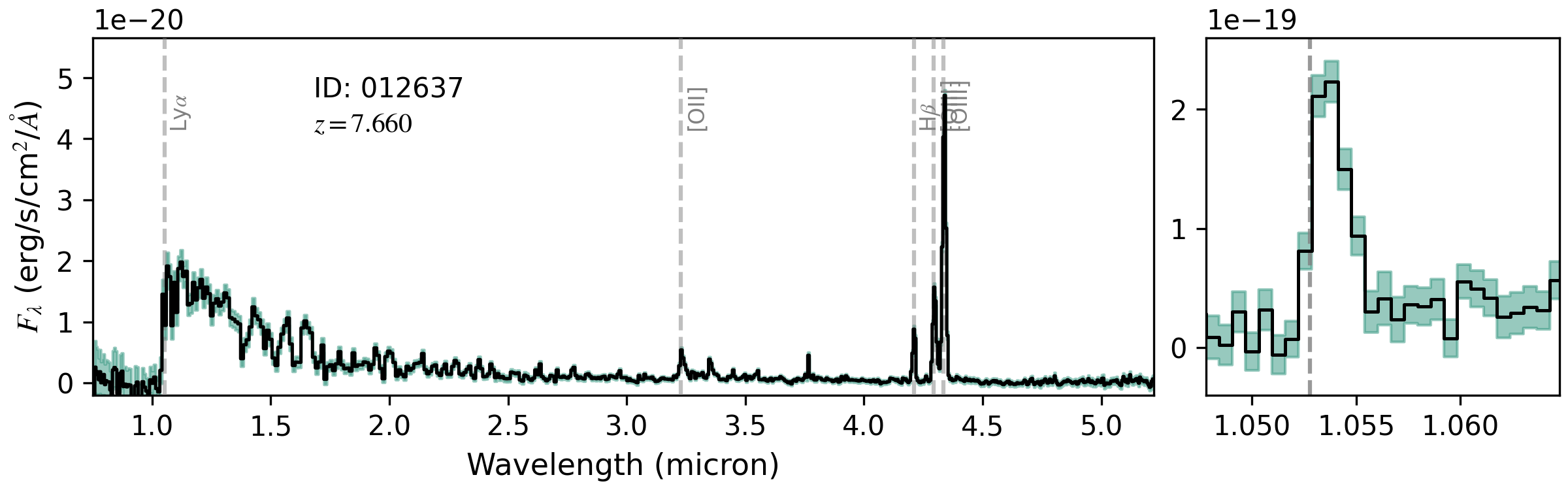}
    \includegraphics[width=\linewidth]{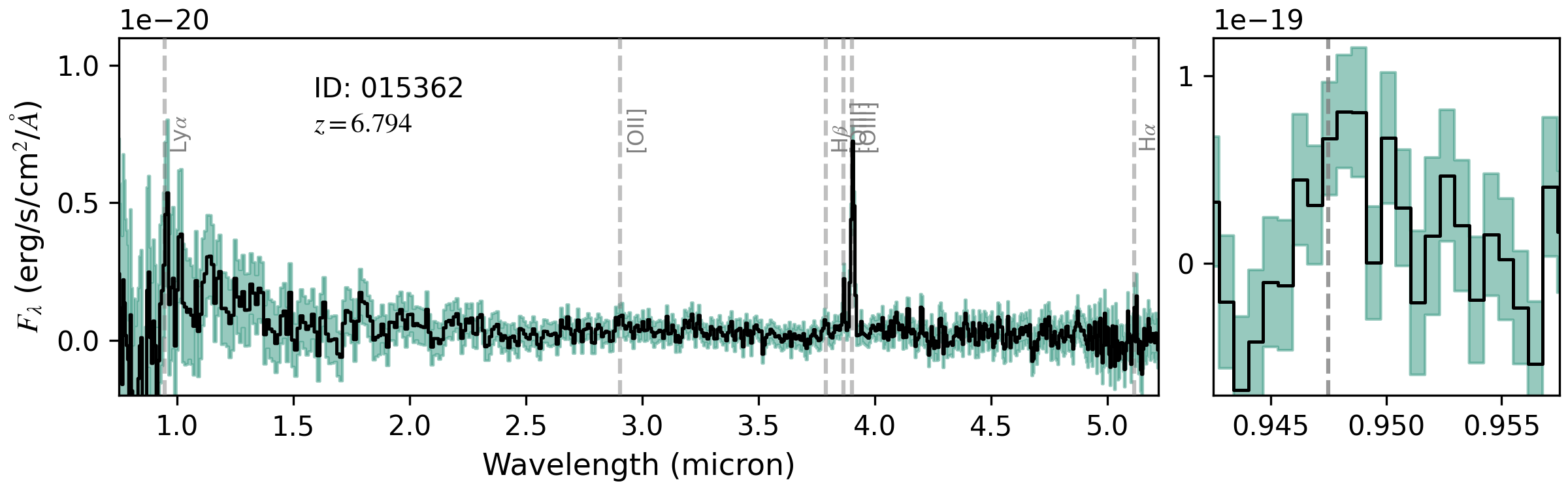}
    \caption[]{continued.}
\end{figure*}

\begin{figure*}
    \centering 
    \ContinuedFloat
    \includegraphics[width=\linewidth]{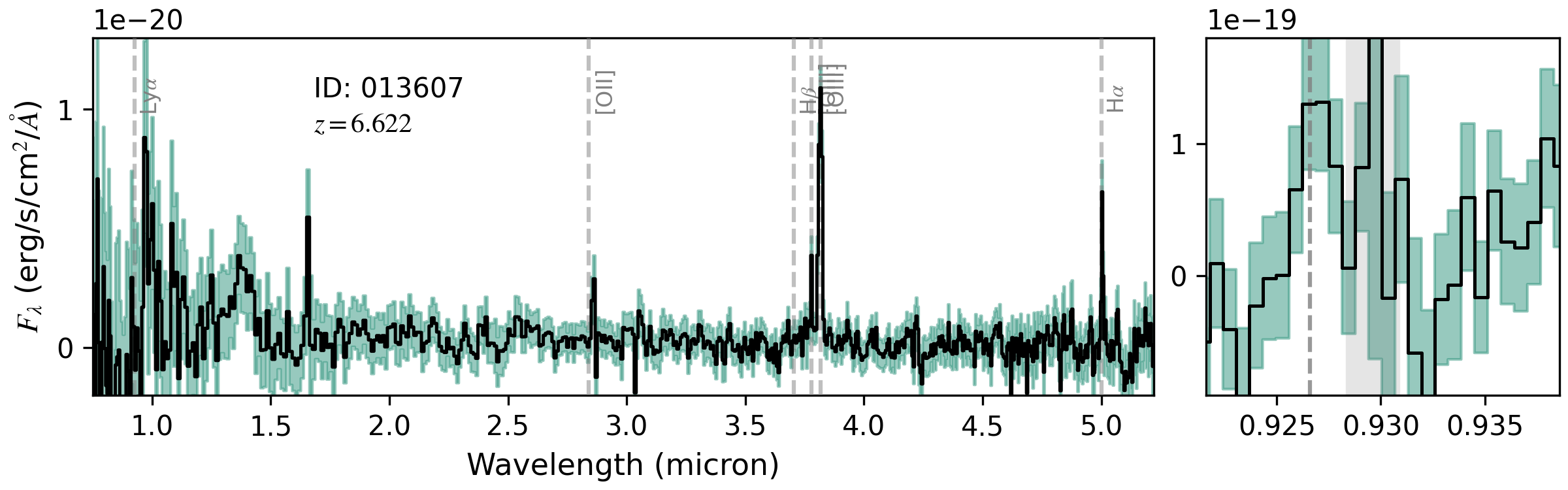}
    \includegraphics[width=\linewidth]{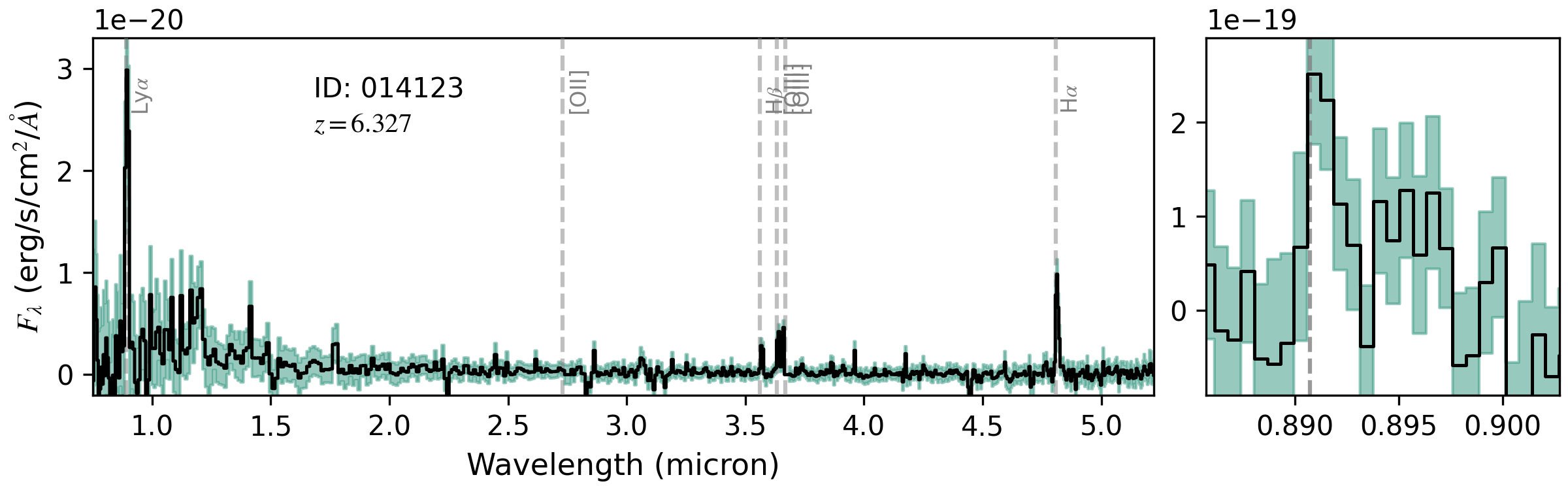}
    \includegraphics[width=\linewidth]{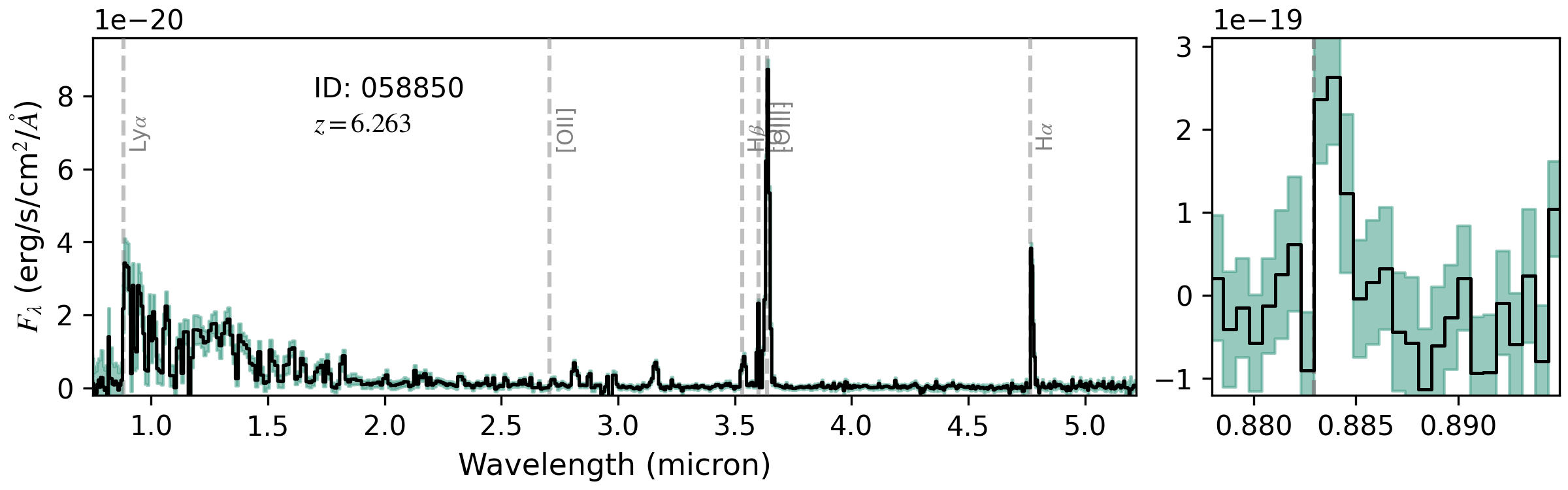}
    \includegraphics[width=\linewidth]{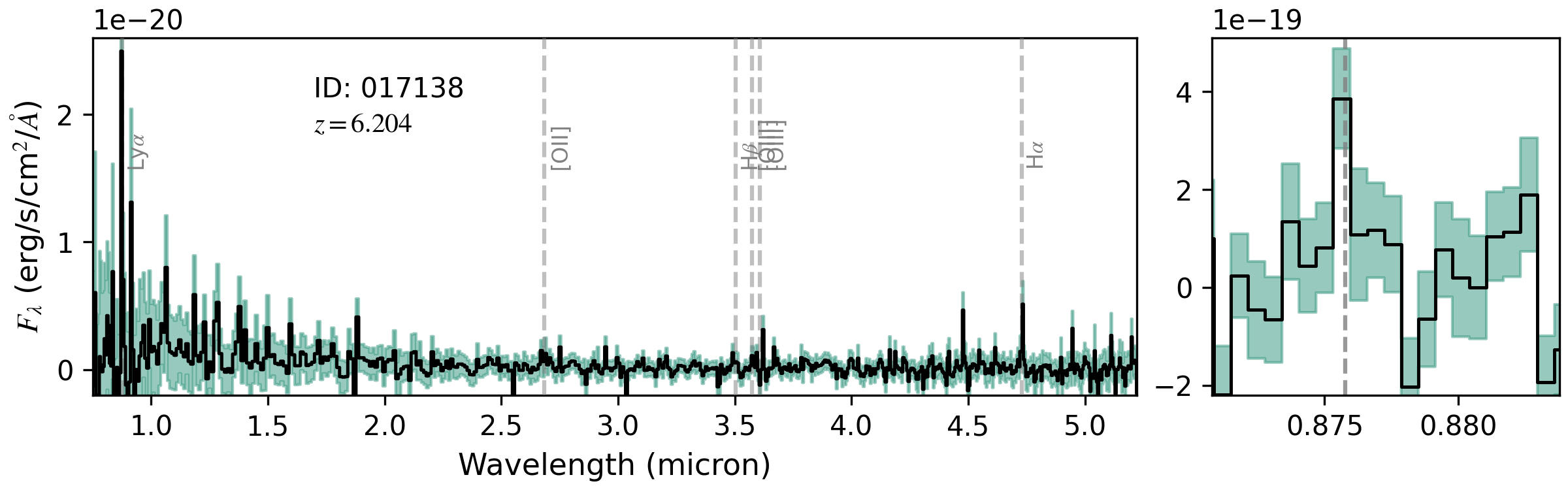}
    \caption[]{continued.}
\end{figure*}

\begin{figure*}
    \centering 
    \ContinuedFloat
    \includegraphics[width=\linewidth]{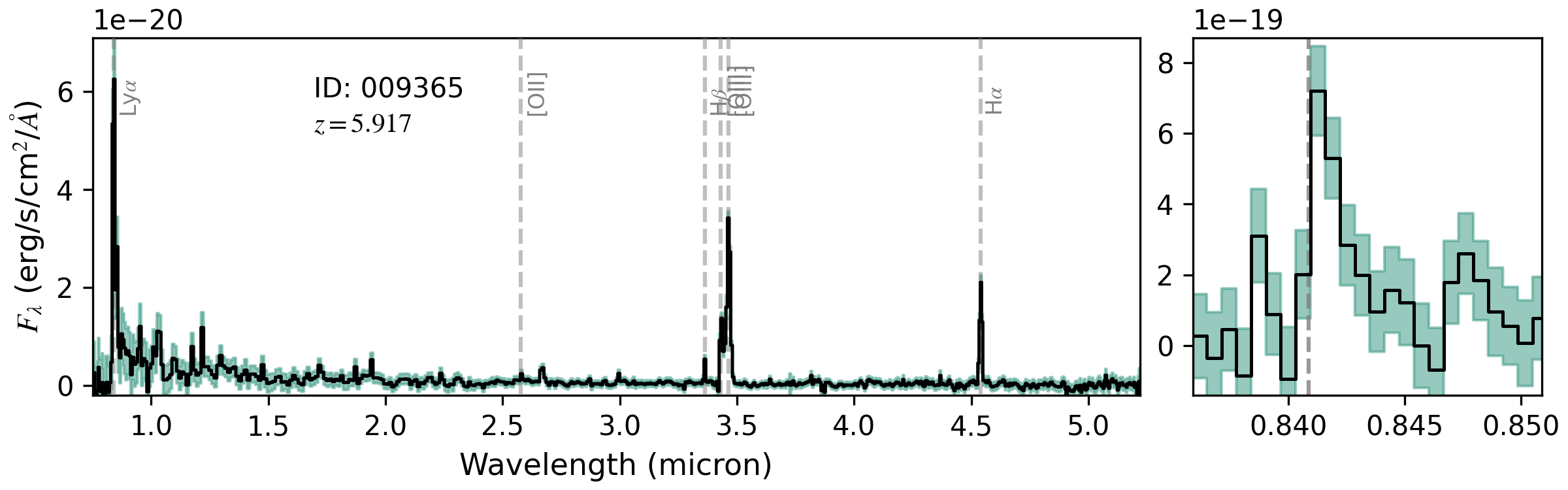}
    \caption[]{continued.}
\end{figure*}

\end{appendix}

\end{document}